\documentclass{aastex}
\usepackage{emulateapj5}
\usepackage{apjfonts}
\usepackage{epsfig}

\shorttitle{Parsec Scale Variability}
\shortauthors{Homan et al.}

\submitted{Accepted by the Astrophysical Journal}

\begin{document}

\title{Parsec-Scale Blazar Monitoring: Flux and Polarization Variability}

\author{Daniel C. Homan\altaffilmark{1}, Roopesh Ojha\altaffilmark{2}, 
John F. C. Wardle\altaffilmark{3}, and David H. Roberts\altaffilmark{3},}
\affil{Physics Department MS057, Brandeis University}
\affil{Waltham, MA 02454} 
\author{Margo F. Aller\altaffilmark{4}, Hugh D. Aller\altaffilmark{4}, and 
Philip A. Hughes\altaffilmark{4}}
\affil{Radio Astronomy Observatory, University of Michigan}
\affil{Ann Arbor, MI 48109} 
\altaffiltext{1}{Current Address: National Radio Astronomy Observatory, 
                 Charlottesville VA 22901; dhoman@nrao.edu}
\altaffiltext{2}{Current Address: CSIRO-ATNF, Marsfield, NSW 2122, Australia; 
                 roopesh.ojha@atnf.csiro.au}
\altaffiltext{3}{wardle,roberts@brandeis.edu}
\altaffiltext{4}{margo,hugh,hughes@astro.lsa.umich.edu}

\begin{abstract}
We present analysis of the flux density and polarization variability of 
parsec scale radio jets from a dual-frequency, six-epoch, VLBA 
polarization experiment monitoring 12 blazars.  The 
observations were made at 15 and 22 GHz at bimonthly intervals 
over 1996.  Here we analyze the flux density, fractional polarization, and
polarization position angle behavior of core regions and jet 
features, considering both the linear trends of these 
quantities with time and more rapid fluctuations about the 
linear trends.  The dual frequency nature of the observations
allows us to examine spectral evolution, to separate Faraday effects
from changes in magnetic field order, and also to deduce empirical 
estimates for the uncertainties in measuring properties 
of VLBI jet features (see the Appendix).  Our main results
include the following:

On timescales $\gtrsim 1$ year, we find that jet features
generally decayed in flux, with older features decaying more
slowly than younger features.  Using the decay rates of jet
features from six sources, we find $I \propto R^{-1.3\pm0.1}$. 
Short term fluctuations in flux
tended to be fractionally larger in core regions than in jet 
features, with the more compact core regions having the larger 
fluctuations. We find significant spectral index changes in the 
core regions of four sources.  Taken together these are consistent 
with an outburst-ejection cycle for new jet components.  Jet 
features from one source showed a significant spectral flattening 
over time.

Jet features either increased in fractional polarization with time 
or showed no significant change, with the smallest observed changes 
in the features at the largest projected radii.  Increasing magnetic field
order explains most of the increasing fractional polarization we 
observed. Only in the case of 3C273 is there evidence of a
feature emerging from behind a Faraday depolarizing screen.

We find a number of significant polarization angle rotations
including two very large ($\gtrsim 180^\circ$) rotations in the 
core regions of OJ287 and J1512$-$09.  In general, polarization angle 
changes were of the same magnitude at both observing bands and 
cannot be explained by Faraday rotation.  The 
observed polarization angle changes most likely reflect underlying 
changes in magnetic field structure.  In jet features, four
of the five observed rotations were in the direction of aligning
the magnetic field with the jet axis, and coupled with the
tendency of jet features to show a fractional polarization 
increase, this suggests increasing longitudinal field order.  
\end{abstract}


\section{Introduction}
\label{s:intro}

We have used the National Radio Astronomy Observatory's (NRAO) 
Very Long Baseline Array (VLBA)\footnote{The National Radio 
Astronomy Observatory is a facility of the National Science Foundation
operated under cooperative agreement by Associated Universities, Inc.} 
to monitor the parsec scale radio jets of twelve active galactic 
nuclei (AGN) in both total intensity and linear polarization.  The 
observations were taken at approximately two month intervals during 1996
at 15 and 22 GHz, and were designed to follow closely the parsec
scale evolution of some of the most currently active sources being monitored
by the University of Michigan Radio Astronomy Observatory (UMRAO).
In \citet{H01}, hereafter Paper I, we published proper motion analysis 
from this monitoring program.  Other results from the program have been 
presented elsewhere (e.g. \citet{WHOR98, HW99, HW00}), and others are in 
preparation. In particular, the entire data-set in the form of images, 
tabular data, and plots of core and jet properties versus time 
will be presented in Ojha et al. (in preparation, hereafter Paper III). 

In this paper we analyze the flux density (hereafter abbreviated
``flux'') and polarization evolution of the
parsec scale radio cores and jets in our monitoring program.  While 
many authors have made detailed studies of the kinematics of jet features 
(e.g. \citet{VC94}), much less attention has been paid to monitoring
their flux and (even more rarely) polarization behavior.
Where such studies have been made, they have focused on individual
objects such as 3C\,273 \citep{U85}, 3C\,345 \citep{BRW94,LZ99},
and 3C\,120 \citep{G00}.  This is in stark contrast to single 
dish monitoring programs, such as that at the UMRAO \citep{AALH85,AAHL99} 
which has tracked the integrated flux and polarization evolution of scores 
of AGN for decades at weekly or bi-weekly intervals.  With the VLBA it is 
now possible to monitor a large number of parsec scale jets at closely 
spaced, regular intervals, and here we present the first analysis of data 
of this type for flux and polarization variability.       

In this paper we focus on the overall variability properties of our
sample to find common trends in the flux and polarization evolution 
of core regions and jet features.
Section \S{\ref{s:obs}} describes our sample, data reduction, and 
model-fitting procedures. Statistical methods used in analyzing the variability
properties of our VLBI observations are discussed in \S{\ref{s:tech}}. 
We present and discuss our results in \S{\ref{s:res}} and summarize 
them in \S{\ref{s:conclude}}. The Appendix explores a by-product of 
our variability analysis -- specifically, empirical estimates of the
uncertainties in the measurement of VLBI component properties. 

In all calculations presented here, we assume a universe with $\Omega_M=0.3$, 
$\Omega_\Lambda=0.7$ and $H_{0}=70$ km s$^{-1}$ Mpc$^{-1}$.  For spectral 
index we follow the convention, $S_\nu \propto \nu^{+\alpha}$.

%%%%%%%%%%%%%
%
%  References \citet{KEY} puts a citation directly  into the text
%             \citep{KEY} puts a parenthetical citation
%
%
%%%%%%%%%%%%

\section{Observations}
\label{s:obs}

\subsection{The Sample}
We used the VLBA to conduct a series of six observations, each of 24
hour duration, at close to two month intervals during the year
1996. The observations were made at 15 GHz ($\lambda$2.0 cm, U-band) 
and 22 GHz ($\lambda$1.3 cm, K-band).
We observed 11 target sources for six epochs and one (J1224+21) for 
only the last five epochs.  These sources are listed in table \ref{t:Sources}.
The epochs of observation during 1996 were the following: January 19th (1996.05),
March 22nd (1996.23), May 27th (1996.41), July 27th (1996.57), September 27th 
(1996.74), and December 6th (1996.93). 

%\placetable{t:Sources}

The sources were chosen from those regularly monitored by the
University of Michigan Radio Astronomy Observatory (UMRAO) in total
intensity and polarization at $4.8$, $8.0$, and $14.5$ GHz.  They were
selected according to the following criteria. (1) High 
total intensity: The weakest sources are about 1 Jy, the most powerful
as much as 22 Jy. (2) High polarized flux: Typically over 50 mJy.
(3) Violently variable: In both total and polarized intensity. Such 
sources are likely to be under-sampled by annual VLBI. 
(4) Well distributed in right ascension: This 
allowed us to make an optimal observing schedule.   
 
Most of the UMRAO sources meet the first three of the above
criteria. The 12 actually selected were the strongest, most violently 
variable sources, subject to the fourth criteria. Clearly these sources
do not comprise a ``complete sample'' in any sense.  

\subsection{Data Calibration} 
\label{s:reduce}

The frequency agility and high slew speeds of the VLBA antennas were used to 
schedule our observations to generate maximal (u,v)-coverage. Scan lengths were
kept short (13 minutes for the first two epochs and 5.5 minutes for the last
four), with a switch in frequency at the end of each scan. In addition, 
scans of neighboring sources were heavily interleaved at the cost of some additional 
slew time.  Each source was observed for approximately 45 minutes per 
frequency at each epoch.  

The data were correlated on the VLBA correlator in
Socorro, NM.  After correlation, the data were distributed on DAT tape to 
Brandeis University where they were loaded into NRAO's Astronomical Imaging 
Processing System (AIPS) \citep{BG94,G88} and calibrated using standard 
techniques for VLBI polarization observations, e.g., \citep{C93,RWB94}.  
For a detailed description of our calibration steps see Paper III.

One point that bears mentioning here is our calibration of the polarization
position angle (also called the Electric Vector Position Angle or EVPA).
Our EVPAs were set at each epoch and at both 15 and 22 GHz by aligning
the strong jet component, U1 (K1), in 3C\,279 to an angle of $67^\circ$.
This orientation is roughly parallel with the structural position angle for this
component and is within $5^\circ$ of the independently calibrated 
observations of \citet{LZD95}, \citet{T98}, and \citet{HW00} whose 
epochs of observation bracket our own.  By examining the other sources
in our sample, we see no evidence that this component in 3C\,279 has 
significant Faraday rotation at these frequencies or varies in EVPA during 
our observations. As a result of this EVPA calibration procedure, our internal 
consistency between epochs is very good with uncertainties 
$\approx 2-3$ degrees on the most robust jet features (see the Appendix).

\subsection{Comparison to Single Dish Monitoring}

\begin{figure}
\figurenum{1}
\begin{center}
\epsfig{file=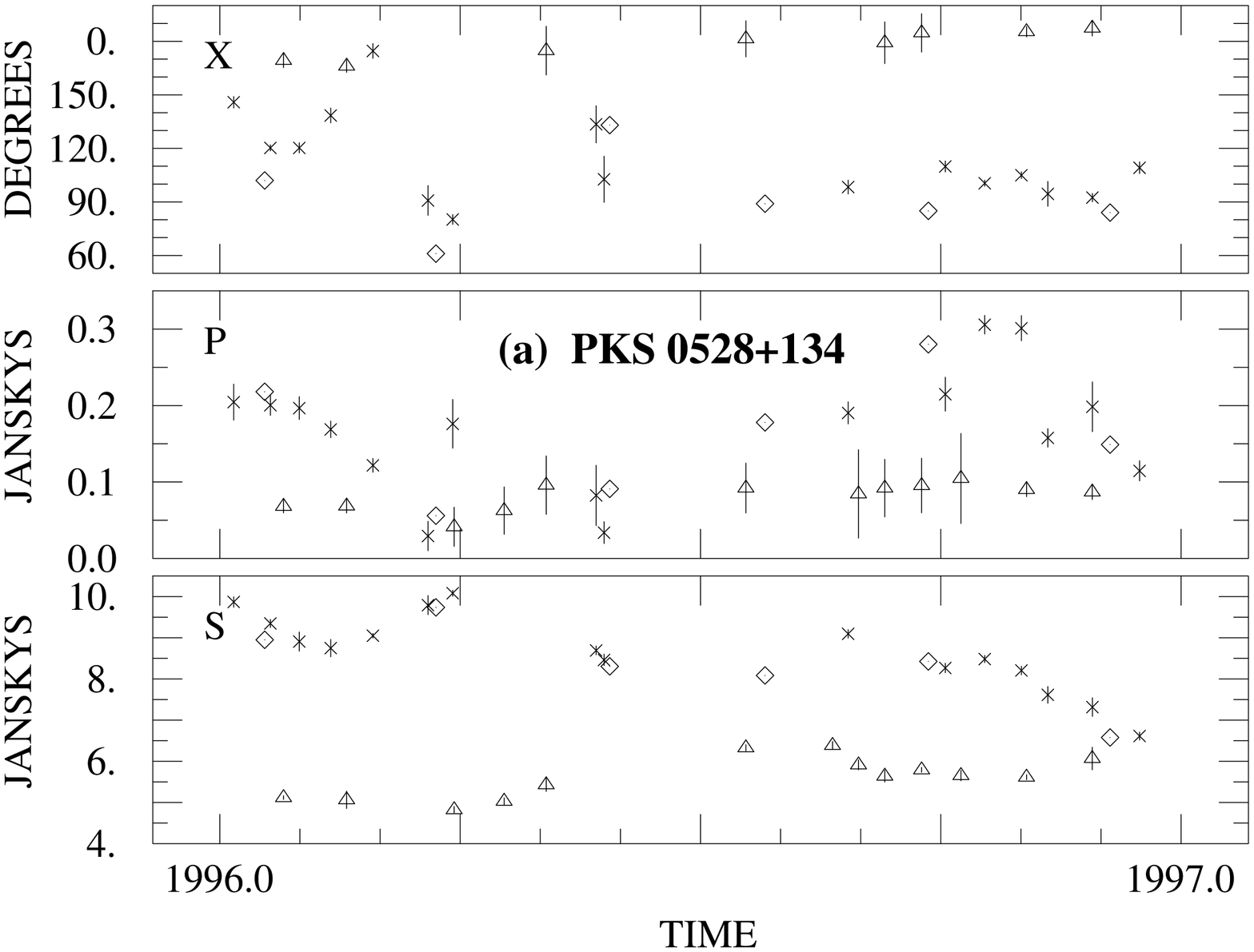,width=3.5in,angle=-90}
\epsfig{file=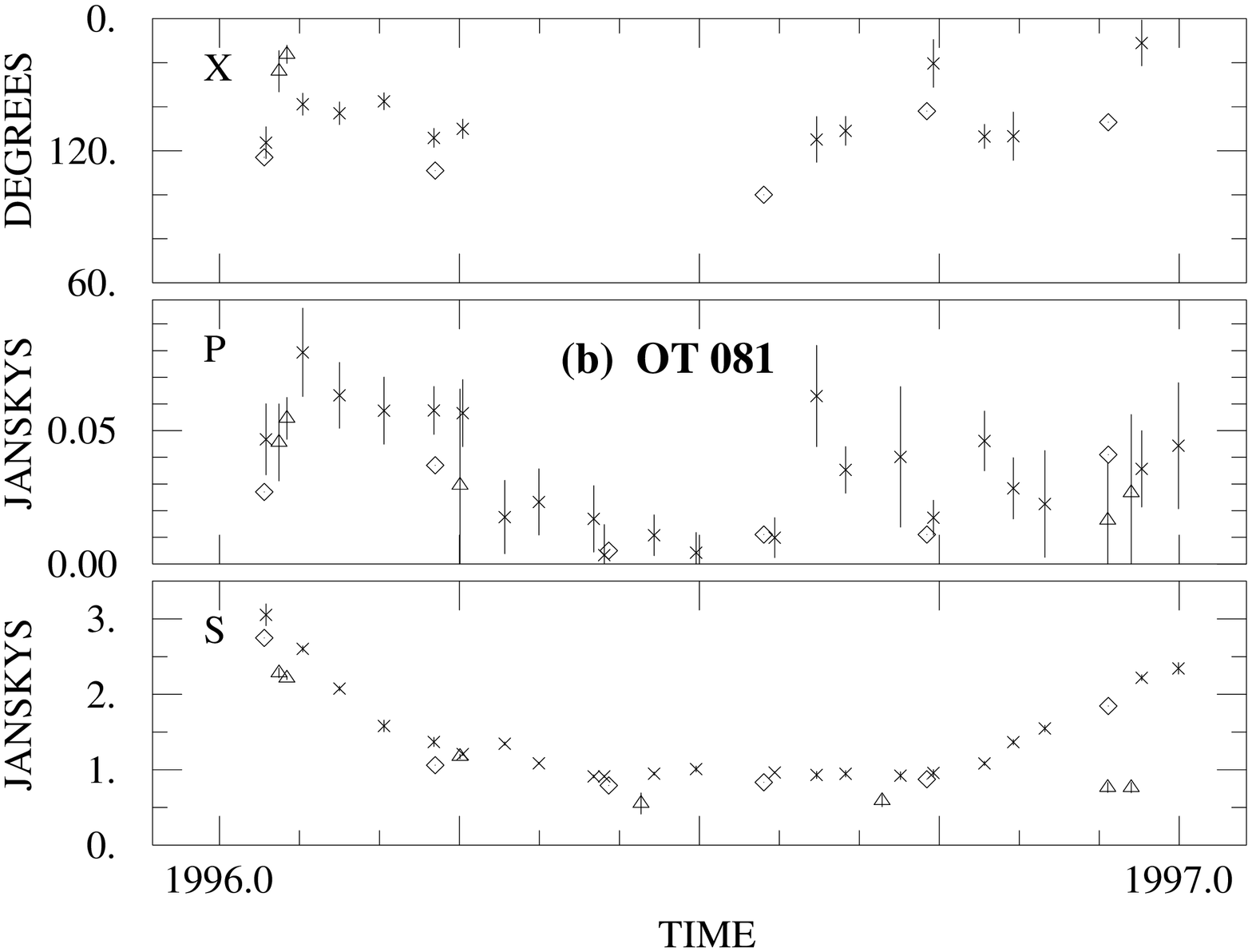,width=3.5in,angle=-90}
\end{center}
\figcaption[f1a.eps,f1b.eps]{\label{f:compare}
Comparison of single dish monitoring by the UMRAO at
14.5 GHz ($\times$ symbols) to the integrated VLBI results 
at 15 GHz (open diamonds).  Total intensity (I), polarized
intensity (P), and polarization position angle (X) are 
plotted for two compact sources, PKS 0528$+$134 (J0530$+$14)
in panel (a), and OT 081 (J1751$+$09) in panel (b). 
UMRAO 5 GHz results (open triangles) are included to 
illustrate opacity effects.
}
\end{figure}

Figure \ref{f:compare} compares our 15 GHz VLBA observations of 
J0530$+$13 (PKS 0528$+$134) and J1751$+$09 (OT 081) to the
single dish monitoring done by the UMRAO at 14.5 GHz.  The
total fluxes and polarizations from our CLEAN maps agree quite
well with the independent UMRAO results in all six of the 
VLBI epochs.  The two sources plotted in figure \ref{f:compare} 
are quite compact, and the VLBA observations account for nearly all
of the single dish flux.  Although a detailed comparison
is difficult, on the more extended sources, such as 3C\,120 
and 3C\,273, we see a nearly constant offset between the 
VLBA and single dish monitoring, and the VLBI core and jet 
seem to account for essentially all of the observed variability
in the single dish monitoring.  Given this, it may be possible
to use frequent single dish monitoring to help
interpolate between the more widely spaced VLBI epochs to
follow the evolution of parsec-scale core and jet features. 

%\placefigure{f:compare}

\subsection{Modeling the data}
\label{s:model}

To parameterize our data for quantitative analysis, we used the model fitting 
capabilities of the DIFMAP software package \citep{SPT94,SPT95} to fit 
the sources with a number of discrete Gaussian components. The fitting was 
done directly on the final, self-calibrated visibility data (i.e., in 
the (u,v)-plane).  

Our procedures for fitting in total intensity (Stokes $I$) 
were described in Paper I.  We fit the polarization in Stokes $Q$ and $U$
by fixing the locations (and sizes) of the $I$ components and allowing the 
fluxes to vary.  This procedure for fitting the polarization forces coincidence
with the $I$ components and does not account for cases where the polarization
may be displaced from the total intensity.  While a close inspection of the CLEAN
images showed a number of cases with small displacements between total
intensity and polarization peaks, our fitting procedure seemed insensitive to 
these and, in general, produced good agreement with the polarized fluxes and
position angles observed in our CLEAN images.  Our full model-fits for each
source will appear in Paper III.  

As discussed in \S{\ref{s:tech}}, obtaining good estimates of the
real ``1 $\sigma$'' uncertainties of VLBI model-fit parameters is a
difficult problem.  Our analysis procedure was designed to extract
robust variability results without a-priori knowledge of these
uncertainties, and a by-product of the analysis was an empirical 
estimate of the accuracy with which we can measure the flux and
polarization of jet features at a single epoch.  These estimates
are presented and discussed in the Appendix.

\subsubsection{Jet Features}
\label{s:sources}

As we noted in Paper I, modeling a jet with Gaussian components
works best with sources dominated by discrete, well separated
structures.  The relative flux densities, positions, and dimensions of
the Gaussian components that make up a fit can be strongly correlated,
particularly when jet features are closely spaced or poorly defined.  
In that paper, our primary concern was deciding which components
were reliable tracers of the motion of jet structures.  Here we are
interested in the components that can be robustly analyzed for
flux and polarization variability.  

For this analysis we discovered that few components have
sufficiently well modeled fluxes in every epoch to be treated 
independently.  Even components that were highly accurate tracers
of jet motion often had one or two epochs (at one or both frequencies)
where their polarizations and/or total intensities were clearly biased 
by a near neighbor.  With only six epochs of observation, one or two
``poorly modeled'' epochs would severely limit our analysis.  To address
this, we have simply summed together the fluxes from closely spaced 
components and performed our analysis on these summed fluxes.  This 
procedure gave us a collection of core and jet {\em features} which 
consist of one or more of our Gaussian model-fit components. 

%\placetable{t:comp}

The features we have chosen for analysis are listed in 
table \ref{t:comp} and illustrated 
below.  To be considered for analysis, a feature must have 
existed at both frequencies for at least five epochs.  For 
polarization analysis, we required the 
polarized flux to be $\geq 10$ mJy and $\geq 0.5$\% of the total 
intensity for at least five epochs at both frequencies.  Of course, a 
feature must represent (to the best of our estimation) the same physical 
structure or pattern from epoch to epoch.  In the case of core 
features, this means the component representing the base of the jet plus
any nearby components whose fluxes cannot be reliably separated out.

In some cases, it is clear that ``sub-components'' of 
the features we analyze behave quite
differently; the core regions of J0530$+$13 and 3C\,279 are
excellent examples of this.  Both of these quasars have
a jet component close to (and emerging from, see Paper I)
the core. While these jet components seem to have distinct 
evolutions, we cannot confidently separate their flux and 
polarization behavior from that of the nearby core.  
We chose a robust analysis that treats 
each core and jet feature on equal footing.  In general,
the core regions we study are a combination of the 
'base-of-the-jet' plus a nearby stationary or emerging 
jet feature, and in this way J0530$+$13 and 3C\,279 are
no different than the other sources in our sample.
 
\begin{figure}
\figurenum{2}
\begin{center}
\epsfig{file=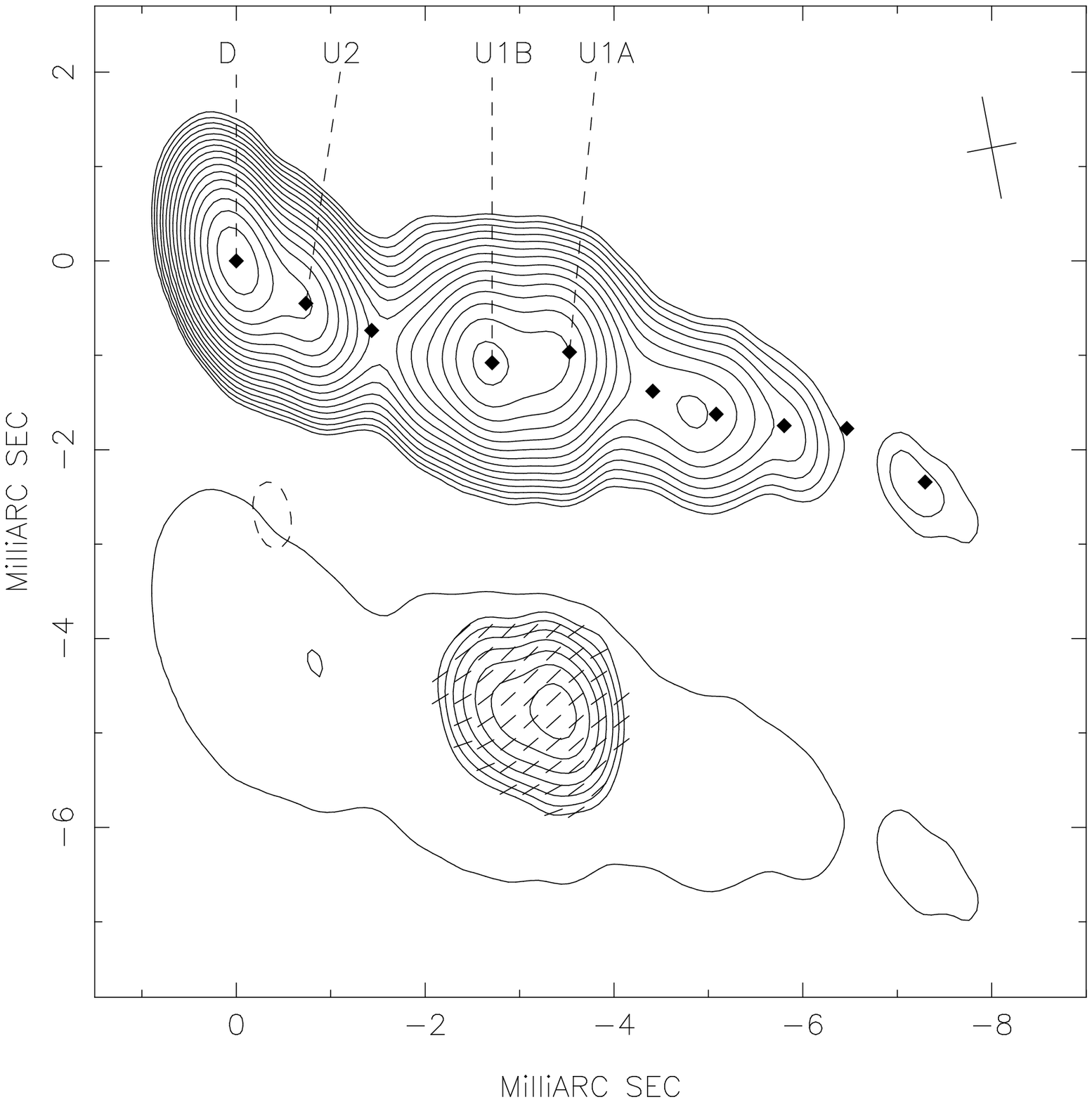,width=3.5in,angle=-90}
\end{center}
\figcaption[f2.eps]{\label{f:3c120} Total intensity and polarization images of
3C\,120 in epoch 1996.57 at 15 GHz.  The total intensity map has a peak flux
of $0.73$ Jy/beam with contour levels starting at $0.003$ Jy/beam and 
increasing in steps of $\times\sqrt{2}$.  The polarization image has a peak 
flux of $0.029$ Jy/beam with contour levels starting at $0.003$ Jy/beam
and increasing in steps of $\times\sqrt{2}$.  Tick marks represent the
polarization position angle.  A single $I$ contour is drawn around the $P$
to show registration.
For variability analysis, 
we consider two jet features: (1) the core region, consisting of the 
components D$+$U2 (D$+$K3$+$K2 at 22 GHz), and (2) the strong jet feature 
consisting of U1A$+$U1B (K1A$+$K1B).  This feature moves with a mean 
apparent speed of $4.3$ times the speed of light (Paper I);
}
\end{figure}

\begin{figure}
\figurenum{3}
\begin{center}
\epsfig{file=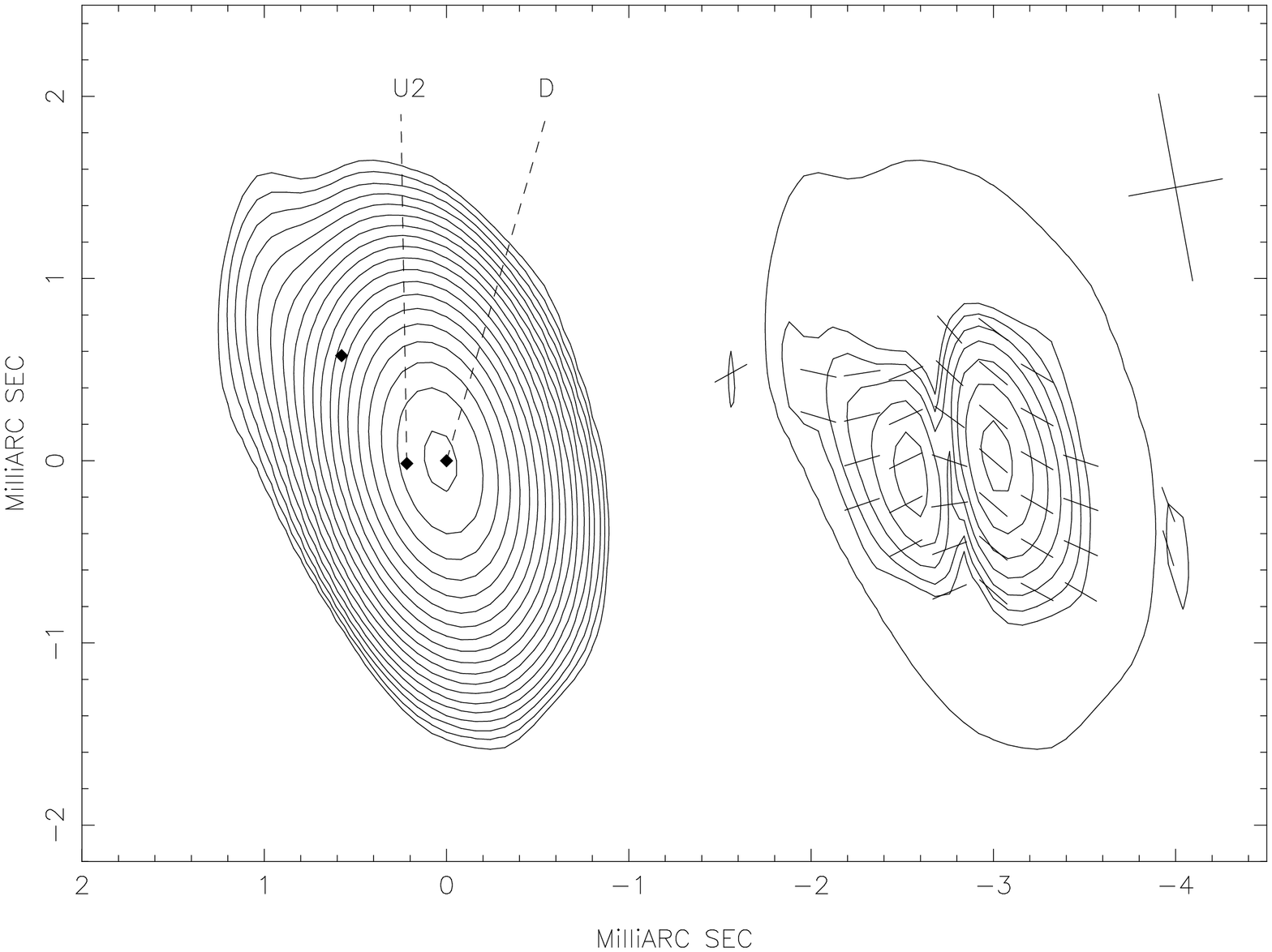,width=2.5in,angle=-90}
\end{center}
\figcaption[f3.eps]{\label{f:j0530} Total intensity and polarization images of
J0530$+$13 in epoch 1996.23 at 15 GHz.  The total intensity map has a peak flux
of $8.20$ Jy/beam with contour levels starting at $0.015$ Jy/beam and 
increasing in steps of $\times\sqrt{2}$.  The polarization image has a peak 
flux of $0.071$ Jy/beam with contour levels starting at $0.008$ Jy/beam
and increasing in steps of $\times\sqrt{2}$.  Tick marks represent the
polarization position angle.  A single $I$ contour is drawn around the $P$
to show registration.
For variability analysis, we consider only the core region, 
consisting of D$+$U2 (D$+$K2 at 22 GHz).
The competing polarizations of D and U2 (K2)
are the most distinct in the displayed epoch.  
}
\end{figure}

\begin{figure}
\figurenum{4}
\begin{center}
\epsfig{file=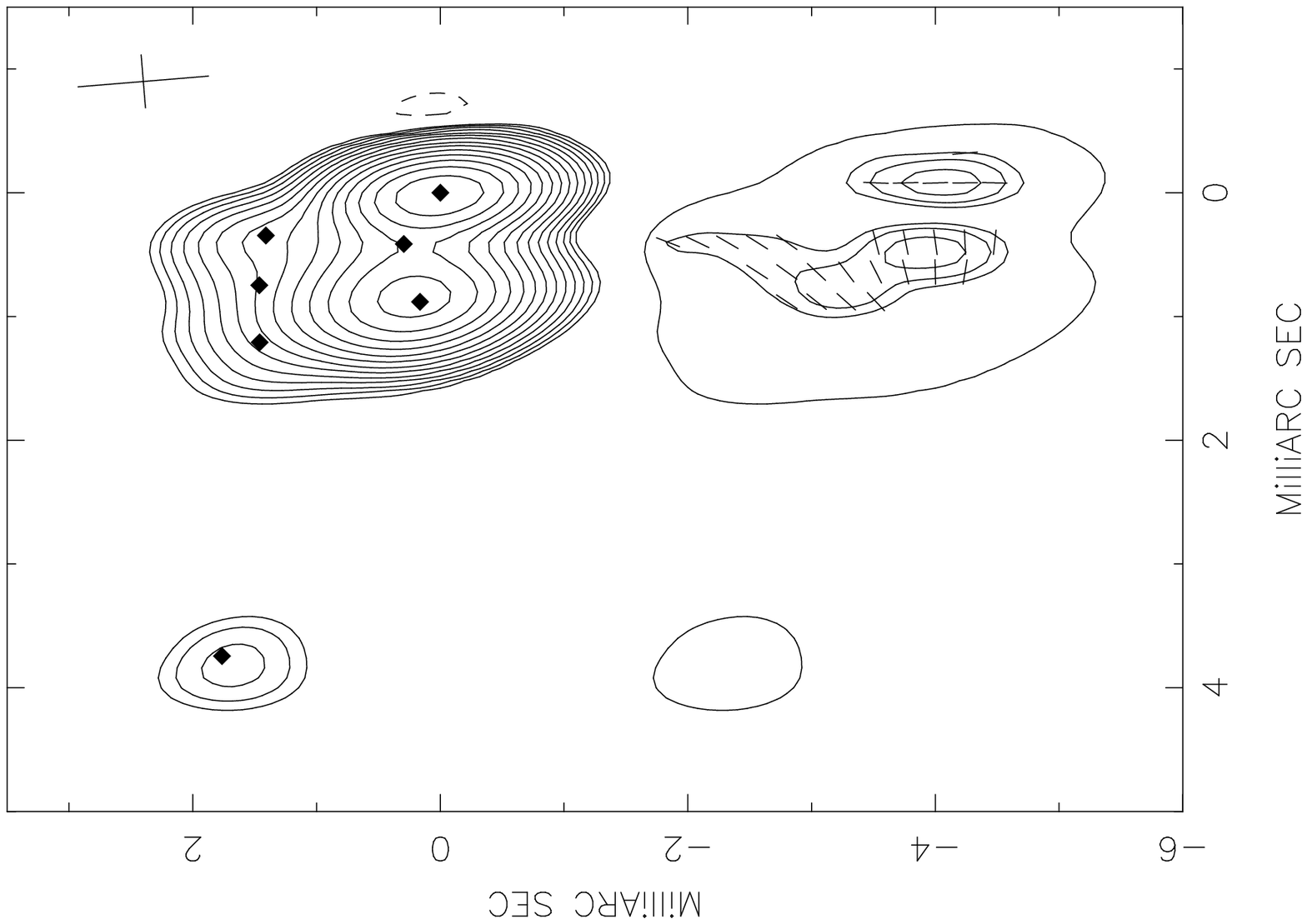,width=3in,angle=-90}
\end{center}
\figcaption[f4.eps]{\label{f:j0738} Total intensity and polarization images of
J0738$+$73 in epoch 1996.74 at 15 GHz.  The total intensity map has a peak flux
of $0.37$ Jy/beam with contour levels starting at $0.003$ Jy/beam and 
increasing in steps of $\times\sqrt{2}$.  The polarization image has a peak 
flux of $0.007$ Jy/beam with contour levels starting at $0.003$ Jy/beam
and increasing in steps of $\times\sqrt{2}$.  Tick marks represent the
polarization position angle.  A single $I$ contour is drawn around the $P$
to show registration.
The jet core
is the western-most feature with jet components extending over a 
range of position angles to the east.  This was a difficult source to 
model consistently across epoch and frequency.  For this reason, we 
have simply summed the entire list of model components 
and analyze the variability of their total.
}
\end{figure}

\begin{figure}
\figurenum{5}
\begin{center}
\epsfig{file=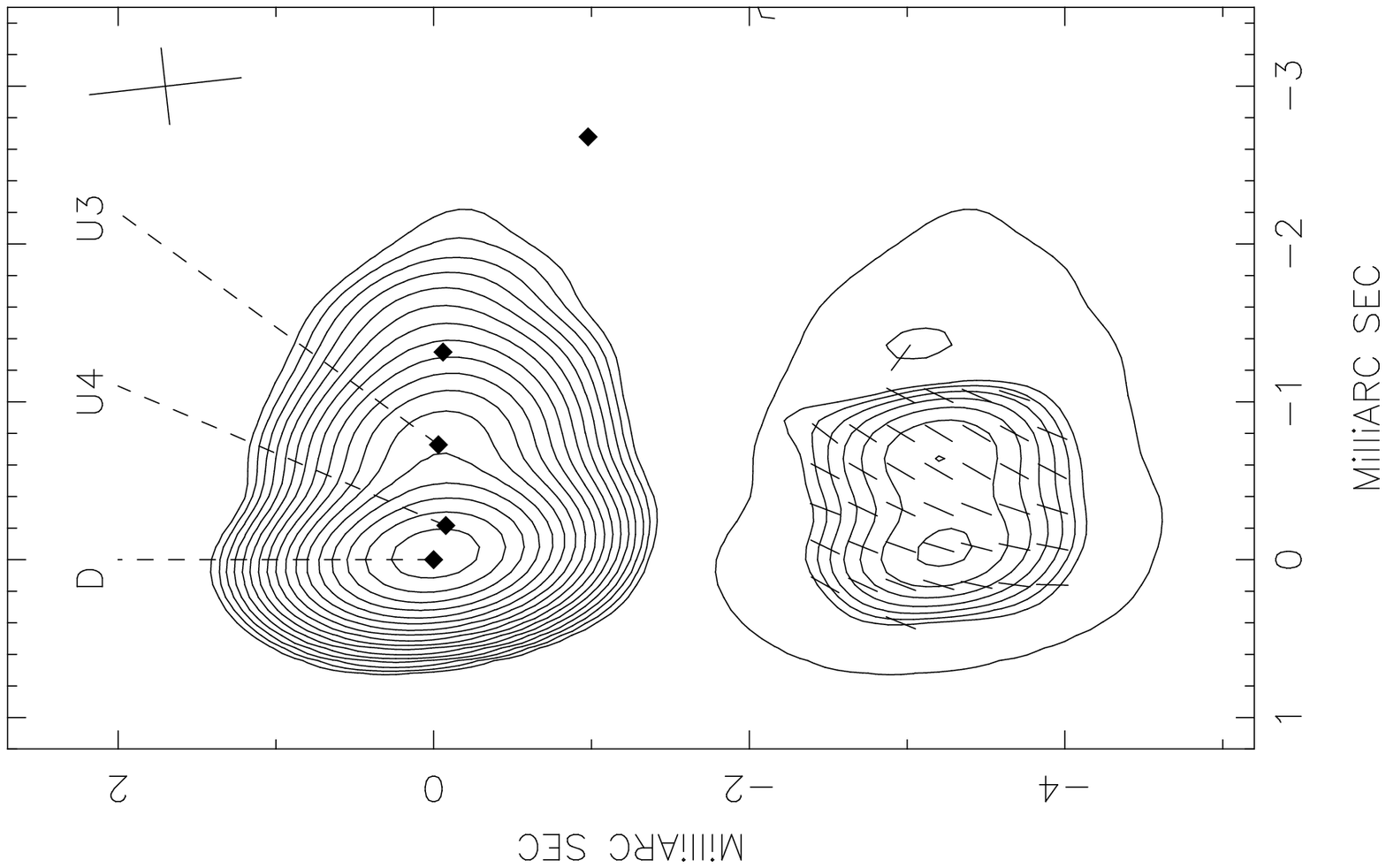,width=3in,angle=-90}
\end{center}
\figcaption[f5.eps]{\label{f:oj287} Total intensity and polarization images of
OJ287 in epoch 1996.57 at 15 GHz.  The total intensity map has a peak flux
of $0.96$ Jy/beam with contour levels starting at $0.003$ Jy/beam and 
increasing in steps of $\times\sqrt{2}$.  The polarization image has a peak 
flux of $0.026$ Jy/beam with contour levels starting at $0.003$ Jy/beam
and increasing in steps of $\times\sqrt{2}$.  Tick marks represent the
polarization position angle.  A single $I$ contour is drawn around the $P$
to show registration.
For variability analysis, we consider two features: 
(1) the core region consisting of D$+$U4 (D$+$K4 at 22 GHz), and (2) the jet 
component U3 (K3), which originates very near the core and moves rapidly
outward during our observations, $\beta_{app} = 19$ (Paper I).  
}
\end{figure}

\begin{figure}
\figurenum{6}
\begin{center}
\epsfig{file=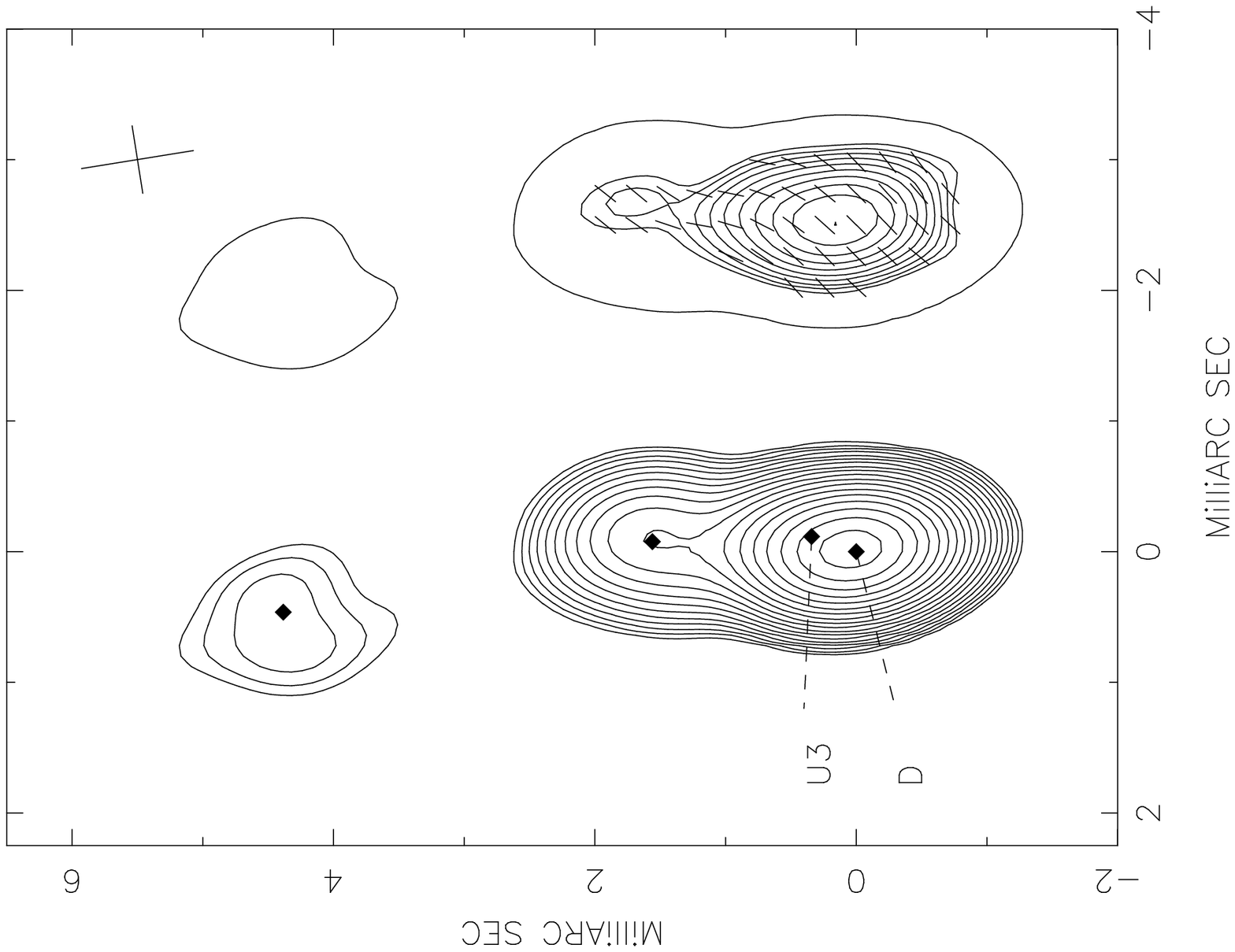,width=4in,angle=-90}
\end{center}
\figcaption[f6.eps]{\label{f:j1224} Total intensity and polarization images of
J1224$+$21 in epoch 1996.57 at 15 GHz.  The total intensity map has a peak flux
of $1.31$ Jy/beam with contour levels starting at $0.003$ Jy/beam and 
increasing in steps of $\times\sqrt{2}$.  The polarization image has a peak 
flux of $0.068$ Jy/beam with contour levels starting at $0.003$ Jy/beam
and increasing in steps of $\times\sqrt{2}$.  Tick marks represent the
polarization position angle.  A single $I$ contour is drawn around the $P$
to show registration.
This source is observed at only five
epochs, and we are able to reliably track only the core region,
consisting of D$+$U3 (D$+$K4$+$K3 at 22 GHz), for variability
analysis.
}
\end{figure}

\begin{figure}
\figurenum{7}
\begin{center}
\epsfig{file=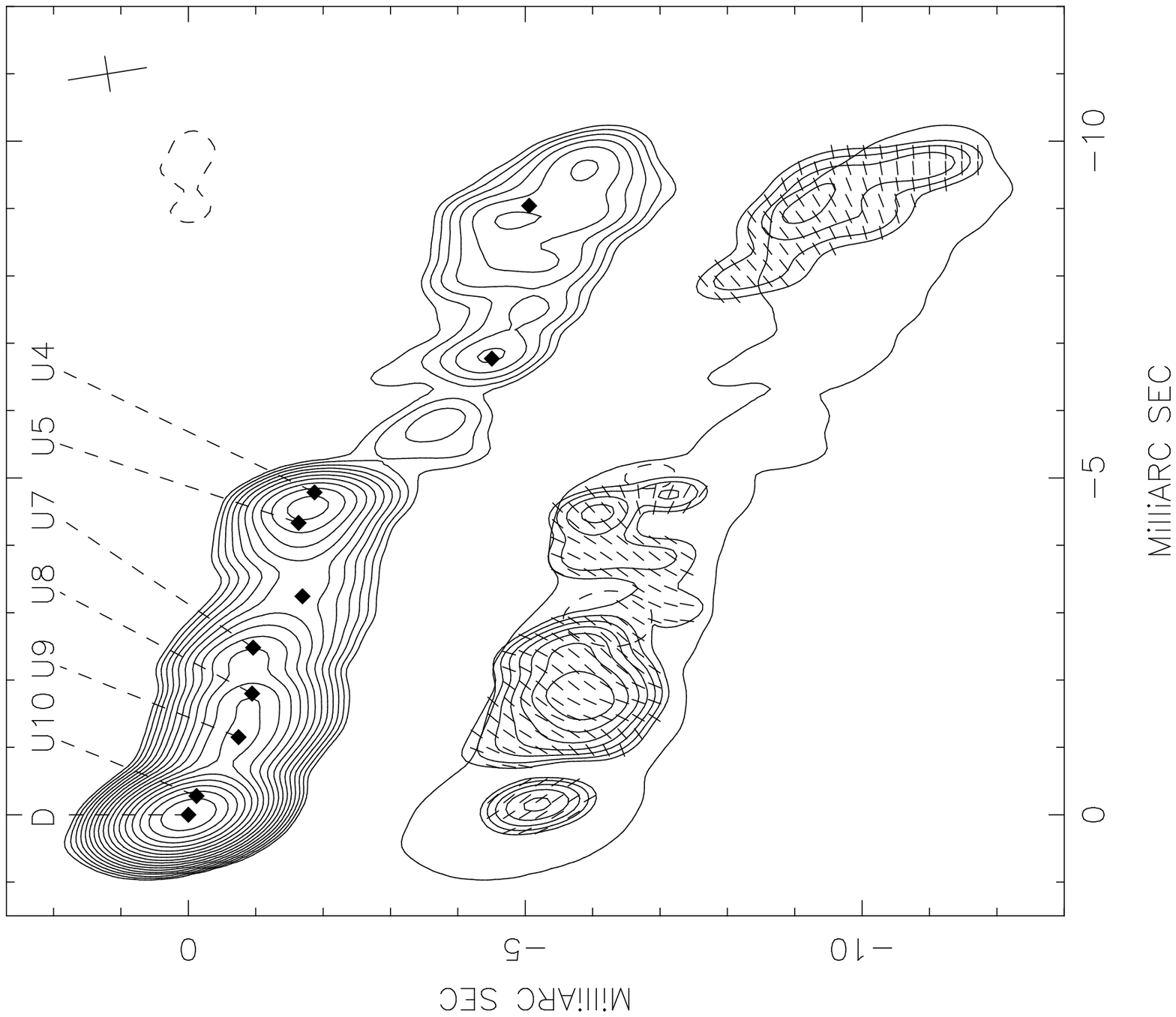,width=4in,angle=-90}
\end{center}
\figcaption[f7.eps]{\label{f:3c273} Total intensity and polarization images of
3C\,273 in epoch 1996.57 at 15 GHz.  The total intensity map has a peak flux
of $10.68$ Jy/beam with contour levels starting at $0.030$ Jy/beam and 
increasing in steps of $\times\sqrt{2}$.  The polarization image has a peak 
flux of $0.166$ Jy/beam with contour levels starting at $0.015$ Jy/beam
and increasing in steps of $\times\sqrt{2}$.  Tick marks represent the
polarization position angle.  A single $I$ contour is drawn around the $P$
to show registration.
For variability analysis,
we consider three features: (1) the core region, consisting of 
D$+$U10 (D$+$K10 at 22 GHz), (2) the first jet region, consisting
of U9$+$U8$+$U7 (K9$+$K8$+$K7), and (3) the bright jet knot, consisting
of U5$+$U4 (K5$+$K4).  Both jet features move with apparent
speeds of approximately $10$ times the speed of light (Paper I).
}
\end{figure}

\begin{figure}
\figurenum{8}
\begin{center}
\epsfig{file=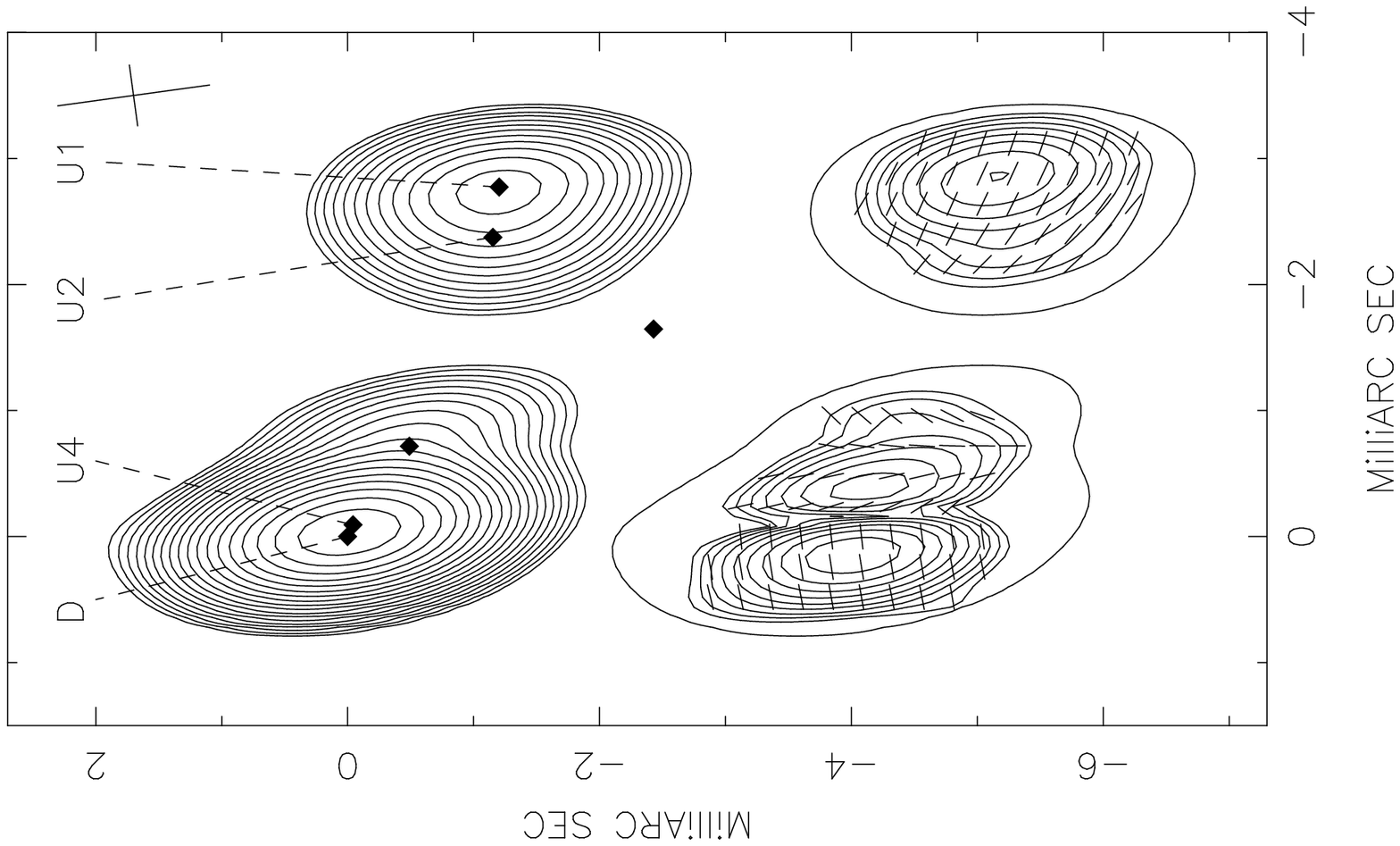,width=3in,angle=-90}
\end{center}
\figcaption[f8.eps]{\label{f:3c279} Total intensity and polarization images of
3C\,279 in epoch 1996.41 at 15 GHz.  The total intensity map has a peak flux
of $14.71$ Jy/beam with contour levels starting at $0.030$ Jy/beam and 
increasing in steps of $\times\sqrt{2}$.  The polarization image has a peak 
flux of $0.317$ Jy/beam with contour levels starting at $0.015$ Jy/beam
and increasing in steps of $\times\sqrt{2}$.  Tick marks represent the
polarization position angle.  A single $I$ contour is drawn around the $P$
to show registration.
For variability analysis, we follow the
core region, D$+$U4 (D$+$K4 at 22 GHz), and the strong jet feature, 
U2$+$U1 (K2$+$K1) which moves at $\beta_{app} = 7.6$ (see Paper I).
The competing polarizations of D and U4 are the most
distinct in the displayed epoch.  
}
\end{figure}

\begin{figure}
\figurenum{9}
\begin{center}
\epsfig{file=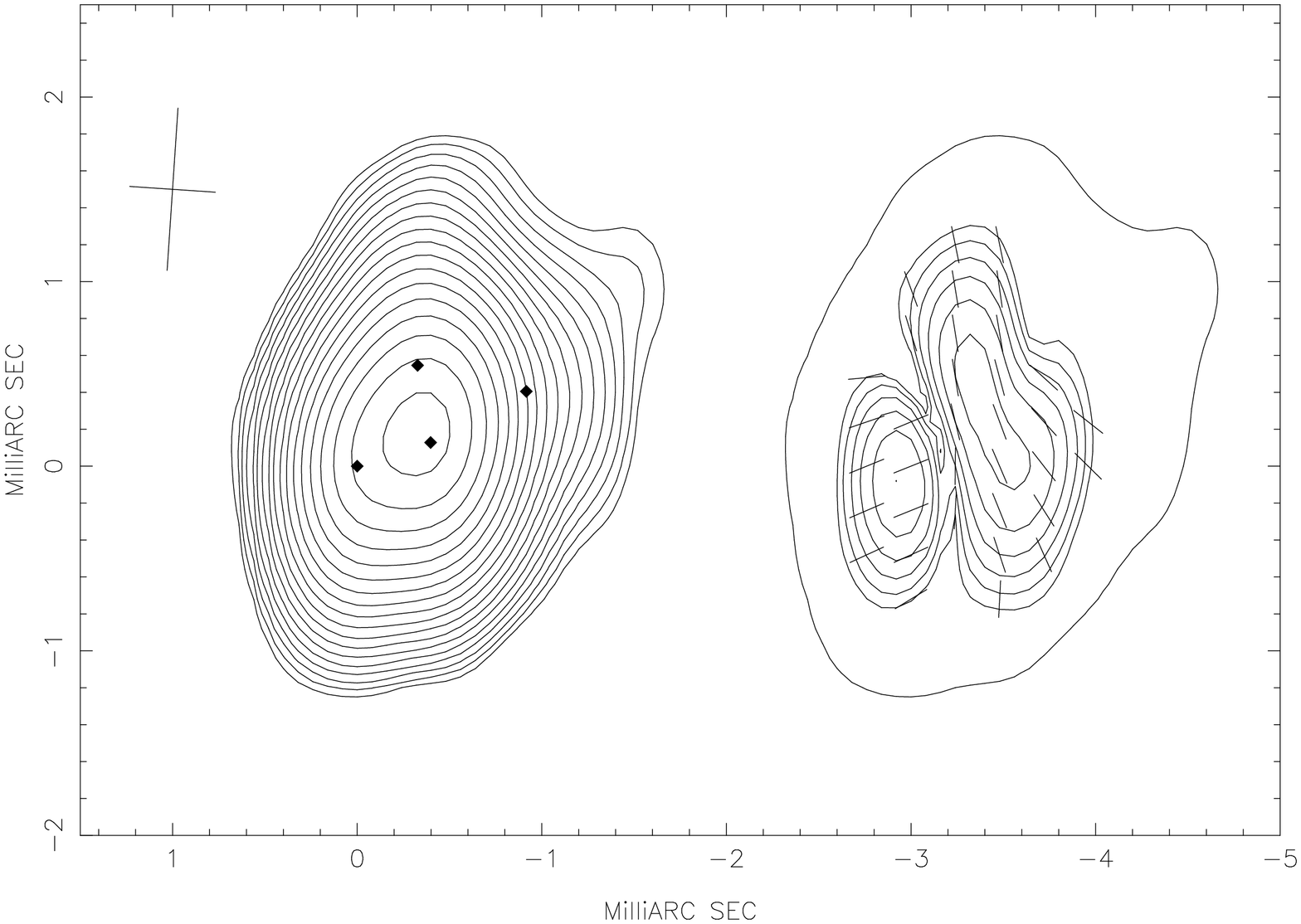,width=2.5in,angle=-90}
\end{center}
\figcaption[f9.eps]{\label{f:j1310} Total intensity and polarization images of
J1310$+$32 in epoch 1996.41 at 15 GHz.  The total intensity map has a peak flux
of $1.70$ Jy/beam with contour levels starting at $0.004$ Jy/beam and 
increasing in steps of $\times\sqrt{2}$.  The polarization image has a peak 
flux of $0.026$ Jy/beam with contour levels starting at $0.004$ Jy/beam
and increasing in steps of $\times\sqrt{2}$.  Tick marks represent the
polarization position angle.  A single $I$ contour is drawn around the $P$
to show registration.
For variability
analysis, we have simply summed all the VLBI model components. 
}
\end{figure}

\begin{figure}
\figurenum{10}
\begin{center}
\epsfig{file=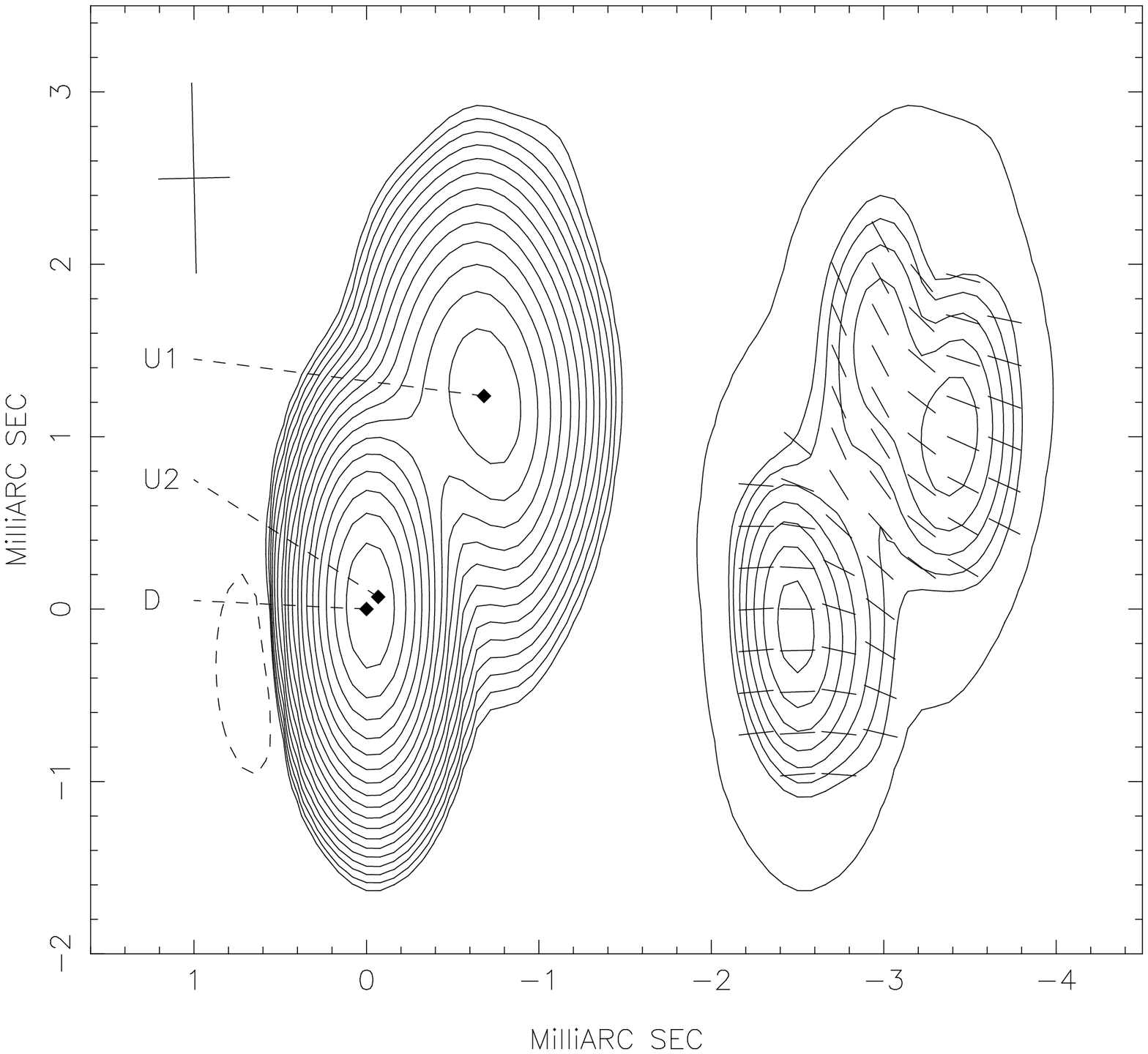,width=3in,angle=-90}
\end{center}
\figcaption[f10.eps]{\label{f:j1512} Total intensity and polarization images of
J1512$-$09 in epoch 1996.41 at 15 GHz.  The total intensity map has a peak flux
of $1.46$ Jy/beam with contour levels starting at $0.003$ Jy/beam and 
increasing in steps of $\times\sqrt{2}$.  The polarization image has a peak 
flux of $0.028$ Jy/beam with contour levels starting at $0.003$ Jy/beam
and increasing in steps of $\times\sqrt{2}$.  Tick marks represent the
polarization position angle.  A single $I$ contour is drawn around the $P$
to show registration.
For
variability analysis, we consider two features: (1) the core region 
consisting of D$+$U2 (D$+$K2 at 22 GHz), and (2) the jet 
component U1 (K1) which has a proper motion of 
$\beta_{app} = 21$ (Paper I).
}
\end{figure}

\begin{figure}
\figurenum{11}
\begin{center}
\epsfig{file=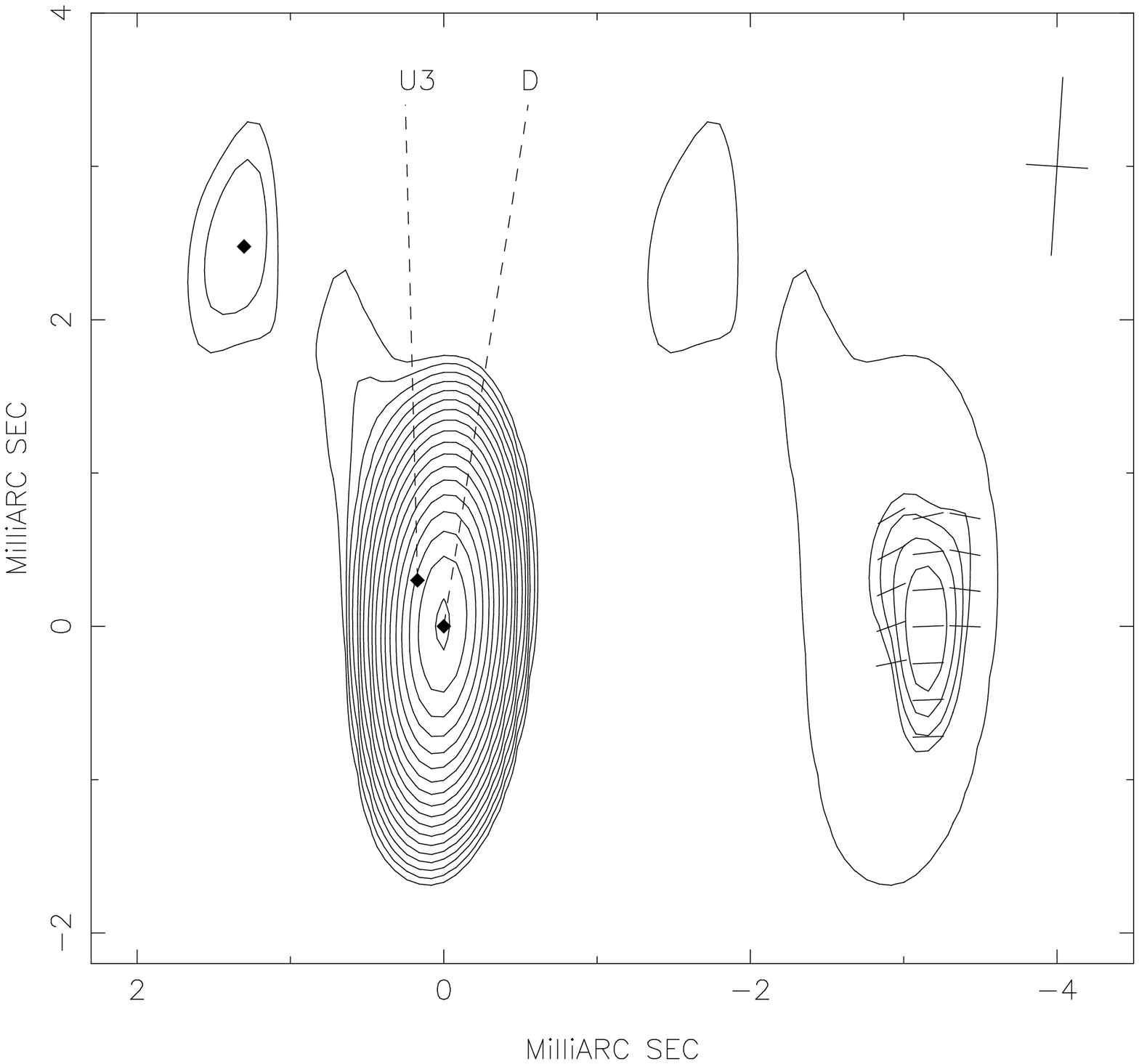,width=3in,angle=-90}
\end{center}
\figcaption[f11.eps]{\label{f:j1751} Total intensity and polarization images of
J1751$+$09 in epoch 1996.57 at 15 GHz.  The total intensity map has a peak flux
of $0.77$ Jy/beam with contour levels starting at $0.002$ Jy/beam and 
increasing in steps of $\times\sqrt{2}$.  The polarization image has a peak 
flux of $0.012$ Jy/beam with contour levels starting at $0.003$ Jy/beam
and increasing in steps of $\times\sqrt{2}$.  Tick marks represent the
polarization position angle.  A single $I$ contour is drawn around the $P$
to show registration.
For variability
analysis, we consider
only the core region, D+U3 (D at 22 GHz).
}
\end{figure}

\begin{figure}
\figurenum{12}
\begin{center}
\epsfig{file=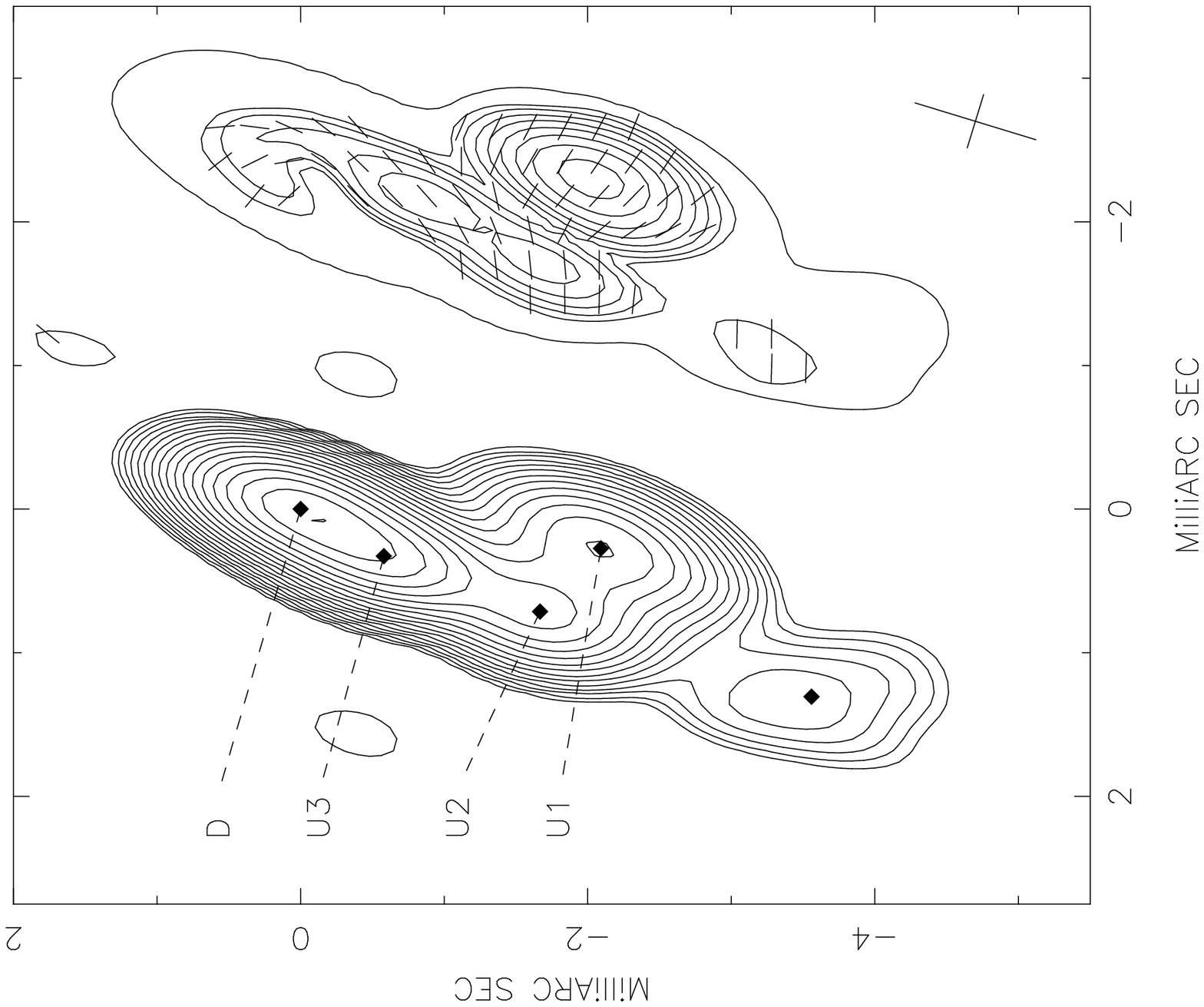,width=3in,angle=-90}
\end{center}
\figcaption[f12.eps]{\label{f:j1927} Total intensity and polarization images of
J1927$+$73 in epoch 1996.57 at 15 GHz.  The total intensity map has a peak flux
of $1.46$ Jy/beam with contour levels starting at $0.004$ Jy/beam and 
increasing in steps of $\times\sqrt{2}$.  The polarization image has a peak 
flux of $0.040$ Jy/beam with contour levels starting at $0.003$ Jy/beam
and increasing in steps of $\times\sqrt{2}$.  Tick marks represent the
polarization position angle.  A single $I$ contour is drawn around the $P$
to show registration.
For variability analysis, we consider three
features: (1) the core region consisting
of D$+$U3 (D$+$K3 at 22 GHz), (2) the first jet component U2 (K2), 
and (3) the second jet component U1 (K1).  These two jet features 
travel on radial trajectories with apparent speeds of 
approximately $4-5$ times the speed of light (Paper I).
}
\end{figure}

\begin{figure}
\figurenum{13}
\begin{center}
\epsfig{file=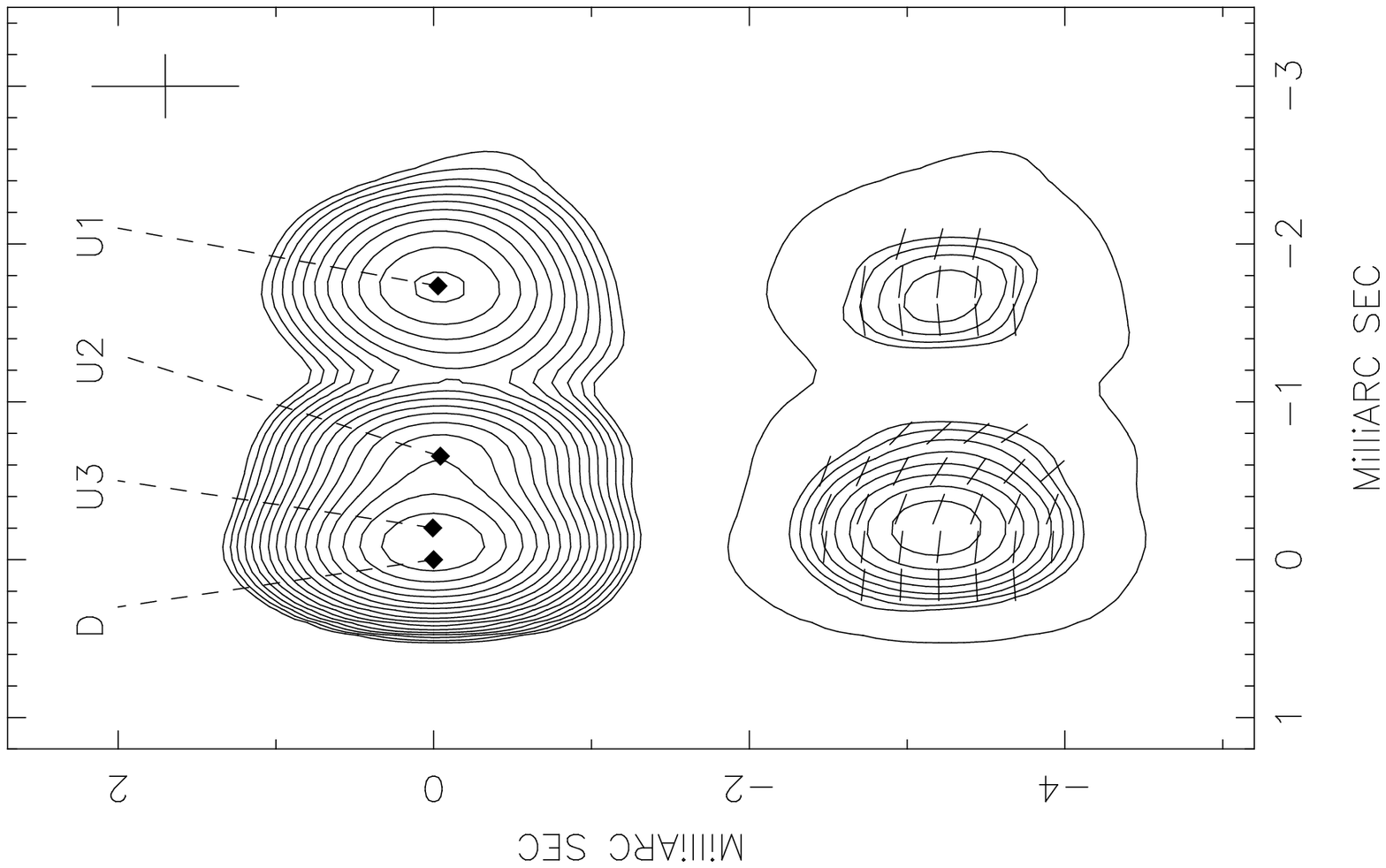,width=3in,angle=-90}
\end{center}
\figcaption[f13.eps]{\label{f:j2005} Total intensity and polarization images of
J2005$+$77 in epoch 1996.23 at 15 GHz.  The total intensity map has a peak flux
of $0.51$ Jy/beam with contour levels starting at $0.002$ Jy/beam and 
increasing in steps of $\times\sqrt{2}$.  The polarization image has a peak 
flux of $0.029$ Jy/beam with contour levels starting at $0.002$ Jy/beam
and increasing in steps of $\times\sqrt{2}$.  Tick marks represent the
polarization position angle.  A single $I$ contour is drawn around the $P$
to show registration.
For variability analysis, we consider two features:
(1) the core region, D$+$U3$+$U2 (D$+$K2 at 22 GHz), and (2) the
jet component U1 (K1), which doesn't move during our observations 
(Paper I).
}
\end{figure}

Figures \ref{f:3c120} through \ref{f:j2005} display a single 15 GHz total
intensity and polarization image of each source in our sample.  Model
components which comprise the core and jet features we follow are labeled on
the images and briefly described in the figure captions.  Mean properties 
of these features are given in table \ref{t:mean}. 

%\placetable{t:mean}
%\placefigure{f:3c120}
%\placefigure{f:j0530}
%\placefigure{f:j0738}
%\placefigure{f:oj287}
%\placefigure{f:j1224}
%\placefigure{f:3c273}
%\placefigure{f:3c279}
%\placefigure{f:j1310}
%\placefigure{f:j1512}
%\placefigure{f:j1751}
%\placefigure{f:j1927}
%\placefigure{f:j2005}

\section{Analysis for Flux and Polarization Variability}
\label{s:tech}

A fundamental problem in the analysis of VLBI data for 
variability is knowing the real ($1\sigma$) measurement 
uncertainties for properties of jet features.  If we knew these 
uncertainties robustly, it would be easy to evaluate
the reality of observed flux or polarization differences 
between epochs.

The long VLBI reduction path includes a-priori 
calibration, opacity corrections, fringe fitting, 
band-pass solutions, self-calibration, feed leakage
corrections, and model-fitting.  While the uncertainties
in many of these steps can be known quite well, others are 
much more difficult to quantify precisely, and it is 
difficult to combine them to obtain 
a robust ``$1\sigma$'' uncertainty on the flux or polarization 
of a particular jet feature in a particular epoch.

\subsection{Linear Trends}

In the absence of robust a-priori uncertainties on the
measured properties of jet features, one can still perform
a linear, least-square regression assuming the data are 
equally weighted.  With six evenly-spaced epochs, we obtained good 
measures (with uncertainties) for the linear 
changes over time in the total intensity ($I$), fractional
polarization ($m$), and polarization position angle ($\chi$) 
of jet features.  We averaged these linear slopes between the
two frequencies to obtain a single average slope over
time for each jet feature\footnote{In total intensity, 
we first divided out the mean flux at each frequency before 
averaging the slopes.}. 

To examine any relative linear changes between the 
frequencies, we also computed the slopes with time 
in spectral index ($\alpha$, $S\propto\nu^{+\alpha}$), 
fractional polarization ratio ($m_{ratio} = m_{15}/m_{22}$), 
and polarization position angle difference 
($\Delta\chi = \chi_{15}-\chi_{22}$). 

\subsection{Fluctuations About the Linear Trends}
\label{s:corr}

Linear changes are interesting, but they represent the
longest time scale ($\gtrsim 1$ yr) changes that our 
observations are sensitive to.  We also wanted to
examine shorter time-scale ($\lesssim 0.5$ yr) fluctuations 
over the six epochs.  To obtain robust results for the fluctuations
about the linear trends in the data, we have exploited the dual 
observing bands in our monitoring program.  Considering the
broad-band nature of synchrotron radiation, the two 
frequencies, 15 and 22 GHz, are closely spaced and changes 
at one frequency should be mirrored at the other.

To quantify this effect, we considered the deviations, $\Delta$,
in every epoch from either the best fit linear 
regressions over time at each frequency or the mean value at
each frequency.  The mean values were used only if the linear 
slopes at {\em both} frequencies were of less than $2\sigma$ 
significance.  We formed a correlation coefficient, 
$r$ from these $\Delta$ values...
\begin{equation}
r = \frac{\sum_j{\Delta_{15j}\Delta_{22j}}}
{\sqrt{\left(\sum_i{\Delta_{15i}^{2}}\right)
\left(\sum_k{\Delta_{22k}^{2}}\right) }}.
\end{equation}

If the $\Delta$ values are deviations from the mean value, $r$ is 
just the standard linear correlation coefficient.  When $r = 1$,
the fluctuations are perfectly correlated between frequencies, $r = 0$
is no correlation, and $r = -1$ is a perfect anti-correlation.  A
useful property of this statistic is that the correlation coefficient
doesn't require the $\Delta$ values at the two frequencies to have the
same magnitude to be correlated.  The fluctuations at the two frequencies
must only be proportional to one another to have a strong correlation. If 
$\Delta_{15j} = 2\times\Delta_{22j}$, for example, the fluctuations will 
be perfectly correlated.  This feature compensates for some
of the spectral differences that exist between the frequencies, such as
a constant spectral index.  A variable spectral index will, of 
course, degrade any correlation between the frequencies; however,  
polarization position angle changes generated by variable Faraday rotation 
{\em are} detectable by this correlation coefficient because the changes 
at 15 GHz will always be $2.1$ times those at 22 GHz.

With only five or six epochs over which we could correlate these
fluctuations, it was important to assess the significances of
the correlations we observed.  For each jet feature, we 
computed $r$ and ran a Monte Carlo simulation to compute the 
probability, $p_r$, that a correlation this
strong could occur by pure chance.  The Monte Carlo simulation randomly
generated $\Delta$ values pulled from a Gaussian distribution with the
same variance as the data but with no intrinsic correlation between 
frequencies.  For the
cases where the original $\Delta$ values had been taken with respect
to the best fit line, the randomly generated $\Delta$ values were 
added to this line.   A new best fit line was then found and
removed from the fake data to make a final set of fake $\Delta$ values 
to correlate. 50,000 fake data sets were generated, and $p_r$ was
the fraction of these with correlation coefficients $\geq r$ (or 
$\leq r$ if $r$ was negative).  For the cases where the $\Delta$
values were taken with respect to the mean at each frequency, our Monte 
Carlo simulation reproduced the expected theoretical 
probabilities \citep{PTVF95}.

\begin{figure}
\figurenum{14}
\begin{center}
\epsfig{file=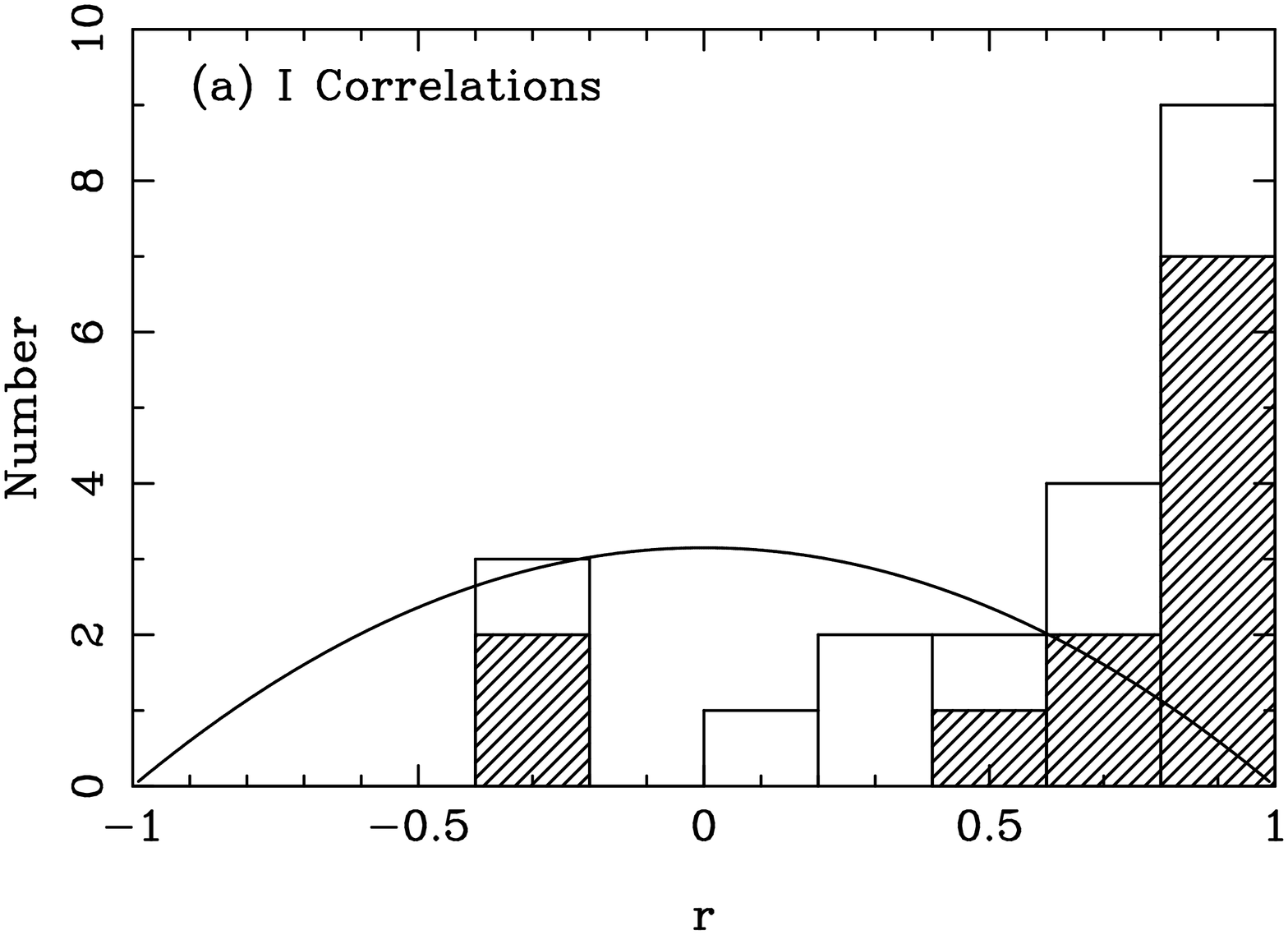,width=2.19in,angle=-90}\\
\epsfig{file=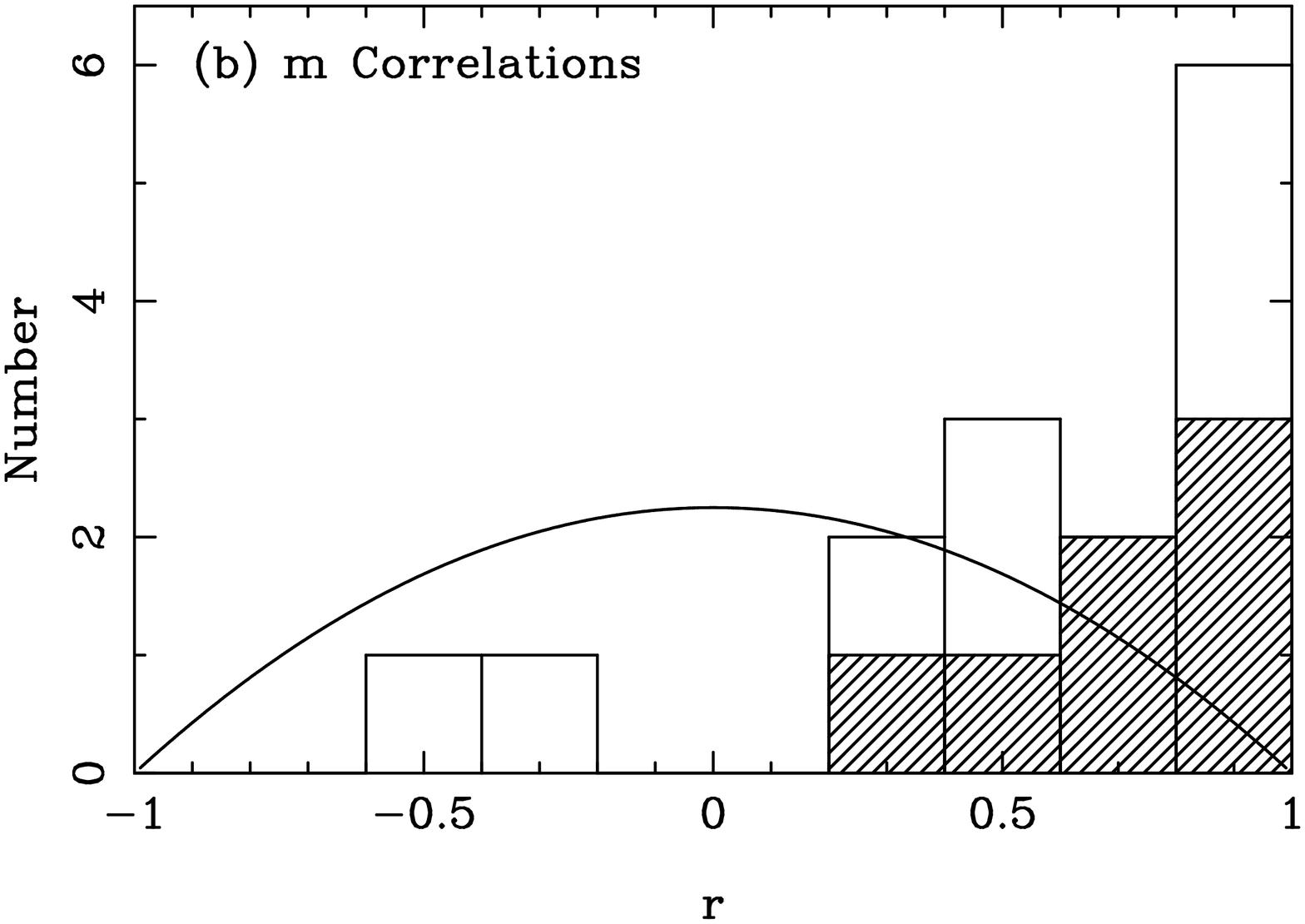,width=2.19in,angle=-90}\\
\epsfig{file=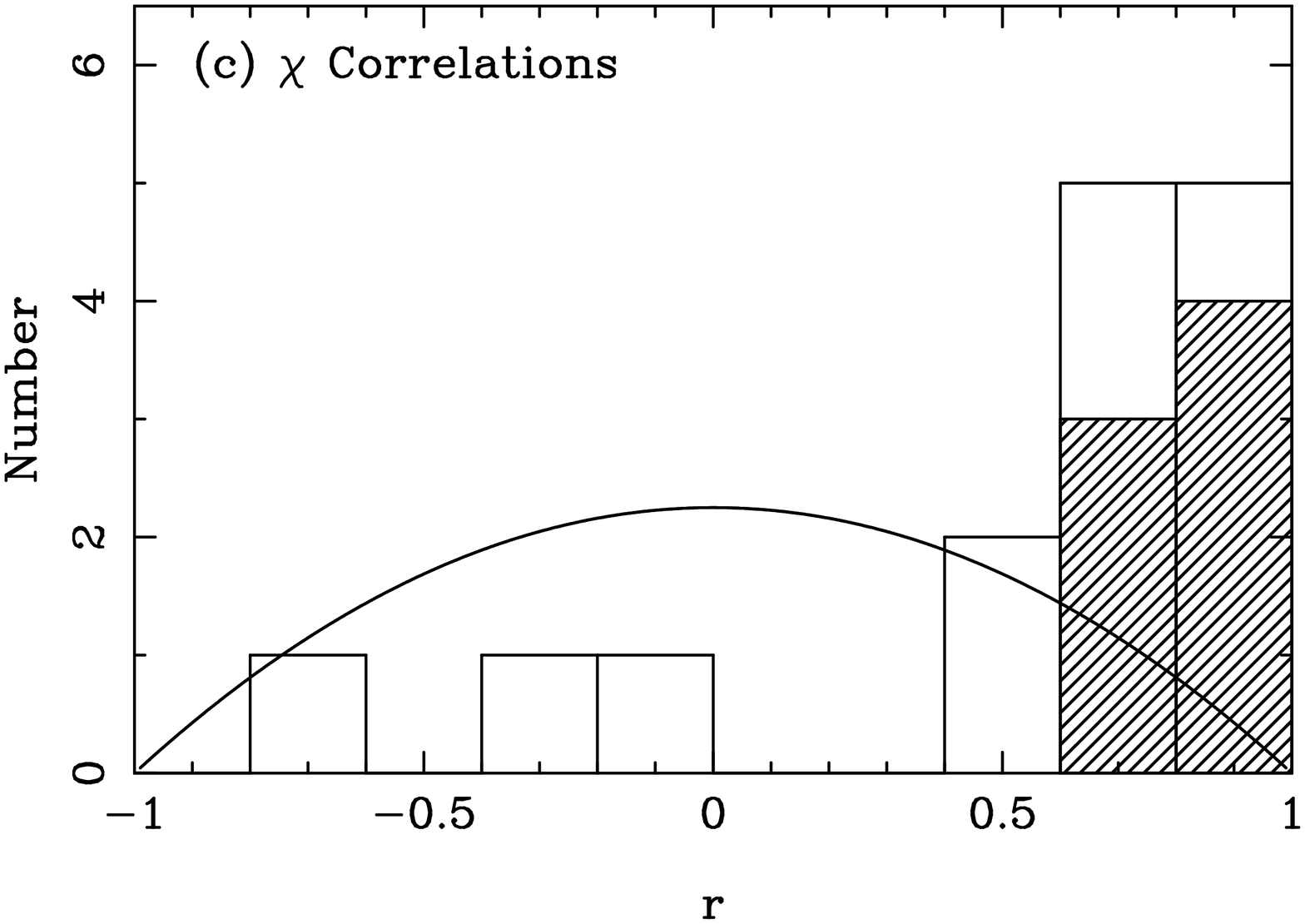,width=2.19in,angle=-90}
\end{center}
\figcaption[f14a.eps,f14b.eps,f14c.eps]{\label{f:hist} Histograms of the correlation, $r$, between
the  fluctuations at 15 and 22 GHz in total intensity (panel (a)), 
fractional polarization (panel (b)), and polarization position angle (panel (c)).
Jet features are represented by empty bars, and hash-filled bars represent
core features.  The solid lines approximate the expected distributions if
there were no intrinsic correlation between the fluctuations at 15 and 22 GHz.  
The large excess of positive correlations shows that real, short term
($\lesssim 0.5$ yr) fluctuations are common throughout our data set.
}
\end{figure}

The correlations, $r$, and their significances, $p_r$,
are given in section \ref{s:res}, and we consider cases with
$p_r \leq 0.05$ to be statistically significant.  It is interesting 
to examine the distribution of correlations over our whole sample, and 
figure \ref{f:hist} shows histograms of $r$ for fluctuations in
flux, fractional polarization, and polarization 
position angle.  Plotted with the histograms are solid lines
indicating the approximate distributions we would expect if 
the feature properties varied independently at each frequency.
The histograms clearly show a very strong bias towards positive
correlations indicating that real shorter term fluctuations 
(which appear at both frequencies) are common throughout our
data set.

%\placefigure{f:hist}

We also computed a measure of the amplitude of the real
correlated fluctuations.  
\begin{equation}
\label{e:corrvar}
\Omega = r\times\frac{\sqrt{\left(\sum_i{\Delta_{15i}^{2}}\right)
\left(\sum_k{\Delta_{22k}^{2}}\right)}}{\left(n-2\right)}, 
\end{equation}
where $\Omega$ is the correlated variance of the deviations
from the best fit lines\footnote{Of course, when the deviations 
are taken with respect to mean values, the factor $\left(n-2\right)$ 
in equation \ref{e:corrvar} should be replaced 
by $\left(n-1\right)$.}.  Here we have simply multiplied the 
correlation coefficient by the geometric mean of the variances at 
the two frequencies.  The square-root of $\Omega$ gives the
correlated standard deviation, a measure of the typical size
of the observed fluctuations.  We computed the standard 
deviation for the correlated fluctuations in each 
quantity\footnote{For negative correlations,
we make the corresponding $\delta$ value negative.}: 
$\delta I = \sqrt{\Omega_I}$, $\delta m = \sqrt{\Omega_m}$, and 
$\delta \chi = \sqrt{\Omega_\chi}$.

To place uncertainty estimates on these values, we ran a second
Monte Carlo simulation similar to the one described above.  In
this simulation, rather than assuming no intrinsic correlation
between frequency in generating the fake data, we specifically
induced an intrinsic correlation of size $r$, matched to that in
the true data.  Next we found the
apparent correlation between 15 and 22 GHz for all the fake 
data sets.  From this distribution of apparent correlations, we
could estimate our uncertainty in measuring a correlation of 
strength $r$.  The uncertainties given in section \ref{s:res}
for the correlated fluctuations reflect the bounds 
in which $70$\% of the 50,000 correlations from the fake data 
fell about their mean.

\section{Results and Discussion}
\label{s:res}

The analysis described in the previous section is sensitive to
two types of variability:

(1) {\em Linear trends} which progress throughout our year 
long window of observations.  These could be a monotonic rise 
or fall in flux, a continuous rotation in polarization position 
angle, or a steady build-up (or decay) of fractional polarization.
Our analysis characterizes these changes using the best fit
linear slopes with time.  The timescale probed by these slopes
are typically $\gtrsim 1$ year.

(2) {\em Fluctuations} about the linear trends which encompass 
variability on timescales $\lesssim 0.5$ year.  These could be 
very short term (epoch to epoch) fluctuations, or they might represent 
distinct changes in the rate or sign of an otherwise continuous 
evolution, e.g. a jet feature that distinctly slows in its flux
decay or a core region that rises then falls in flux.
Our analysis quantifies the fluctuations by looking
at the deviations about the best fit linear trends,
correlating them between our two observing bands, 
and computing the standard deviation of the correlated 
fluctuations.

\subsection{Flux Variability}

Our results for linear trends in flux and fluctuations about 
the linear trends are summarized in table \ref{t:i_var}, and 
they are plotted and discussed in the following sections.

%\placetable{t:i_var}

\subsubsection{Linear Flux Trends}
\label{s:flux_slope}

\begin{figure}
\figurenum{15}
\begin{center}
\epsfig{file=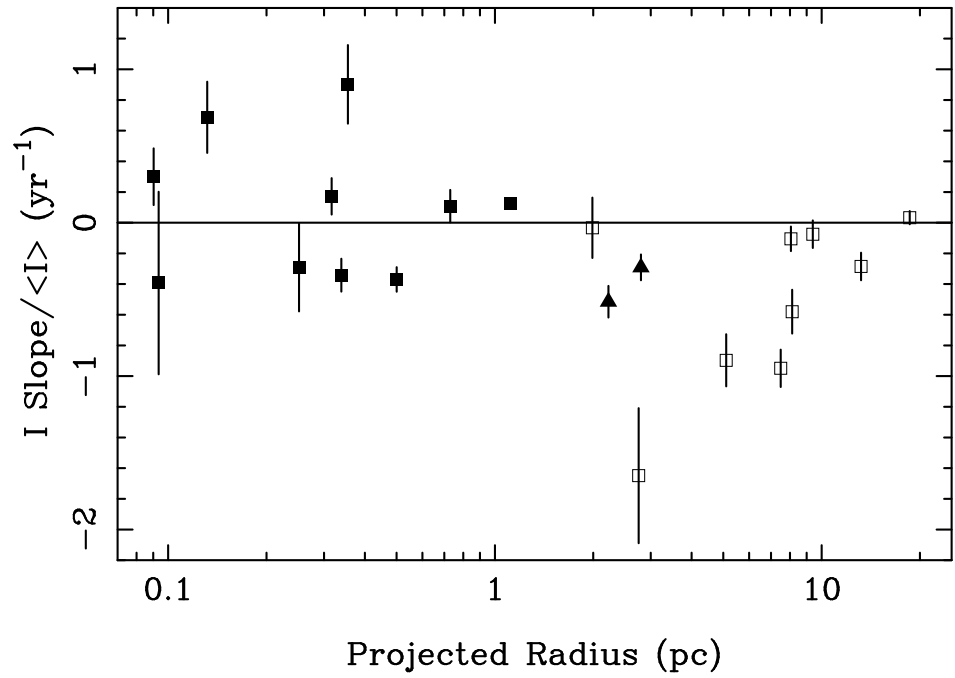,width=3.5in,angle=-90}
\end{center}
\figcaption[f15.eps]{\label{f:flux_slope}
Fractional flux slope, $\langle\dot{I}/\langle I \rangle\rangle$ (see table
\ref{t:i_var}), plotted against projected
radius in parsecs.  Filled squares represent core regions, and open squares 
represent jet features.  J0738$+$17 and J1312$+$32, where we have analyzed the
total VLBI flux of the source, are represented by filled triangles.
}
\end{figure}

Figure \ref{f:flux_slope} shows fractional flux slopes with time 
plotted against projected radius in the source frame.  Both 
core and jet features are plotted, and 
while they have fractional slopes of similar magnitude, the
distributions of the signs of the slopes are quite different.
Core regions either increase or decay in flux over the year, 
while jet features only decay (or show no significant change) 
on that time-scale.   

%\placefigure{f:flux_slope}

It is interesting that the fractional decay rates of the core 
regions do not exceed $-0.5$ yr$^{-1}$ and cluster about $\sim -0.4$ 
yr$^{-1}$, while nearly half the jet features have faster 
decay rates than any core region.  These slopes represent the average 
flux trends over nearly a
years time, and faster decays in core regions may simply be muted 
by subsequent outbursts.  An excellent example of this is
J1751$+$09 (see figure \ref{f:compare}) which decays sharply in 
flux (from $\sim 2.8$ Jy to $\sim 1.0$ Jy) between the first two 
epochs and then rises sharply
(from $\sim 1.0$ Jy to $\sim 2.1$ Jy) between the last two epochs. 
These kind of fast fluctuations are examined in \S{\ref{s:nl_flux}}.

%\placefigure{f:3c120i}

\begin{figure}
\figurenum{16}
\begin{center}
\epsfig{file=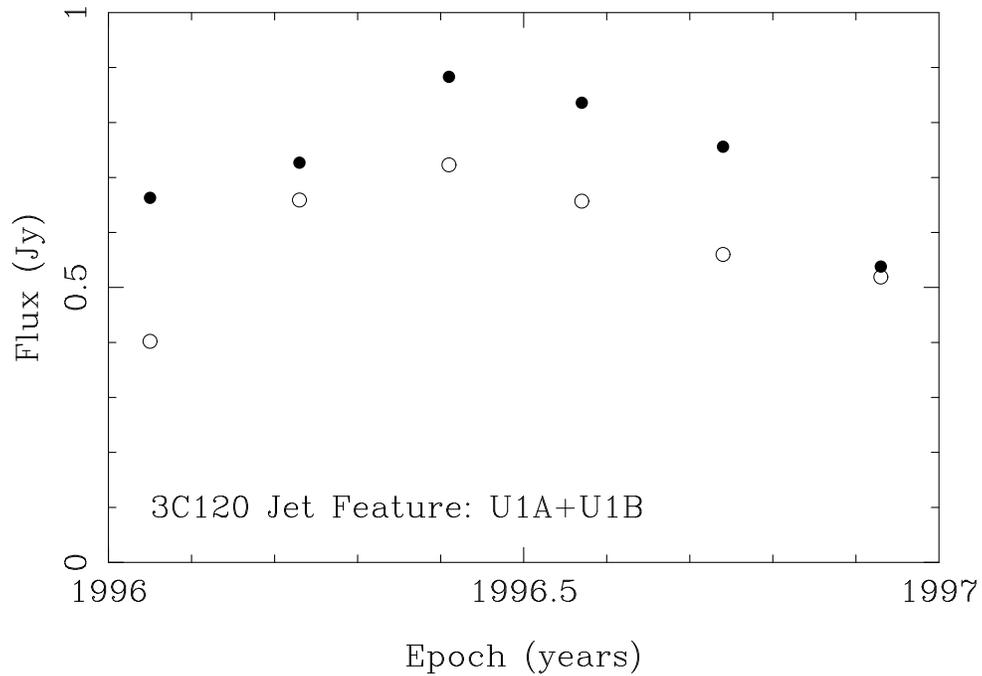,width=3.5in,angle=-90}
\end{center}
\figcaption[f16.eps]{\label{f:3c120i}
Flux of the jet feature, U1A$+$U1B (K1A$+$K1B)
in 3C\,120, plotted against epoch.  Filled and open circles 
represent measurements at 15 and 22 GHz, respectively. 
}
\end{figure}

\begin{figure}
\figurenum{17}
\begin{center}
\epsfig{file=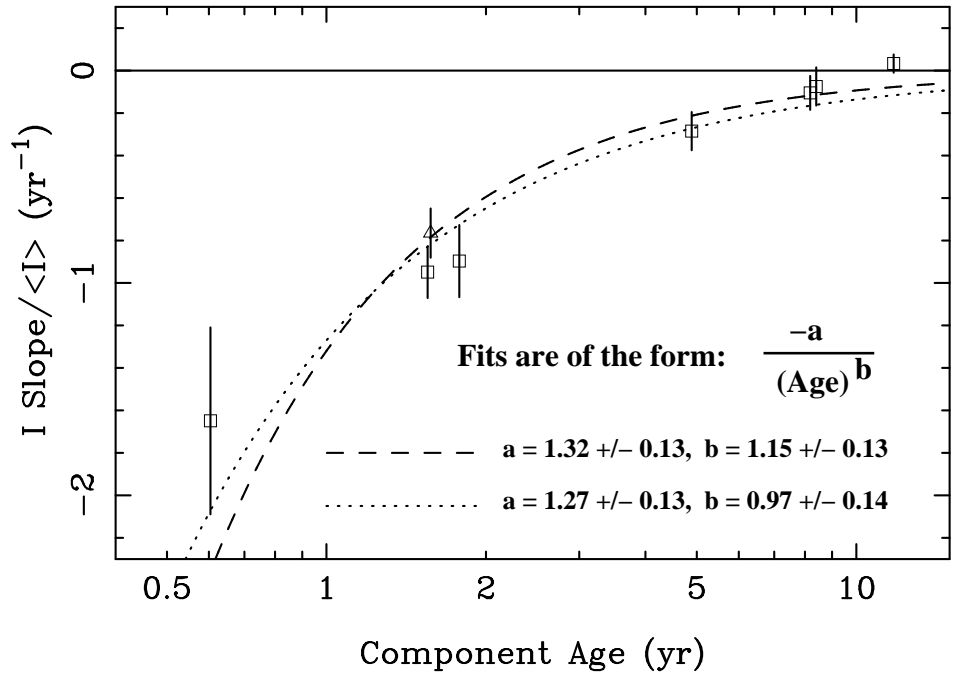,width=3.5in,angle=-90}
\end{center}
\figcaption[f17.eps]{\label{f:decay}
Fractional flux slope, $\langle\dot{I}/\langle I \rangle\rangle$, 
for jet features plotted 
against the apparent age of the feature in our frame.  Fits to the
model, $(dI/dt)/\langle I \rangle = -a/T_{age}^b$, are plotted and discussed
in the text, \S{\ref{s:flux_slope}}.  The open triangle represents the
jet feature U1A$+$U1B (K1A$+$K1B) from 3C\,120 for only the last four
epochs when its flux decayed (see the text).
}
\end{figure}

The fractional decay rates of jet features are smaller with
larger projected radius.  The exception to this trend is
the jet feature in 3C\,120 at a mean projected radius of 2 parsecs.
This feature rose in flux from 1996.05 to 1996.41 then showed a
steady decay over the last four epochs from 1996.41 to 1996.93 with 
a slope of $(dI/dt)/\langle I \rangle = -0.77\pm0.12$ 
yr$^{-1}$ (see figure \ref{f:3c120i}).  
The trend of smaller decay rates at larger 
radii may simply reflect the fact that longer lived features
must decay more slowly, or we would not observe them.  Figure 
\ref{f:decay} plots fractional decay
rate versus the age of the jet feature, $T_{age}$.  The ages are 
in our frame and computed by simply dividing the mean angular
radius, $<R>$, by the proper motion, $\mu = dR/dt$, given 
in Paper I.   For the jet feature in 3C\,120 we have plotted
the decay slope over the last four epochs with an open triangle.
The plot shows a remarkable relation between fractional decay rate, 
$(dI/dt)/\langle I \rangle$, and the logarithm of $T_{age}$.

%\placefigure{f:decay}

We can construct a simple phenomenological model for the
relation in figure \ref{f:decay}.  If we assume that the flux
of a jet feature depends only on its distance along the 
jet axis, $I \propto R^{-a}$, it follows that the fractional
decay rate is inversely proportional to $T_{age}$:
\begin{equation}
(dI/dt)/I = -a\times(dR/dt)/R = -a/T_{age}.
\end{equation}
We performed two fits of a generalized
version of this model: $(dI/dt)/I = -a/T_{age}^{b}$, and these
fits are plotted in figure \ref{f:decay}.  The first fit (dashed 
line) includes all the points for a reduced 
chi-squared value of 2.2.  We noticed that most of the 
chi-squared was due to a single point on 3C\,279 with a slightly
positive flux slope.  The second fit (dotted line) excludes this 
point for a reduced chi-squared value of 0.9.  Both fits 
find $a = 1.3\pm0.1$ and a value of $b$ close to unity.

We note that the jet feature in J2005$+$77 does not appear
in figure \ref{f:decay} because it has no detectable motion
(see Paper I), although it does have a very significant flux
decay.  The simple geometrical model described above cannot
account for cases like this, nor can it account for the flux 
rise seen early in our observations of the jet feature in 3C\,120.
Another example is the feature in 3C\,279, which had
a slightly positive (but not significant) flux slope.  
By the very end of 1997 (a year after the observations presented here),
that feature had risen $\sim 50$\% in flux and dropped in fractional
polarization by approximately a factor of two \citep{HW00}.  
Many factors can influence the flux evolution of a jet feature
in unpredictable ways: variable Doppler factors due to speed 
or trajectory changes, interactions with the external medium,
and catching up with previous slower moving features.  

Figure \ref{f:decay} contains eight data points from six different 
sources, and given the above considerations, it is remarkable
that the flux decays fit so well to a simple geometrical model, 
$I \propto R^{-1.3\pm0.1}$.  Models of this form have been fit
to the flux profiles of VLBI jets by many authors (e.g. 
\citet{WBU87,UW92,X00}) with power law indexes typically falling 
between $-1$ and $-2$, very similar to the power-law index we 
obtain from the decay rates of superluminal jet features.
VLA scale jets also show similar power laws with increasing 
jet width (presumably proportional to radius) with typical 
indexes between $-1.2$ and $-1.6$ \citep{BP84}.  It is tempting
to draw parallels between the physical processes that maintain
the brightness of kiloparsec scale jets and those which operate on
superluminal jet features.  However, we note that selection criteria
probably play some role in the range of observed power law 
indexes, and that in all of these cases, including our measurement 
from the dynamical decays of jet features, the power law index 
represents only the mean behavior and averages over a wide
range of jet micro-physics (A. H. Bridle, private communication).

\subsubsection{Fluctuations in Flux}
\label{s:nl_flux}

\begin{figure}
\figurenum{18}
\begin{center}
\epsfig{file=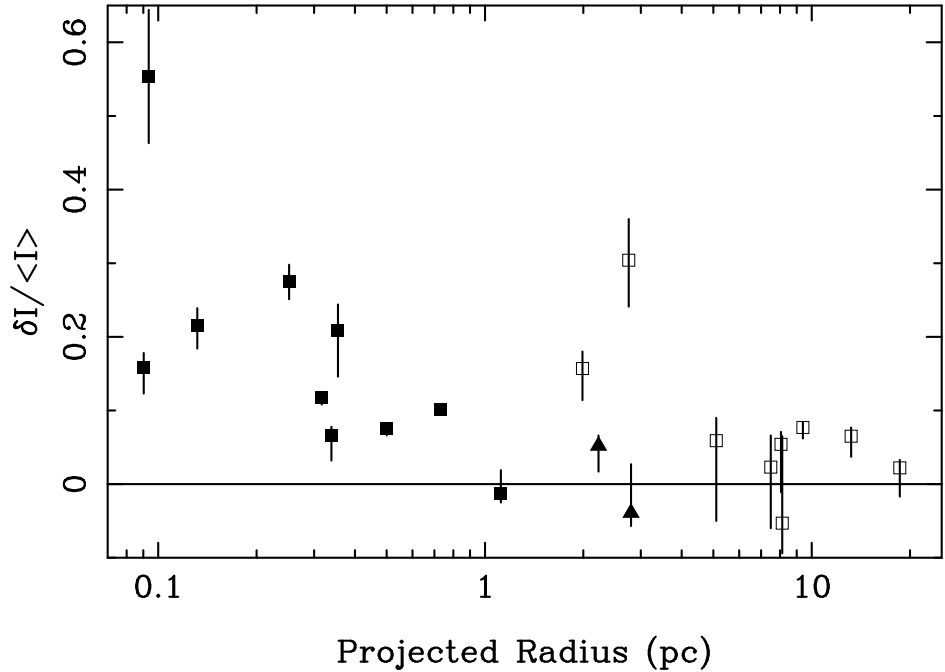,width=3.5in,angle=-90}
\end{center}
\figcaption[f18.eps]{\label{f:flux_nl}
Standard deviation of the correlated fluctuations in flux divided by
the mean flux, $\delta I/\langle I \rangle$, 
plotted against projected radius in parsecs.  Filled squares represent 
core regions, and open squares represent jet features.  J0738$+$17 
and J1312$+$32, where we have analyzed the total VLBI flux of 
the source, are represented by  
filled triangles.
}
\end{figure}

Figure \ref{f:flux_nl} shows the standard deviation of the 
correlated fluctuations in flux divided by the mean flux, 
$\delta I/\langle I\rangle$ (see \S{\ref{s:corr}}), for each core and 
jet feature plotted against projected radius.  Core regions have much 
larger fractional fluctuations on average than do jet features.  
In general, core regions with very small projected radii have
the largest fluctuations about the linear trends.  These core regions 
have smaller
projected radii because they are less biased by nearby, barely 
resolved jet components which are included in their sum (see
\S{\ref{s:sources}}).

%\placefigure{f:flux_nl}

The fluctuations displayed by core regions fall
into three broad categories: (1) a gradual rise and plateau 
(or fall) in flux (3C\,273, 3C\,279, and J1512$-$09 primarily
display this behavior.), (2) sudden, large changes in flux that
occur between neighboring epochs (The fluctuations in 
3C\,120 and J1751$+$09 are dominated by this
behavior.), and (3) smaller epoch to epoch fluctuations 
(J0530$+$13 and OJ287 display this behavior).  
J2005$+$77 displays the opposite behavior from 
category (1), with a gentle fall then rise in flux. 

The jet feature with the largest fluctuations about the 
linear trend is U3 (K3) in OJ287.  This feature also has a 
large average flux decay, and the 30\% standard deviation 
is mainly due to a change in the slope of this decay in the 
middle of our observations.  The other jet feature with large
fluctuations is U1A$+$U1B (K1A$+$K1B) in 3C\,120.  The flux of 
this feature rises then decays during our observations as 
illustrated in figure \ref{f:3c120i}.
Beyond a projected radius of 5 pc, no jet feature has
fluctuations with a standard deviation larger than 10\%, and 
only one of these, U1 (K1) in J1927$+$73, is significant.  
The fluctuations in this component appear to be genuine 
epoch-to-epoch changes.  

\subsection{Polarization Variability}

Our results for linear trends in polarization and the 
fluctuations about the linear trends 
are summarized in tables \ref{t:m_var} and \ref{t:chi_var}, and 
they are plotted and discussed in the following sections.

%\placetable{t:m_var}
%\placetable{t:chi_var}

\subsubsection{Linear Trends in Fractional Polarization} 
\label{s:m_line}

\begin{figure}
\figurenum{19}
\begin{center}
\epsfig{file=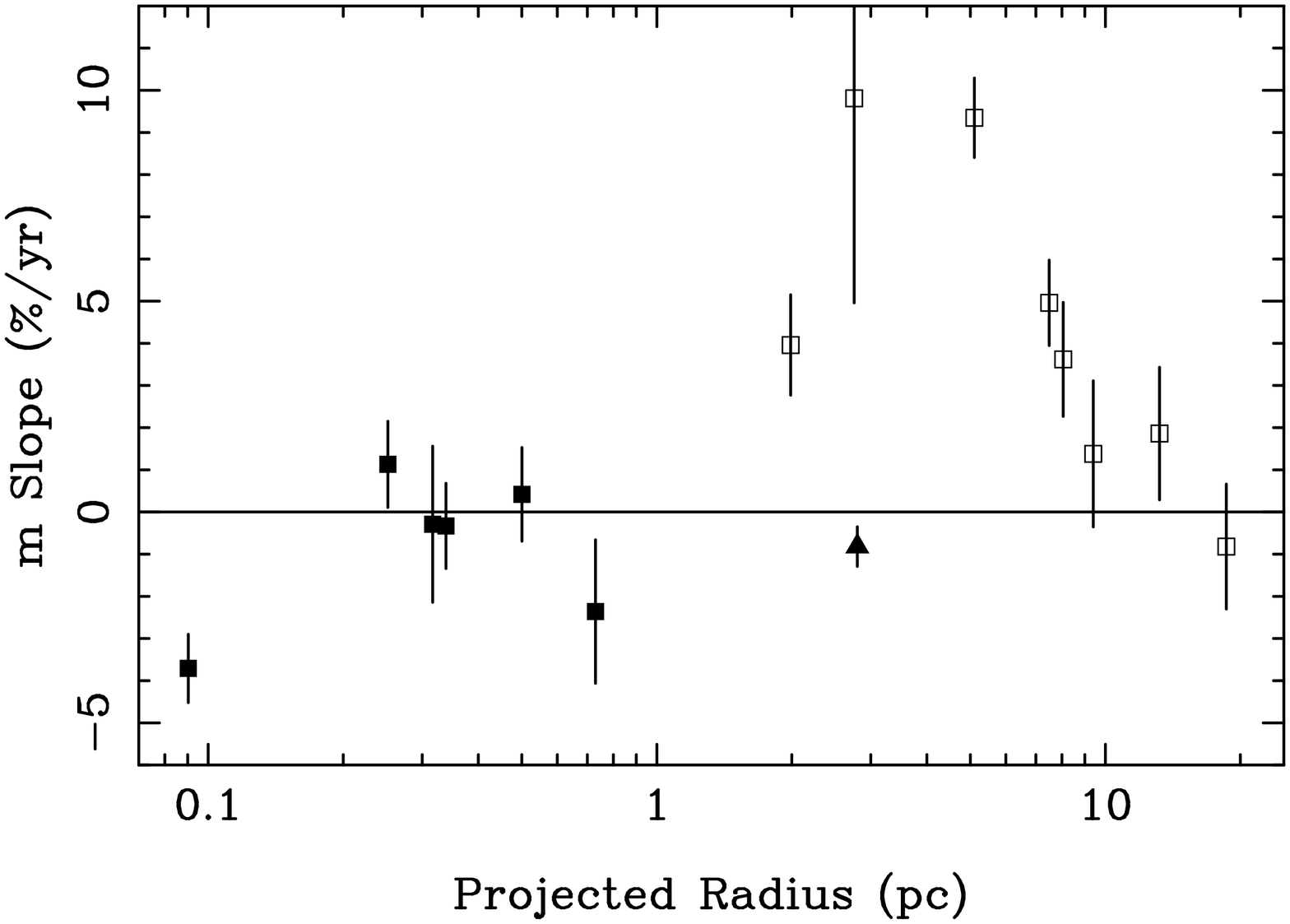,width=3.5in,angle=-90}
\end{center}
\figcaption[f19.eps]{\label{f:m_slope}
Fractional polarization slope, $\langle\dot{m}\rangle$ (see table
\ref{t:m_var}), 
plotted against projected
radius in parsecs.  Filled squares represent core regions, and open squares 
represent jet features.  J1312$+$32, where we have analyzed the
total VLBI flux of the source, is represented by a filled triangle.
}
\end{figure}

Figure \ref{f:m_slope} plots the linear slopes in fractional
polarization over our year long observations against
projected radius.  As with flux
evolution, there are distinct differences between core 
regions and jet features.  Only one core region shows a 
significant linear trend in fractional polarization, and this is 
OJ287 with a decay of $3-4$\% over the year.

%\placefigure{f:m_slope}

Several jet features have significant increases
in fractional polarization, and no jet feature shows a 
significant decrease.  
This implies that jet features either experience
a growth in magnetic field order or emerge from behind
Faraday depolarizing screens.  Growth in magnetic
field order explains most of the increasing
fractional polarization observed.  We discuss spectral
changes, such as changes in any Faraday depolarization,
in section \ref{s:spectral}, and with the exception of
one feature in 3C\,273, we find little evidence of 
components emerging from behind depolarization screens.

It is interesting that the three jet features at the largest 
radii have the smallest linear trends in fractional
polarization, implying that the degree of magnetic 
field order changes slowly beyond a projected 
radius of $\sim10$ parsecs.

\subsubsection{Fluctuations in Fractional Polarization}

\begin{figure}
\figurenum{20}
\begin{center}
\epsfig{file=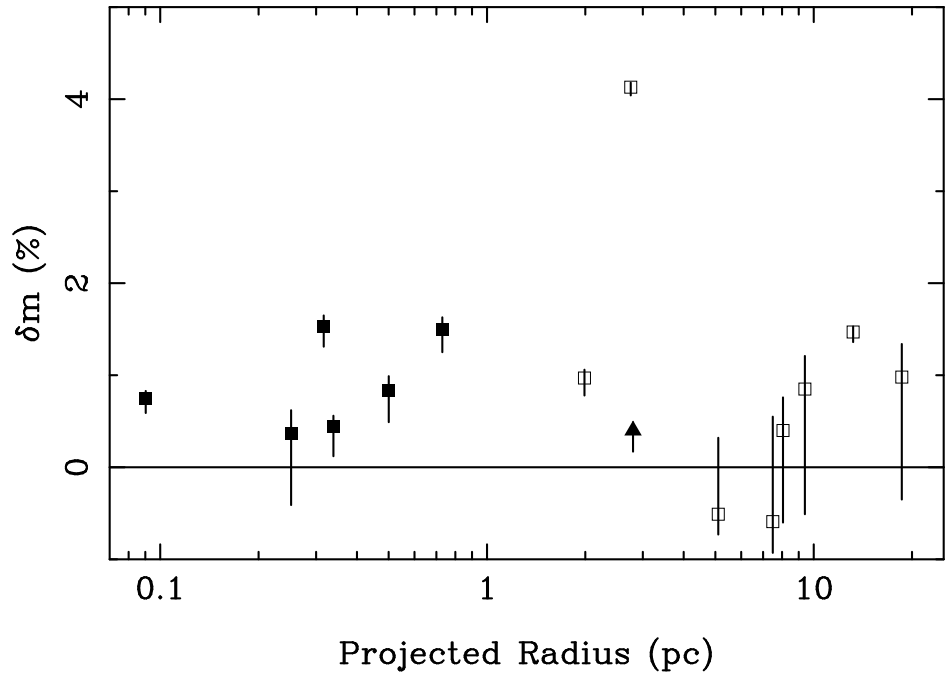,width=3.5in,angle=-90}
\end{center}
\figcaption[f20.eps]{\label{f:m_nl}
Standard deviation of the correlated fractional polarization fluctuations, $\delta m$, 
plotted against projected radius in parsecs.  Filled squares represent core 
regions, and open squares represent jet features.  J1312$+$32, where we have 
analyzed the total VLBI flux of the source, is represented by a filled triangle.
The high point in the figure at 4\% is the jet feature, U3, from OJ287.
}
\end{figure}

Figure \ref{f:m_nl} shows the standard deviation of the
correlated
fractional polarization fluctuations for each core and jet
feature plotted against projected radius.  
There are no clear trends or distinct differences
between core and jet features.  The high point at four percent
is the jet component, U3 (K3), in OJ287 which rises and then falls
in fractional polarization during our observations.

%\placefigure{f:m_nl}

\subsubsection{Linear Trends in Polarization Angle}
\label{s:chi_slope}

\begin{figure}
\figurenum{21}
\begin{center}
\epsfig{file=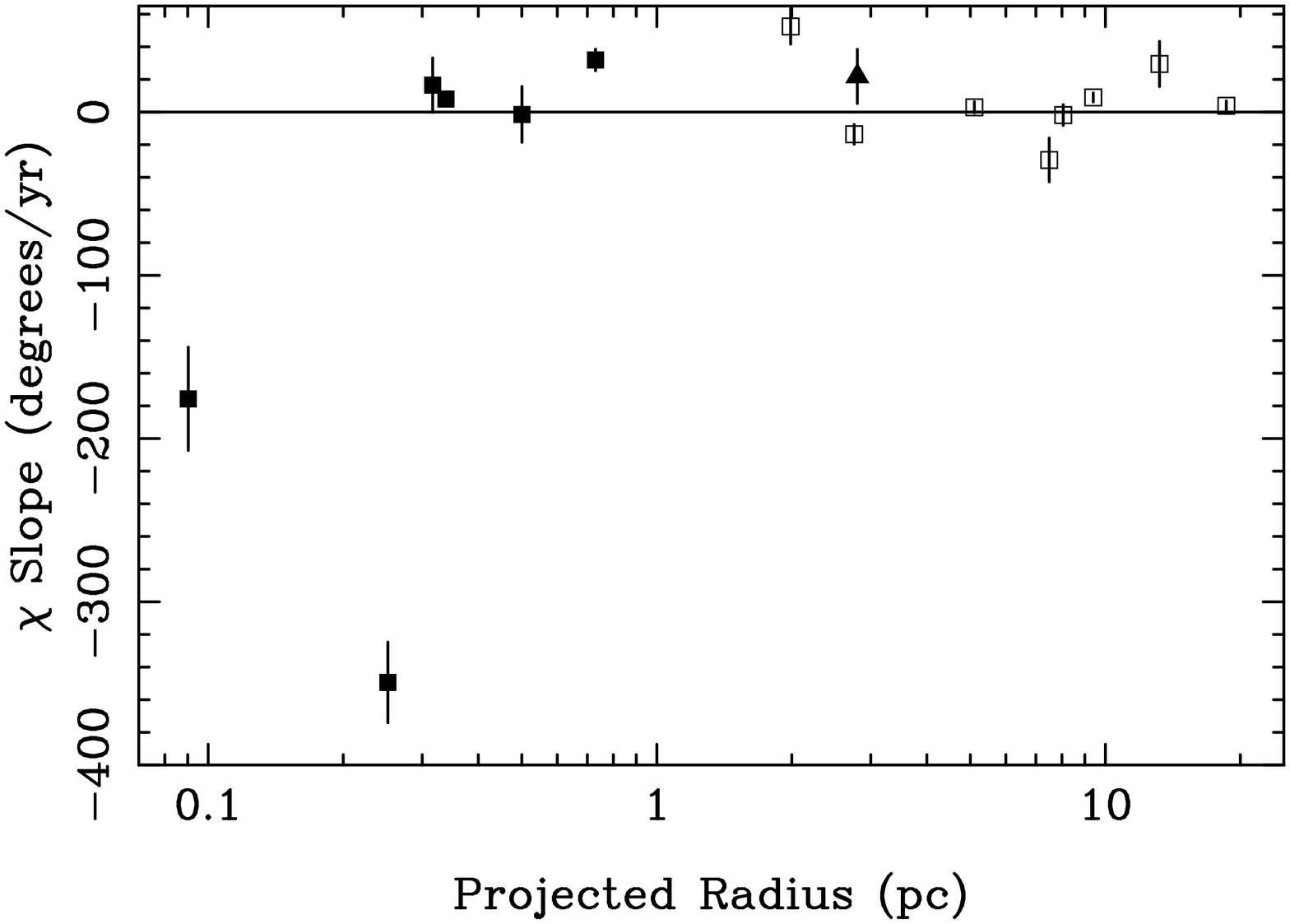,width=3.5in,angle=-90}
\end{center}
\figcaption[f21.eps]{\label{f:chi_slope}
Polarization angle slope, $\langle\dot{\chi}\rangle$ (see table
\ref{t:chi_var}), 
plotted against projected
radius in parsecs. 
Filled squares represent core regions, and open squares 
represent jet features.  J1312$+$32, where we have analyzed the
total VLBI flux of the source, is represented by a filled triangle.
}
\end{figure}

Figure \ref{f:chi_slope} plots the linear slopes with
time of the polarization position angles, $\chi$, of core
and jet features.  While there are several significant 
rotations in polarization angle, the core regions of OJ287
and J1512$-$09 stand out with slopes of $-175$ and $-350$
degrees per year respectively. 

%\placefigure{f:chi_slope}

\begin{figure}
\figurenum{22}
\begin{center}
\epsfig{file=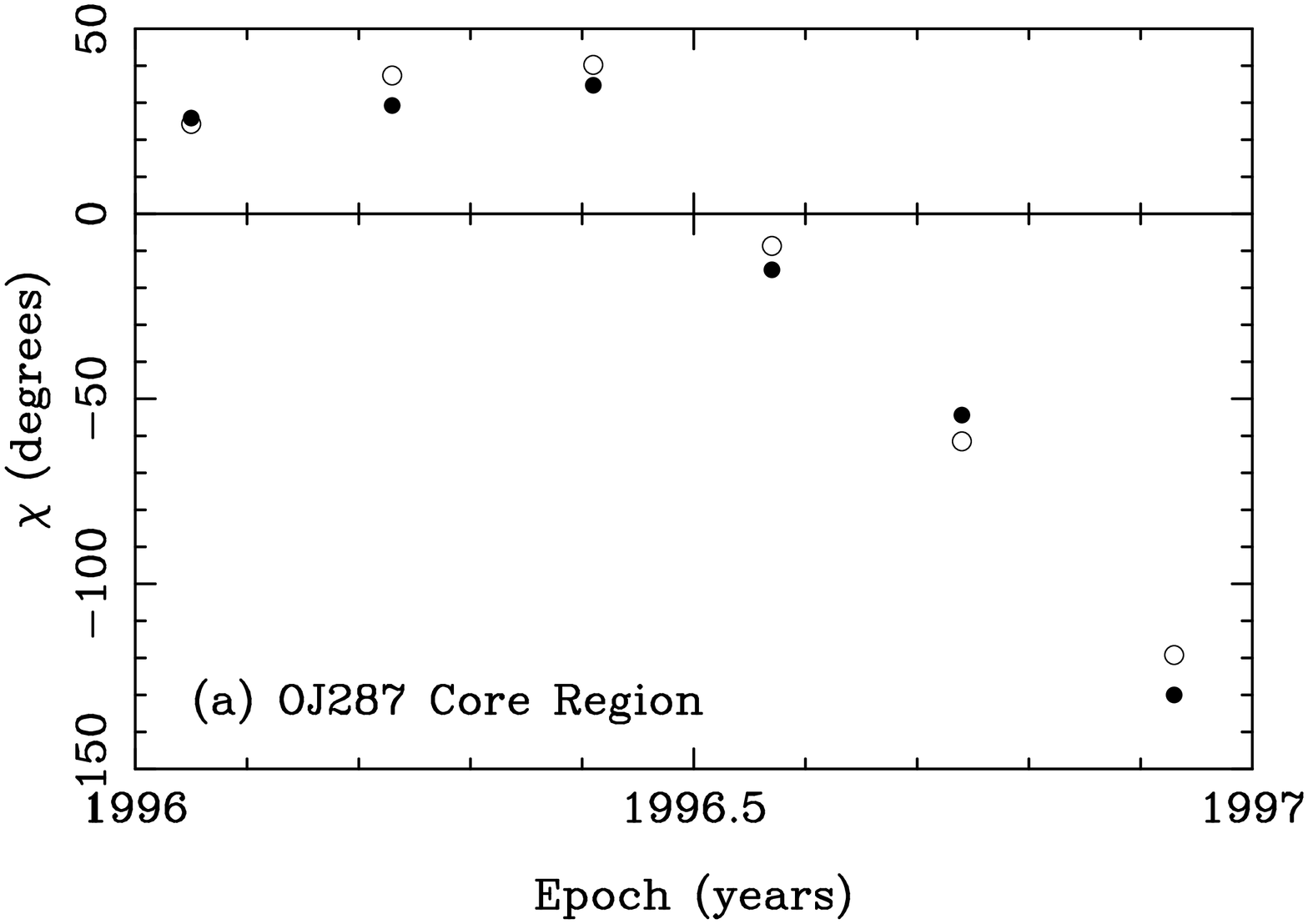,width=2.53in,angle=-90}
\epsfig{file=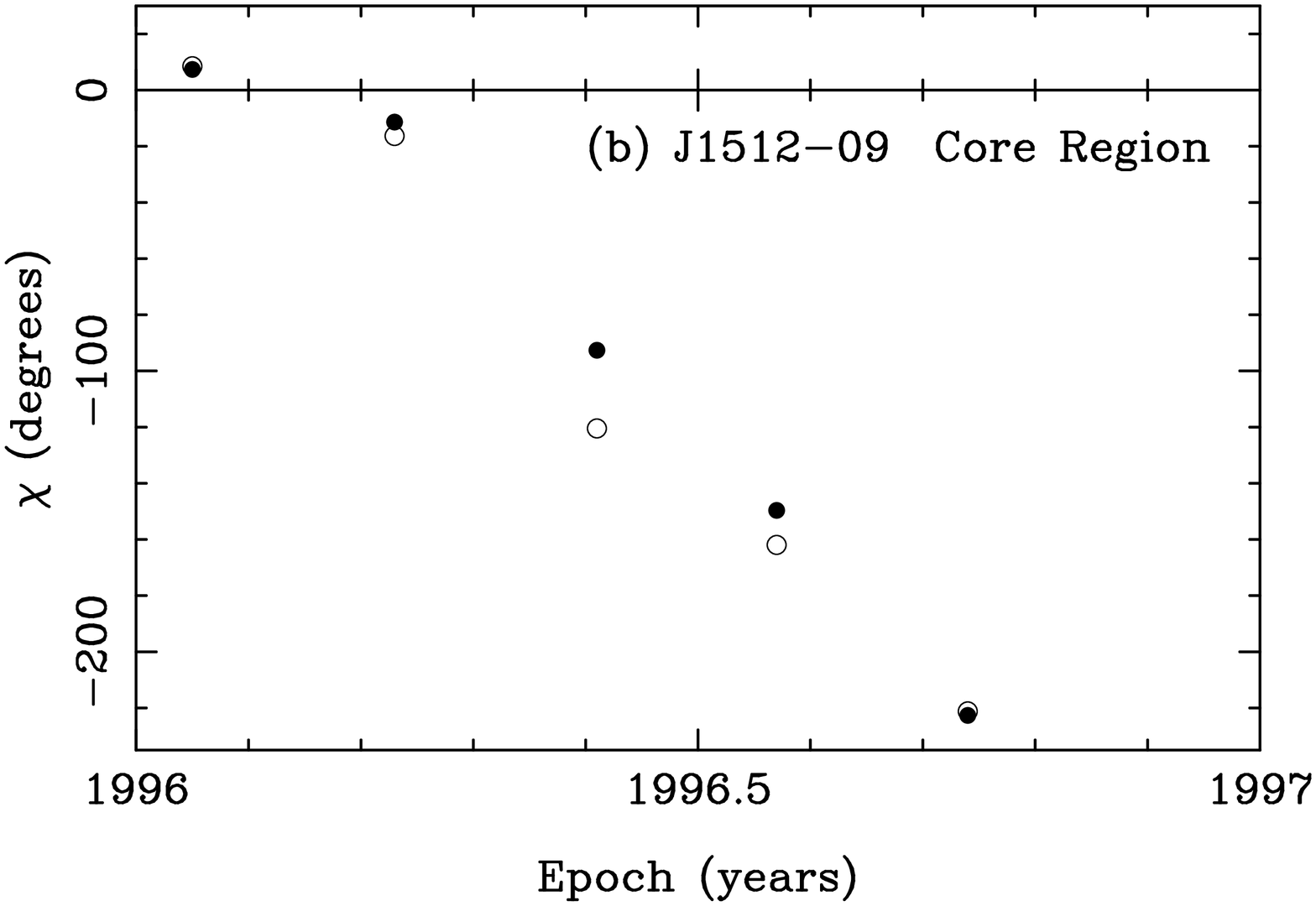,width=2.53in,angle=-90}
\end{center}
\figcaption[f22a.eps,f22b.eps]{\label{f:rotations}
Polarization angle, $\chi$, plotted against epoch for the core regions of OJ287  
(panel (a)) and J1512$-$09 (panel (b)).  Filled and open circles 
represent measurements at 15 and 22 GHz, respectively. 
}
\end{figure}

Figure \ref{f:rotations} shows plots of polarization angle 
versus epoch for the core regions of OJ287 and J1512$-$09.
With such large changes, the
$n\pi$ ambiguity in assigning the polarization angles
is an issue here.  We resolved this ambiguity by simply assigning 
the angles to give the smallest change between epochs.  
With this criterion, the only ambiguous
case was the jump between epochs 1996.23 and 1996.41 on J1512$-$09,
where the change was less than $90^\circ$ at 15 GHz and more than 
$90^\circ$ at 22 GHz.  The choice we made in this case (as 
displayed in the figure) was consistent with the sign of rotation 
between the other epochs.

%\placefigure{f:rotations}

The very large rotations in both of these objects are extremely
interesting.  Unlike the large polarization angle rotation
observed in 3C\,120 by \citet{G00}, the rotations are the same at the two 
frequencies and cannot be due to Faraday effects.  The rotations
must reflect changes in the observed net magnetic field direction which 
can be generated either by structural changes in the magnetic field,
trajectory changes of a highly polarized sub-component, or changes in
flow speed or angle which will change the (aberrated) angle of
observation.

The fact that the changes are much larger than $90$ degrees rules out
opacity effects and simple shock or shear models which would align the 
magnetic field
either perpendicular or parallel to the jet axis.  The regular nature
of the rotations is very curious, as is the fact that there is no
clear relationship between the rotations and either the total intensity
or fractional polarization behaviors.  The fractional polarization 
of the core region of OJ287 steadily declines over the last four epochs 
(during the rotation) from 4\% to 2\% while the flux has small, irregular, 
epoch to epoch fluctuations.  For J1512$-$09, the fractional 
polarization of the core region fluctuates irregularly between 1\% and 
3\% during our observations while the flux shows a single large 
rise and fall.  Although the polarized flux in the core
region of J1512$-$09 falls below our analysis threshold in 
the final epoch, 1996.93, we note that the 15 GHz
polarization image in 1996.93 shows nearly the same
polarization angle for the core as the 1996.74 epoch, indicating
that the large core rotation had abruptly stopped.

Polarization ``rotators'' have been observed in integrated 
polarization observations for over 20 years, e.g. 
\citep{LA79,A80,AAH81}.  The
acceleration-aberration model of \citet{BK79} can not 
account for rotations larger than 180 degrees, and either 
a true physical rotation or a quasi-circular motion of the 
emitting region seems to be at work \citep{AHA81}.
\citet{J85} interpreted observations of polarization rotations
in terms of a random walk model when the 
magnetic field evolution is dominated by turbulence.
More recently, \citet{HAA98} performed a wavelet analysis on two 
decades of UMRAO observations of OJ287.  Their observations were
not consistent with the random walk model, and the analysis 
suggested a small amplitude, cyclic fluctuation in the 
flow direction.  Our observations cannot easily distinguish
between these models, but they do provide some constraints.
The rotations we observed occurred in the compact core 
region ($\lesssim 0.3$ pc, in projection).  As noted above, the 
rotations were large ($\gtrsim 180^\circ$) and had no clear 
relation with the flux and fractional polarization evolution 
of the core region.  Additionally, the 
observed rotations are {\em not} seen in the integrated UMRAO 
monitoring observations that accompany our VLBI observations.
In the integrated measurements, these rotations are muted by
contributions from polarized jet features, suggesting that 
polarization rotations may be even more common 
than previously revealed by integrated monitoring.

%\placefigure{f:chi_theta}

We noted above that these two large polarization angle rotations
cannot be due to Faraday rotation as they have the same magnitude
at both frequencies.  
The following analysis of polarization rotations observed in
jet features is made with the assumption that changes
in the net magnetic field direction primarily drive all of 
the polarization angle changes we observe.  This assumption 
is supported in section \ref{s:farad}. 

\begin{figure}
\figurenum{23}
\begin{center}
\epsfig{file=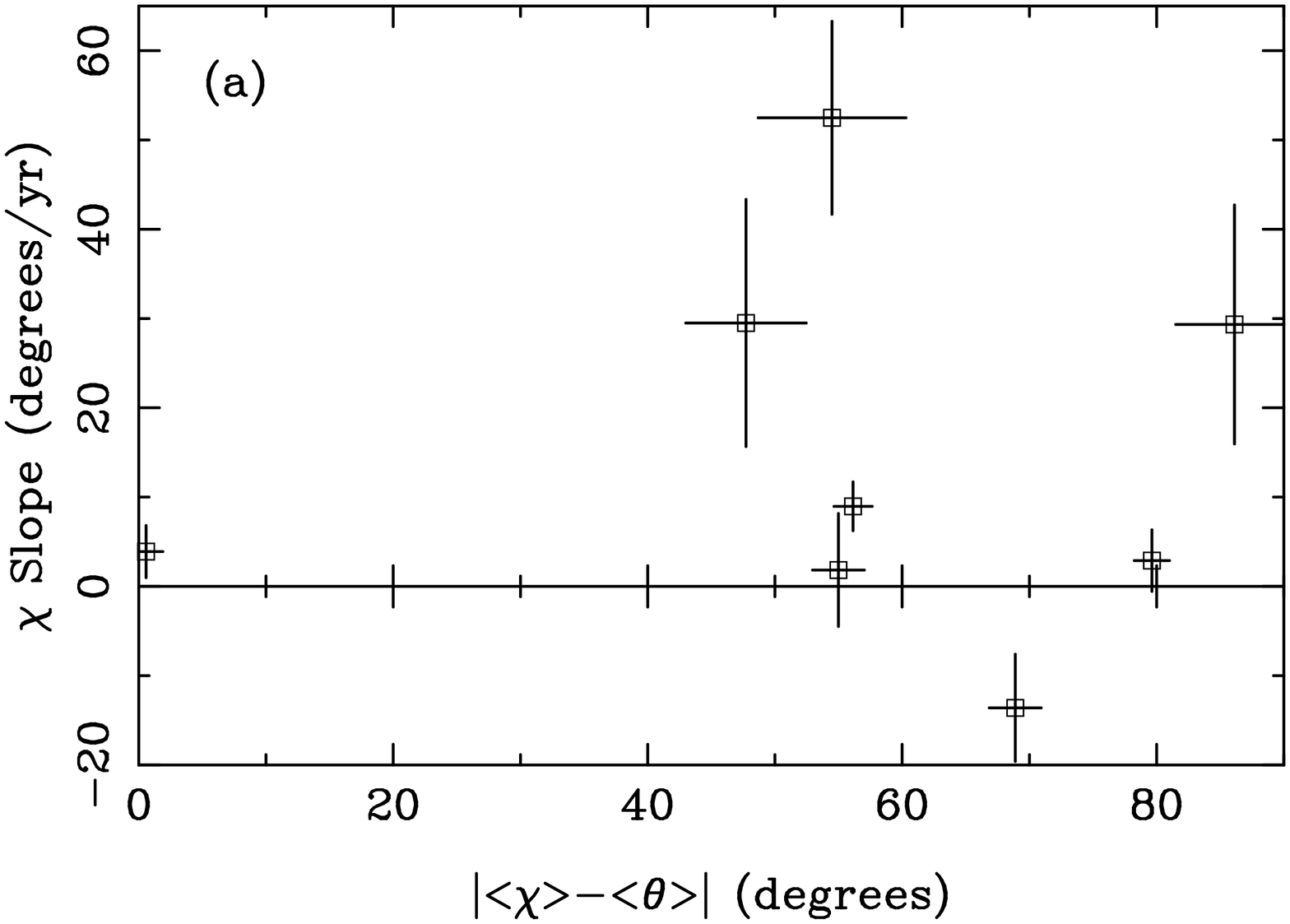,width=2.6in,angle=-90}
\epsfig{file=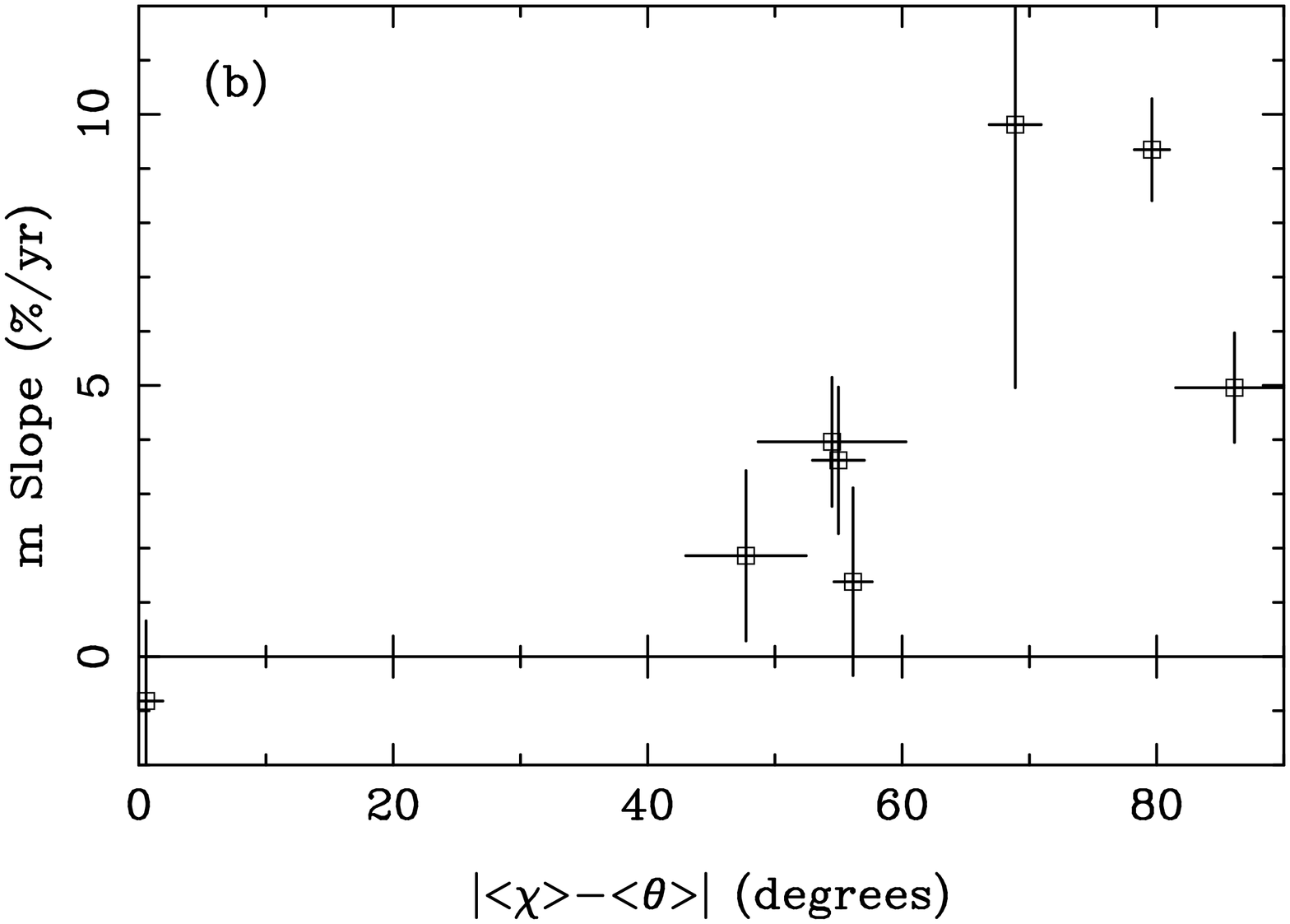,width=2.6in,angle=-90}
\end{center}
\figcaption[f23a.eps,f23b.eps]{\label{f:chi_theta}
Plots of polarization angle slope, $\langle\dot{\chi}\rangle$, (panel
(a)) 
and fractional polarization slope, $\langle\dot{m}\rangle$, (panel
(b)) 
versus the ``mis-alignment'' between mean polarization
angle, $\langle\chi\rangle$, and mean structural position angle, 
$\langle\theta\rangle$, of jet features.   
$|\langle\chi\rangle-\langle\theta\rangle| = 90^\circ$
indicates magnetic field aligned parallel to the structural position angle.
In panel (a), the polarization angle slopes are positive if they are rotations
in the direction of aligning the magnetic field with the structural position 
angle.
}
\end{figure}

It is interesting to examine the relationship between the linear
trends in polarization angle and the alignment
with the structural position angle of the jet feature, $\theta$.  Panel (a)
of figure \ref{f:chi_theta} shows polarization angle rotation as a 
function of the difference between the mean polarization angle and
the mean structural position angle.  In this plot, a positive slope
indicates a rotation of the polarization angle such as to align
the magnetic field more closely with the mean structural
position angle, and a negative slope indicates a rotation away from 
magnetic field alignment.\footnote{
In the absence of strong Faraday effects the polarization
position angle for synchrotron radiation is perpendicular to the
net magnetic field direction assuming the optical depth is 
$\lesssim 7$.}  
It is interesting that both $\geq 3 \sigma$ slopes and two of 
three $\geq 2 \sigma$ slopes are in the direction of aligning 
the magnetic field along the jet axis (assumed to be parallel
to the mean structural position angle).  

Panel (b) of figure \ref{f:chi_theta} plots fractional polarization
slope against the same 
$|\langle\chi\rangle-\langle\theta\rangle|$ axis as in panel (a).
As discussed in section \ref{s:m_line}, the only significant 
linear trends we see in the fractional polarizations of jet 
features are increases with time. With the possible exception
of a jet feature from 3C\,273, we believe the positive slopes
reflected increasing magnetic field order in jet features.  Taken
with the observation from panel (a) that the magnetic field of jet
features tends to rotate in the direction of alignment with the
jet axis, we have a tentative picture suggesting increasing longitudinal
field order, which is (1) produced in the jet features themselves
as they propagate, perhaps through shear, and/or (2) part of the underlying 
flow which is sampled (and enhanced) by the passing jet features.  

While the above picture regarding increasing longitudinal
field in the jets is attractive, we note that in only one 
jet feature (U1A$+$U1B in 3C\,120) do we clearly observe 
{\em both} an increase in magnetic field order and a rotation towards 
magnetic field alignment with the jet axis. (\citet{GMA99} observe 
qualitatively
the same behavior in two subsequent jet features, their F and G, 
in 3C\,120.)  We also note that our sample of jet 
features is dominated by magnetic fields within $45^\circ$ of alignment 
with the jet axis; a collection of jet features with magnetic fields 
oriented at larger (``shock-like'') angles to the jet axis might 
behave quite differently.  At 5 GHz, \citet{CWRG93} found a trend of 
increasing longitudinal magnetic field order with jet radius 
in quasars, which tended to have magnetic fields aligned with
the jet axis.  They found this trend by plotting single epoch 
observations of many sources on one plot.  Here we 
find a similar result by observing the dynamic evolution 
of the magnetic field of individual jet features.

\subsubsection{Fluctuations in Polarization Angle}

\begin{figure}
\figurenum{24}
\begin{center}
\epsfig{file=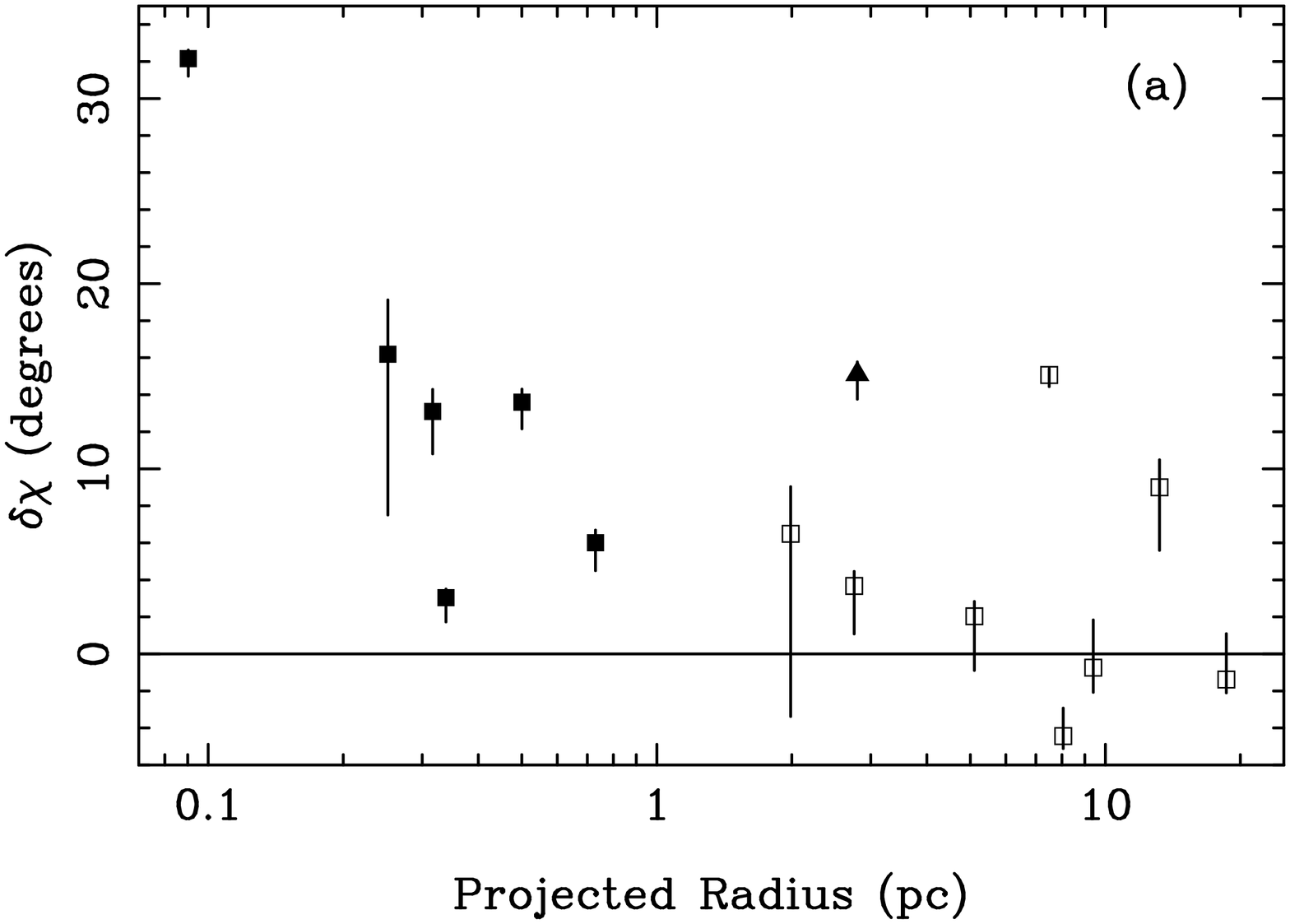,width=2.6in,angle=-90}
\epsfig{file=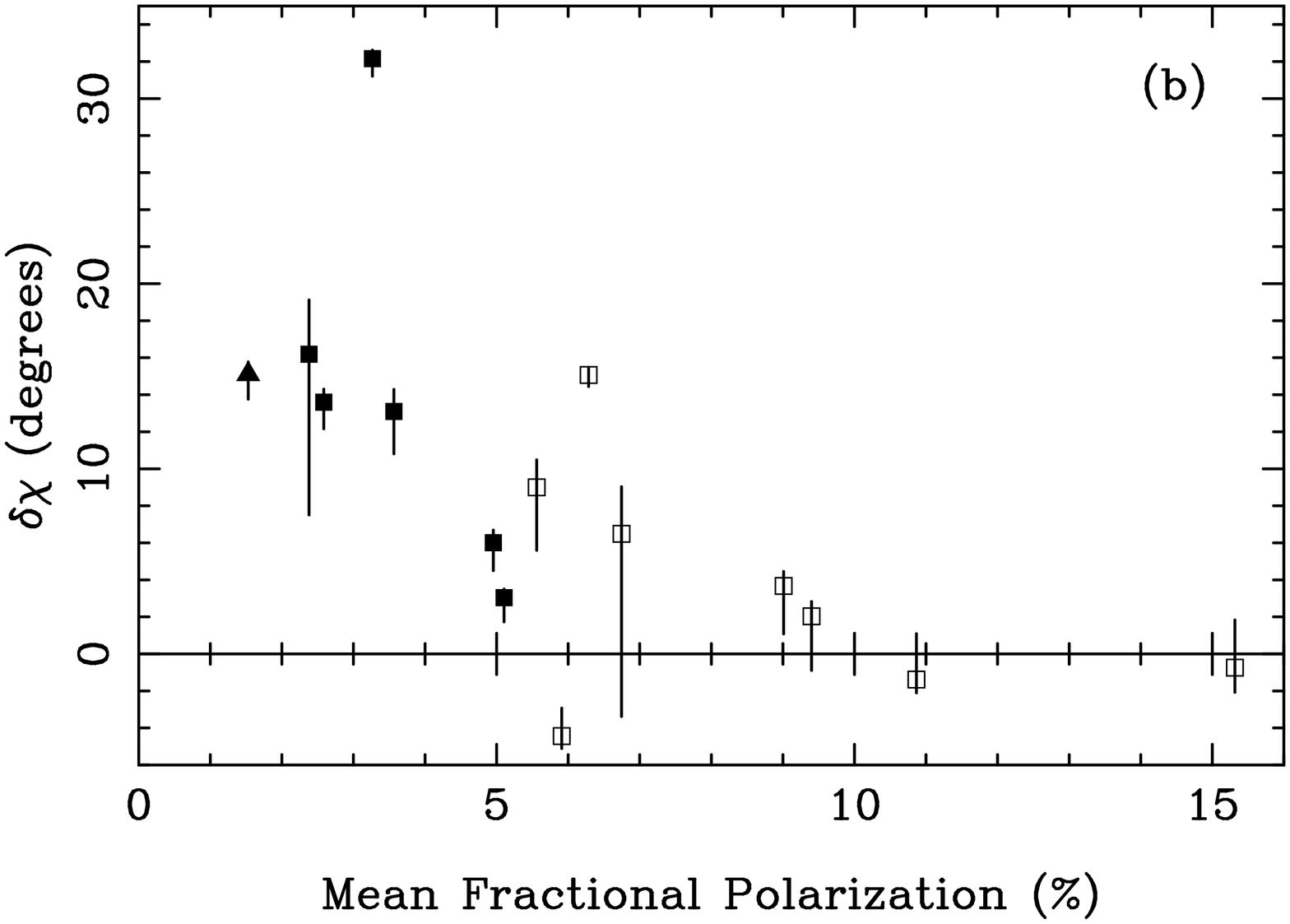,width=2.6in,angle=-90}
\end{center}
\figcaption[f24a.eps,f24b.eps]{\label{f:chi_nl}
Standard deviation of the correlated polarization angle fluctuations, 
$\delta\chi$, plotted against projected radius in parsecs in panel (a).  
In panel (b), the standard deviation is plotted against mean fractional
polarization.  Filled squares represent core regions, and open squares 
represent jet features.  
J1312$+$32, where we have analyzed the total VLBI flux of the source, 
is represented by a filled triangle.
}
\end{figure}

Figure \ref{f:chi_nl} plots the standard deviation of the 
correlated polarization angle fluctuations against projected radius (panel a)
and mean fractional polarization (panel b).  While we see very
little relationship with projected radius, the polarization angle 
fluctuations are smaller with larger fractional polarization.  This
may imply that polarization angle fluctuations are 
generated by relatively small scale orderings or re-orderings of
the magnetic field.  For highly polarized components, such small
scale changes would have little effect on the net polarization
angle direction.  

%\placefigure{f:chi_nl} 

\subsubsection{Are Polarization Angle Changes due to Faraday Rotation?}
\label{s:farad}

\begin{figure}
\figurenum{25}
\begin{center}
\epsfig{file=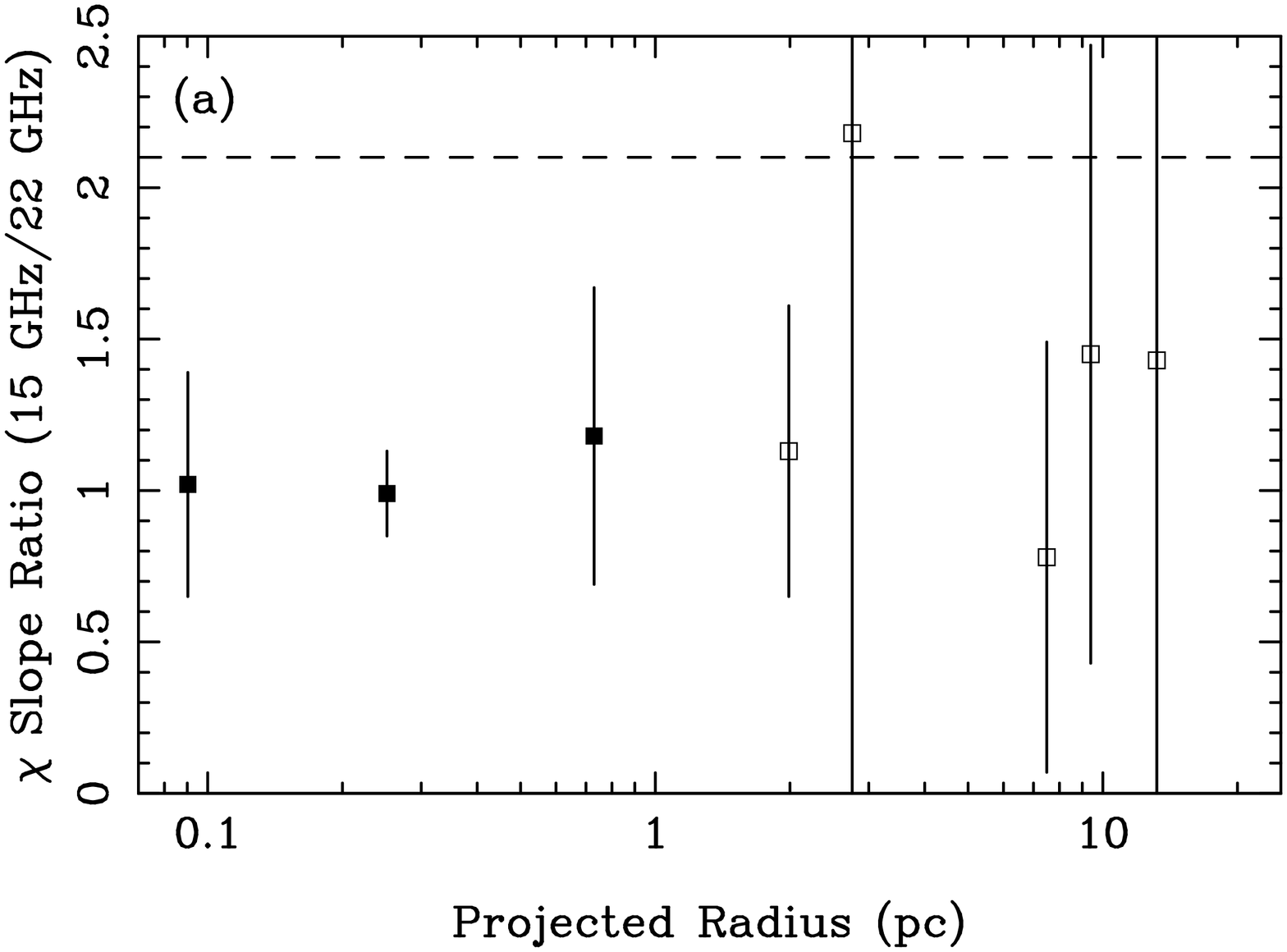,width=2.6in,angle=-90}
\epsfig{file=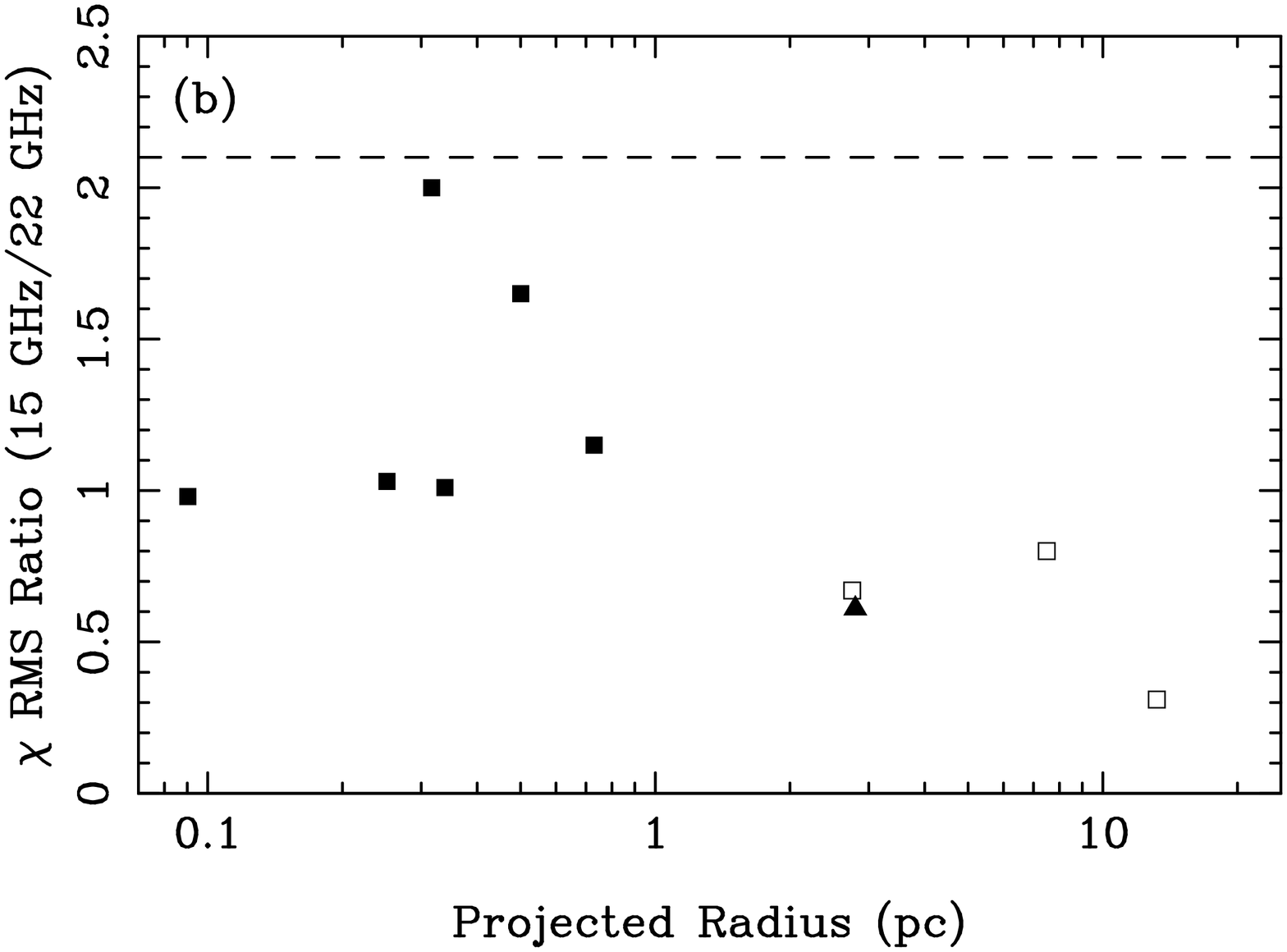,width=2.6in,angle=-90}
\end{center}
\figcaption[f25a.eps,f25b.eps]{\label{f:chi_farad}
Ratio of $\chi$ changes at 15 GHz to those at 22 GHz plotted against
projected radius. Panel (a) displays the ratio of linear $\chi$ slopes 
at the two frequencies.  Only features with a mean $\chi$ slope greater
than $2\sigma$ are plotted.  Panel (b) displays the ratio of the RMS
fluctuations at the two frequencies.  Only features where the fluctuations
are correlated at the $r > 0.5$ level are plotted.
Filled squares represent core regions, and open squares represent 
jet features.  J1312$+$32, where we have analyzed the total VLBI flux 
of the source, is represented by a filled triangle.
The dashed lines indicate where the points should cluster if Faraday
rotation was primarily responsible for the observed polarization 
angle changes.
}
\end{figure}

We evaluate this question by plotting the ratio of the polarization 
angle changes at 15 GHz to those at 22 GHz in figure
\ref{f:chi_farad}.  If Faraday rotation 
is primarily responsible for the
changes we observe, the changes at 15 GHz should be 2.1 times
those at 22 GHz.  Panel (a) of figure \ref{f:chi_farad} plots
the linear slope ratio, and panel (b) plots the ratio of the 
RMS fluctuations.  Low signal to noise points have been filtered
out by only plotting features with a mean slope
of greater than $2\sigma$ significance in panel (a) and by only
plotting features where the fluctuations are correlated 
at the $r > 0.5$ level in panel (b).  While there is still some 
scatter in the plots, the data clearly cluster around ratios of 1.0
rather than 2.1.  A ratio of 1.0 is precisely what we expect
if the observed polarization angle changes reflect changes 
in the net magnetic field direction. 

%\placefigure{f:chi_farad}

We can investigate this further by examining the mean position
angle difference, $\langle\Delta\chi\rangle$, between frequencies 
in table \ref{t:comb}, and 
we see little evidence for large mean Faraday rotations in the 
core regions and jet features we follow.  The difference between
the two frequencies is typically a few degrees with error bars 
comparable to or larger than the difference.  These results
are consistent with \citet{T98,T00} who finds observed rotation measures  
of $\sim 1000$ rad/m$^2$ ($\Delta\chi_{15-22}\approx 10^\circ$)
for quasars cores and falling off to $< 100$ rad/m$^2$ 
($\Delta\chi_{15-22} \lesssim 1^\circ$) beyond a projected radius
of 20 pc.  Rotation measures affecting individual features 
would need to change by $\geq 100\%$ (in several cases many 
times this amount) in less than a year to explain the 
significant polarization angle changes that we observe. 
\citet{ZT01} recently reported a change of $800$ rad/m$^2$ over 
1.5 years in the core of 3C\,279; they also observed spatial
variations spanning $4000$ rad/m$^2$ in the inner 10 pc of 
3C\,273. 

While large rotation measure changes are possible, particularly
in AGN cores, we do not find evidence that such variability
in the Faraday screens is responsible for the polarization 
angle changes that we observe.  In general, the polarization 
angle changes are the same magnitude at both 
frequencies, and are best explained by changes in the net
magnetic field direction. In examining the data for individual 
features in detail, no case stands out as having significant
polarization angle changes driven by Faraday rotation, although we 
cannot Faraday rotation as the cause in a few cases.  

\subsection{Spectral Variability}
\label{s:spectral}

Table \ref{t:comb} lists average spectral properties
(spectral index, $\alpha$, polarization ratio, 
$m_{ratio}$, and polarization angle difference, $\Delta\chi$)
and their respective linear slopes with time over our
year-long window of observation.  Here we are 
primarily interested in the spectral evolution of
core and jet features as revealed by these linear
slopes.

%\placetable{t:comb}

All of the $\geq 3\sigma$ slopes are in spectral index.
The core regions of 3C\,273 and 3C\,279 both have a negative
spectral index slope indicating a shift towards more optically
thin radiation. Both of these core regions undergo outbursts
during our observations with a distinct rise and plateau (with 
perhaps a small fall) in total core flux, and in both cases we
clearly see the development of new jet components in the
core region (see Paper I).  In 3C\,273 the spectral changes in
the core are linked with the sudden appearance of circular 
polarization in the middle of 1996 \citep{HW99}.

The core regions of J1751$+$09 and J2005$+$77 have distinctly
positive spectral index slopes indicating a shift towards more
optically thick radiation.  Both of these core regions show
a flux dip early in our year-long window of observations followed
by a rise in flux at the end of the year.  In J1751$+$09, both 
dip and rise occur sharply as discussed in \S{\ref{s:nl_flux}}, and
in J2005$+$77 the dip and rise are more gradual.  The shift
towards higher optical depth may be connected with particle
injection at the start of a new outburst.  Together with the
observations of 3C\,273 and 3C\,279, we may be seeing different
phases of a single outburst/component ejection cycle 
(e.g. \citep{1999ApJS..120...95V}) similar to that observed 
in UMRAO single dish monitoring \citep{AAH96} with a shift 
towards higher opacity very early in the cycle as the outburst 
is developing and a shift towards lower opacity as the outburst 
peaks and a new component is ejected. 

We also see a positive spectral index slope in
the jet feature U1 (K1) of J1927$+$73.  In the same source we detect 
a similar positive slope at the $2.3\sigma$ level on the nearby 
component U2 (K2).  These two components are nearly side by
side in the jet and move on radial trajectories (see Paper I).
Their simultaneous shift towards flatter spectral index may be
related to passing through local jet conditions such as a mild 
standing shock where particle re-acceleration is occurring.  
It is interesting to note that the feature U1 (K1) also has 
significant $\sim 10$\% fluctuations in its flux.

\begin{figure}
\figurenum{26}
\begin{center}
\epsfig{file=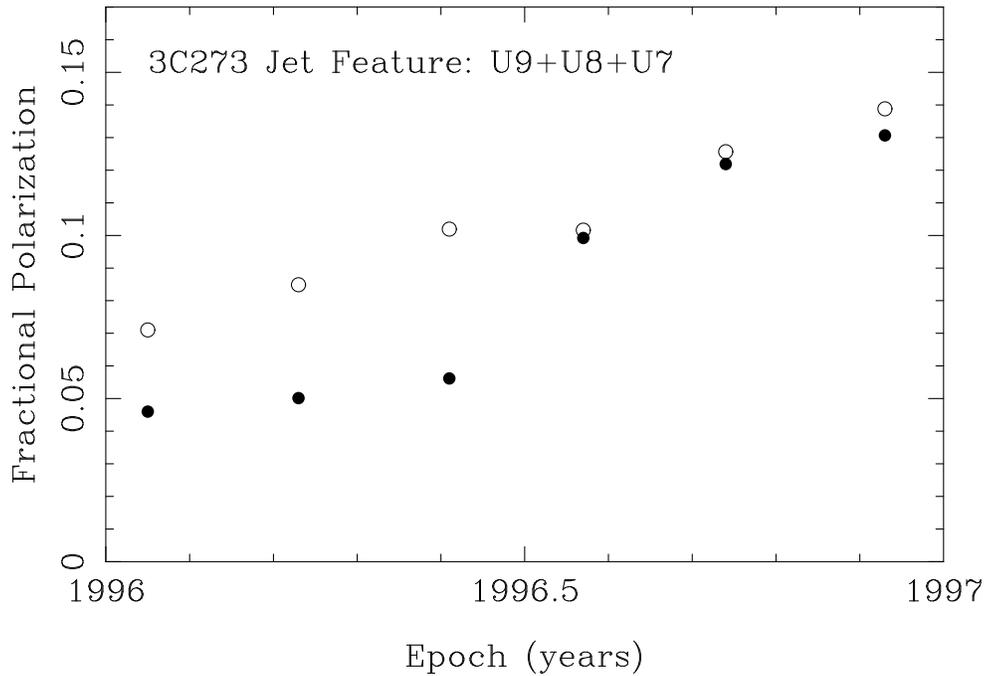,width=3.5in,angle=-90}
\end{center}
\figcaption[f26.eps]{\label{f:3c273m}
Fractional polarization of the jet feature, U9$+$U8$+$U7 (K9$+$K8$+$K7)
in 3C\,273, plotted against epoch.  Filled and open circles 
represent measurements at 15 and 22 GHz, respectively. 
}
\end{figure}

While there are no $\geq 3\sigma$ slopes (and only two $\geq 2\sigma$)
in either fractional polarization ratio or polarization angle
difference between the frequencies, there are two individual
variability events in fractional polarization that are interesting
to consider.  The first is on the jet feature U9$+$U8$+$U7 in 3C\,273,
and figure \ref{f:3c273m} shows fractional polarization at both 
15 and 22 GHz plotted against epoch.  The fractional polarization
ratio makes a distinct shift in the middle of the observations from
being strongly depolarized at 15 GHz relative to 22 GHz to having 
roughly equal amounts of polarization at the two frequencies.  
The fractional polarization of this feature increases continually
during our observations, and a more detailed examination of 
sub-component behavior leads us to believe that much of the fractional
polarization increase in this feature is due to emergence from
behind a Faraday depolarizing screen (J. F. C. Wardle et al.,
in preparation).  We do not see similar evidence for emergence
from behind Faraday depolarizing screens in other jet features.

%\placefigure{f:3c273m}
%\placefigure{f:3c120m}

\begin{figure}
\figurenum{27}
\begin{center}
\epsfig{file=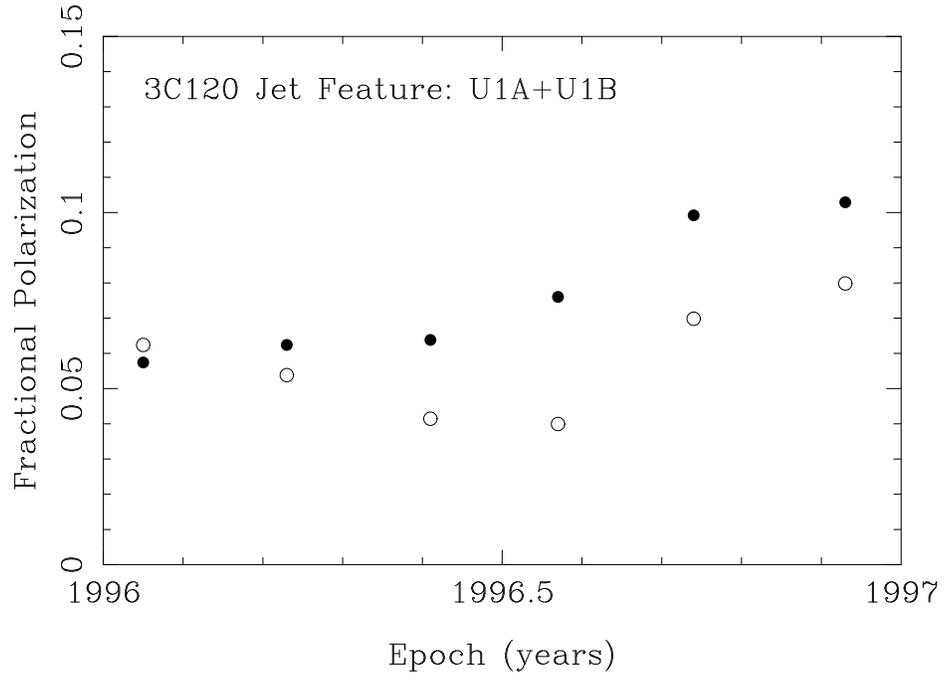,width=3.5in,angle=-90}
\end{center}
\figcaption[f27.eps]{\label{f:3c120m}
Fractional polarization of the jet feature, U1A$+$U1B (K1A$+$K1B)
in 3C\,120, plotted against epoch.  Filled and open circles 
represent measurements at 15 and 22 GHz, respectively. 
}
\end{figure}

Jet feature U1A$+$U1B (K1A$+$K1B) in 3C\,120 also has interesting
fractional polarization changes between 15 and 22 GHz, and its 
fractional 
polarization over time is plotted in figure \ref{f:3c120m}.   After
the second epoch, we observe a distinct separation in the fractional
polarization at the two frequencies with fractional polarization at
15 GHz being noticeably {\em larger}.  Interestingly, the two frequencies
track one another in the jump in fractional polarization from 
epochs 1996.57 to 1996.74.  The higher levels of polarization at
15 GHz are clearly visible in our maps as well as our model-fits, and
cannot be explained by traditional Faraday depolarization which 
would have less fractional polarization at 15 GHz.  Satisfactory
explanations for the observed effect require at least a two component
model, such as a spine-sheath structure, where one component has a
more highly ordered field and a steeper spectral index.  The required 
difference in spectral index between
sub-components is large if the two components have
similar polarizations or even if one of the components is unpolarized.  
We can reduce the required spectral index difference considerably if 
the two sub-components have significant polarization cancellation 
between them.
In \S{\ref{s:chi_slope}} we noted this jet feature showed both 
an increase in fractional polarization and a rotation of the 
polarization angle indicating increasing magnetic field order 
along the jet axis.  Perhaps we are seeing an increasing
contribution from ordered magnetic field in a shear layer
which has a distinctly steeper spectrum than the bulk of the
jet.

\section{Summary}
\label{s:conclude}

We have analyzed the flux and polarization evolution of 
twelve parsec-scale radio jets over a single year.   These objects 
were monitored with the VLBA
at 15 and 22 GHz for six epochs, spaced at approximately two month intervals, 
during 1996.  We analyzed the flux, fractional polarization, and
polarization position angle behavior of both core regions and jet features. 
Our analysis considered both the linear trends of these quantities
with time and the fluctuations about the linear trends.  With dual
observing frequencies, we were able to examine spectral trends
and distinguish between Faraday effects and changes in the net magnetic
field directions of core regions and jet features.  The closely spaced 
frequencies were also extremely 
helpful in assessing the reality of the fluctuations we observed.

The key results of our analysis include the following:

(1) Jet features generally decayed in flux, with older features decaying
more slowly than younger features.  There was a distinct relation between 
the age of a feature, as measured by its position and proper motion, 
and its rate of flux decay.  We found that this relationship could be
explained if the flux of a jet feature depended only upon its position in 
the jet: $I \propto R^{-1.3\pm0.1}$. 
(We note that at least one jet feature did show 
significant brightening between some epochs, and another jet feature 
had a significant flux decay with no apparent proper motion.)

(2) We observed significant fluctuations in the flux of both 
core regions and jet features.  Core regions tended to have larger 
fractional fluctuations than jet features, with more compact core 
regions having larger fluctuations.  

(3) Jet features either had a significant increase in fractional polarization 
or showed no change, with the smallest changes in the features at the largest
projected radii.   With the exception of a single jet feature
(U9$+$U8$+$U7 in 3C\,273), we saw no evidence for features emerging from 
behind Faraday depolarizing screens, and increasing magnetic field order 
explains most of the increasing fractional polarization we observed.

(4) Changes in the net magnetic field direction were the primary cause
of the polarization angle changes we observed.   The linear rotations and 
 fluctuations in polarization angle were of the same size at 
15 and 22 GHz, and could not easily be explained by Faraday rotation which 
would have required the changes at 15 GHz to be twice the size of the 
changes at 22 GHz.

(5) We observed large ($\gtrsim 180^\circ$) rotations in the polarization
position angles of two core regions.  These rotations were the same at both
observing frequencies and could not be due to Faraday effects.  
The rotations were similar to those that have been observed in integrated 
measurements for over twenty years, and we could not easily distinguish
between previously proposed models for this phenomenon.

(6) The magnetic field of jet features tended to become increasingly
longitudinal.  Four out of five polarization angle 
rotations were in the direction of aligning the magnetic field with 
the jet axis.  This observation, coupled with the tendency towards 
increasing magnetic field order described above, suggests
increasing longitudinal field order.  However, we note that we 
clearly observed {\em both} a rotation towards alignment and 
increasing field order in only one jet feature, U1A$+$U1B in 3C\,120.
In this feature there was also an interesting ``negative depolarization'' 
effect with 15 GHz more highly polarized than 22 GHz, suggesting an
increasing contribution from a longitudinally ordered field in a shear
layer that has a steeper spectral index than the bulk of the flow.

(7) Polarization angle fluctuations decrease in amplitude with increasing 
fractional polarization.  Features with more highly ordered magnetic
fields would be less affected by small scale orderings or re-orderings 
of the field which might have driven the  polarization angle 
fluctuations.

(8) We observed significant spectral index trends in the core regions 
of four sources and in the jet features of one object.  For the core regions,
the spectral index moved towards more optically thin radiation as
core outbursts in two sources came to a peak and plateaued (or declined
slightly), and in two other sources the spectral index moved 
towards more optically thick radiation in the gap between outbursts.  
In the jet of J1927$+$73, there was a shift to flatter spectral
indexes in two features that are moving side-by-side in the jet
on nearly parallel tracks.  These features may have experienced the same
local conditions in the jet, such as a mild standing shock where
particle re-acceleration was occurring.

\section{Acknowledgments}

This work has been supported by NASA Grants NGT-51658 and NGT5-50136
and NSF Grants AST 91-22282, AST 92-24848, AST 94-21979,   
AST 95-29228, AST 98-02708, and  AST 99-00723.
We thank C. C. Cheung and G. Sivakoff for their assistance in the 
initial organization of the modelfit data, and Alan Bridle for helpful 
discussion regarding jet flux profiles.  This research has made use of
the NASA/IPAC Extragalactic Database (NED) which 
is operated by the Jet Propulsion Laboratory, California Institute of 
Technology, under contract with the National Aeronautics and Space 
Administration. This research has also made use of NASA's Astrophysics 
Data System Abstract Service.

%%%%%%%%%%%%%%%%%%%%%%
%%                  %%  
%%    Appendix      %%
%%                  %%
%%%%%%%%%%%%%%%%%%%%%%

\appendix

\section{Empirical Estimates of the Uncertainties in 
Measuring VLBI Component Properties}

In section \ref{s:tech} we discussed the difficulties in 
deducing robust ``1 $\sigma$'' uncertainty estimates for
the fluxes and polarizations of core and jet features 
observed in VLBI jets.  Lack of good a-priori uncertainty
estimates led us to correlate observed fluctuations about the
linear trends 
between our two observing bands (15 and 22 GHz), as described
in \S{\ref{s:corr}}, to assess the reality of these fluctuations.
In this appendix, we take the methods in \S{\ref{s:corr}} a 
step further to obtain empirical estimates of the uncertainties
in measuring VLBI component properties.

Here we are interested in the un-correlated part of the
 fluctuations discussed in \S{\ref{s:corr}}.
\begin{equation}
\Omega_{15} = (1-r)\times\frac{\sum_i{\Delta_{15i}^{2}}}{n-2} \qquad 
\Omega_{22} = (1-r)\times\frac{\sum_i{\Delta_{22i}^{2}}}{n-2}. 
\end{equation}
Note that these expressions are very similar to equation 
\ref{e:corrvar} with $r$ replaced by $(1-r)$, so here $\Omega$ is 
the {\em un-correlated} part of the variance of the deviations 
at each frequency.  For negative values of $r$, we set $r=0$ for
the purposes of this calculation.  The square-root of $\Omega$
gives the standard deviation of the un-correlated fluctuations, 
and we have computed this for each quantity
and frequency: $\sigma I_{15} = \sqrt{\Omega_{I\,15}}$, etc.  
We used the second Monte Carlo simulation described in 
\S{\ref{s:corr}} to place rough uncertainties on these
$\sigma$ values.  Table \ref{t:err} presents these numbers for
each core and jet feature appearing in our analysis.

%\placetable{t:err}
   
These un-correlated fluctuations consist of 
(1) real spectral changes in core and jet feature 
properties, and (2) any measurement, calibration, {\em and}
model-fitting errors that are not systematic between 
epoch or frequency.  Because we have no way of separating
out the real spectral changes, we take these estimates
of the standard deviation of the un-correlated fluctuations 
as conservative estimates on the total (non-systematic) 
uncertainty in measuring VLBI flux and polarization of 
core and jet features in a single epoch. 

\begin{figure}
\figurenum{28}
\begin{center}
\epsfig{file=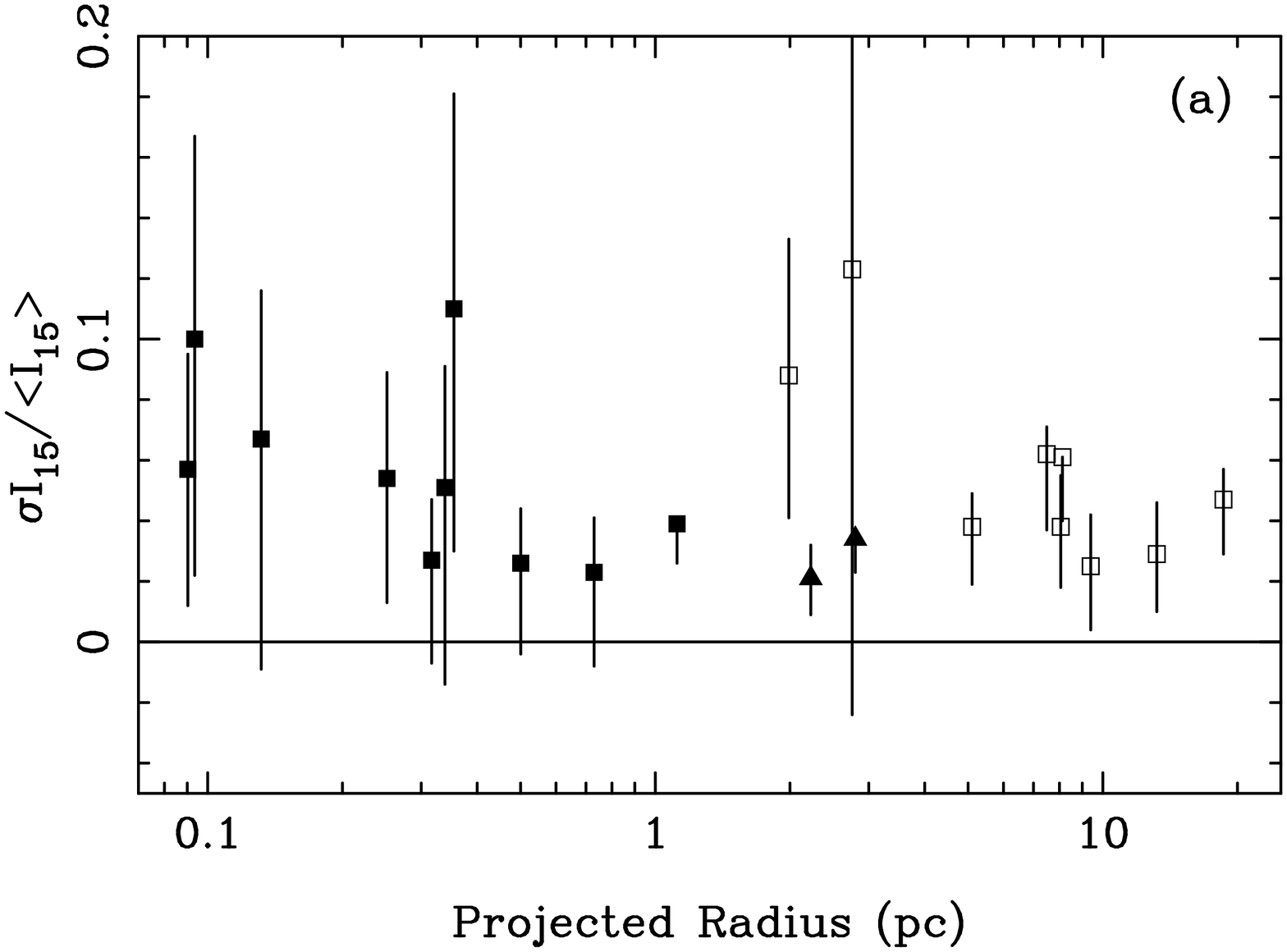,width=2.6in,angle=-90}
\epsfig{file=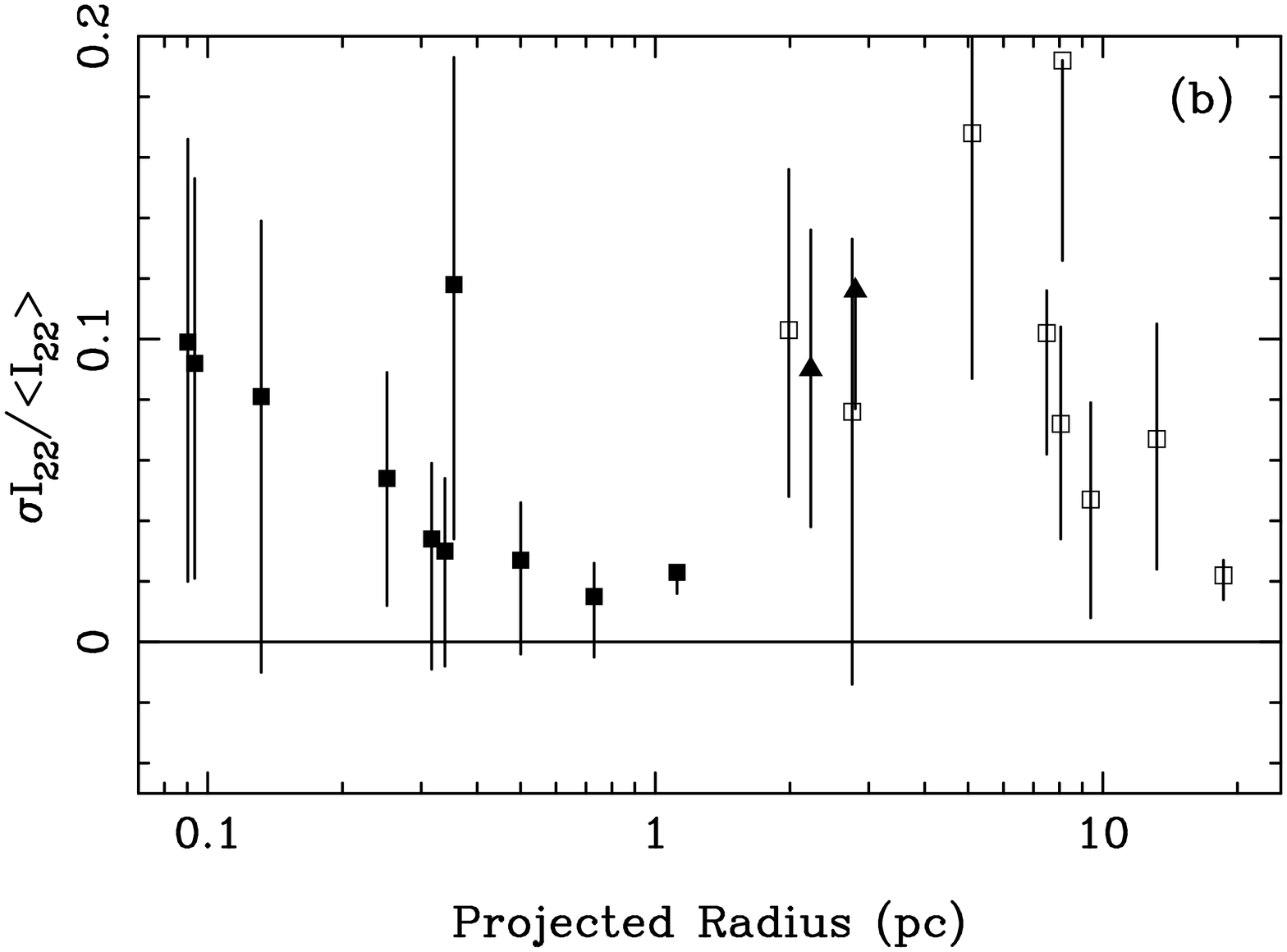,width=2.6in,angle=-90}
\end{center}
\figcaption[f28a.eps,f28b.eps]{\label{f:i_err}
Fractional flux uncertainty estimates at 15 GHz (panel (a)) and
22 GHz (panel (b)) plotted against projected radius.
Filled squares represent core regions, and open squares 
represent jet features.  J0738$+$17 and J1312$+$32, where we have analyzed the
total VLBI flux of the source, are represented by filled triangles.
}
\end{figure}

\begin{figure}
\figurenum{29}
\begin{center}
\epsfig{file=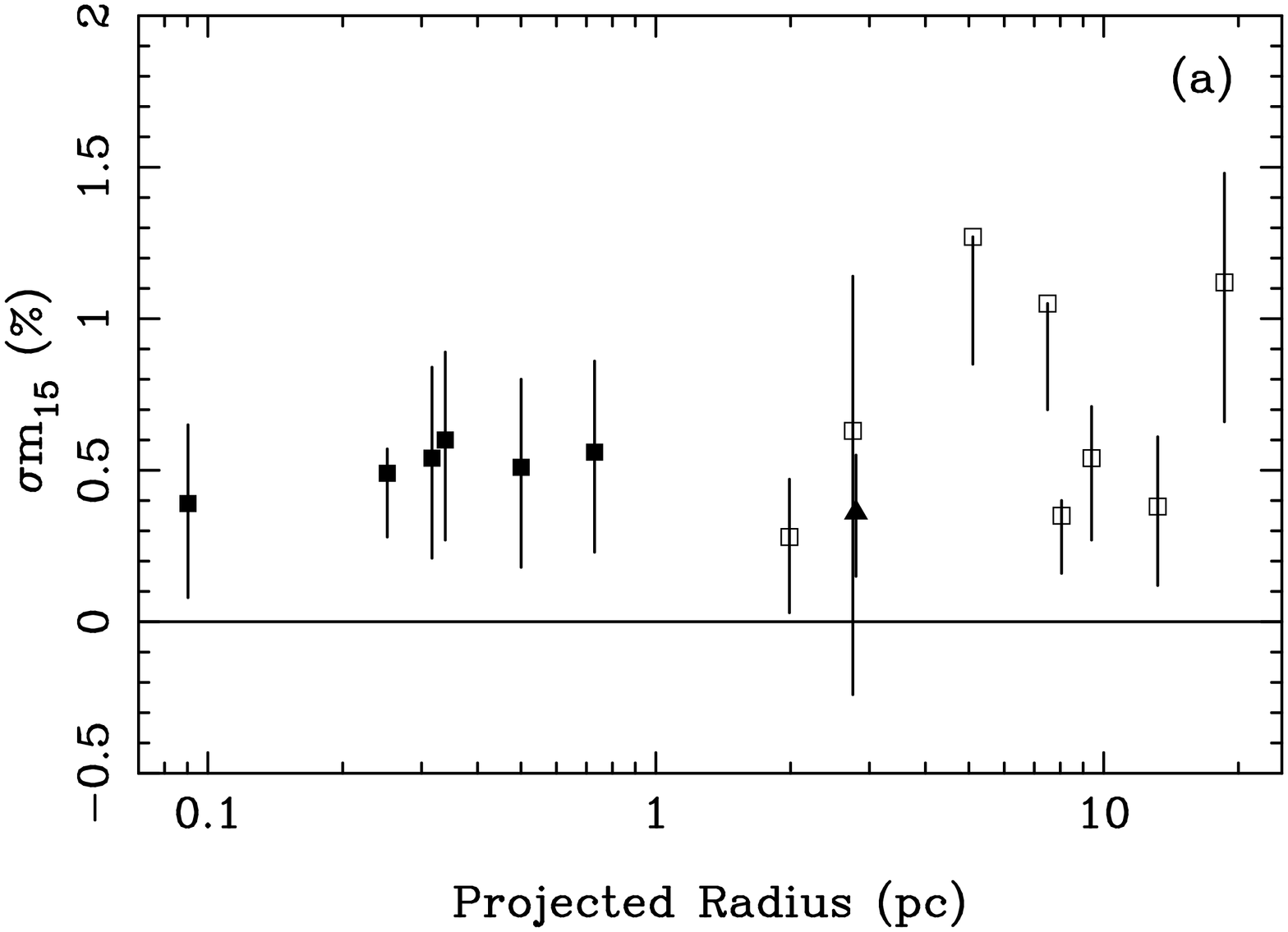,width=2.6in,angle=-90}
\epsfig{file=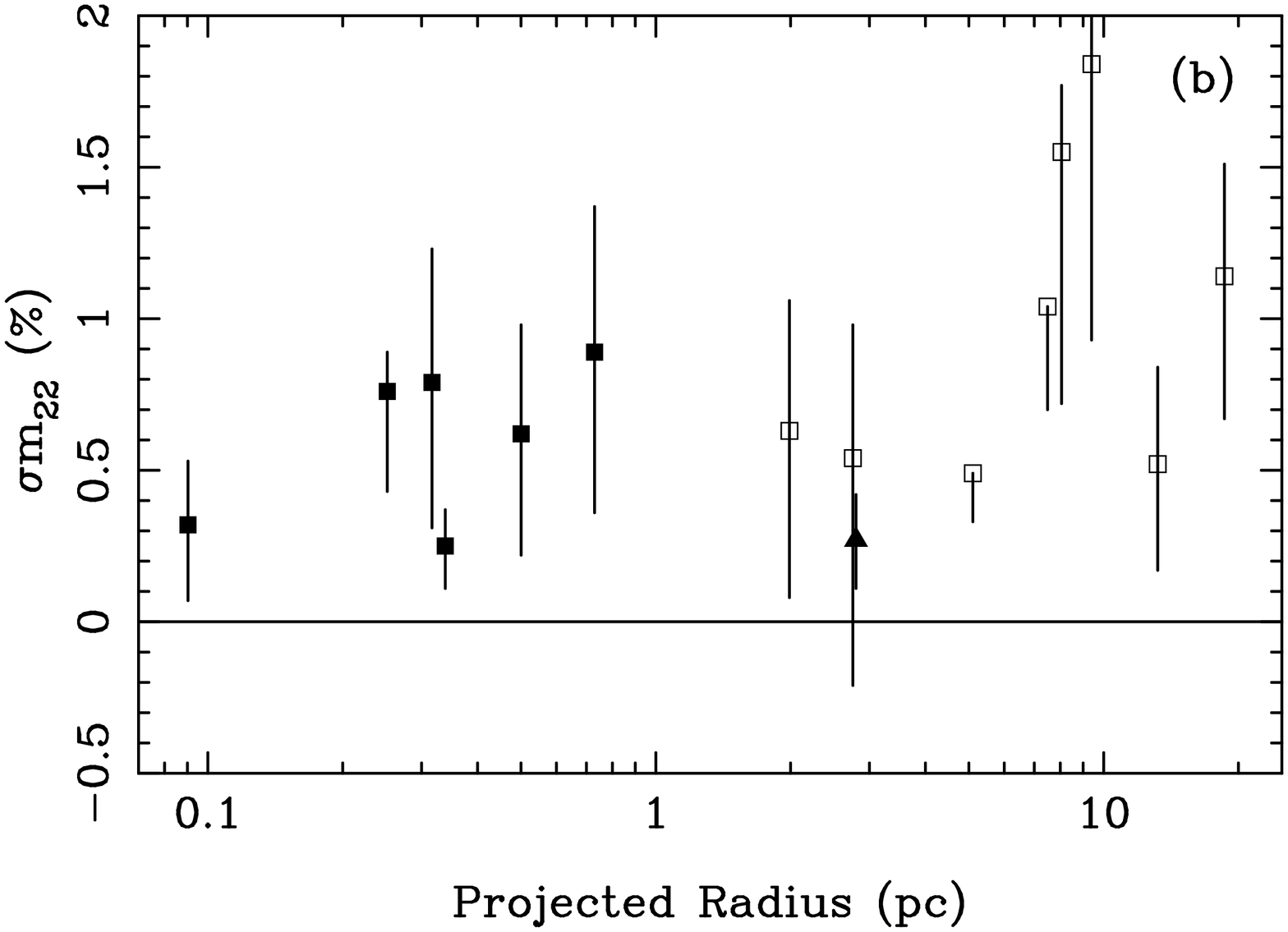,width=2.6in,angle=-90}
\end{center}
\figcaption[f29a.eps,f29b.eps]{\label{f:m_err}
Fractional polarization uncertainty estimates at 15 GHz (panel (a)) and
22 GHz (panel (b)) plotted against projected radius.
Filled squares represent core regions, and open squares 
represent jet features.  J1312$+$32, where we have analyzed the
total VLBI flux of the source, is represented by a filled triangle.
}
\end{figure}

\begin{figure}
\figurenum{30}
\begin{center}
\epsfig{file=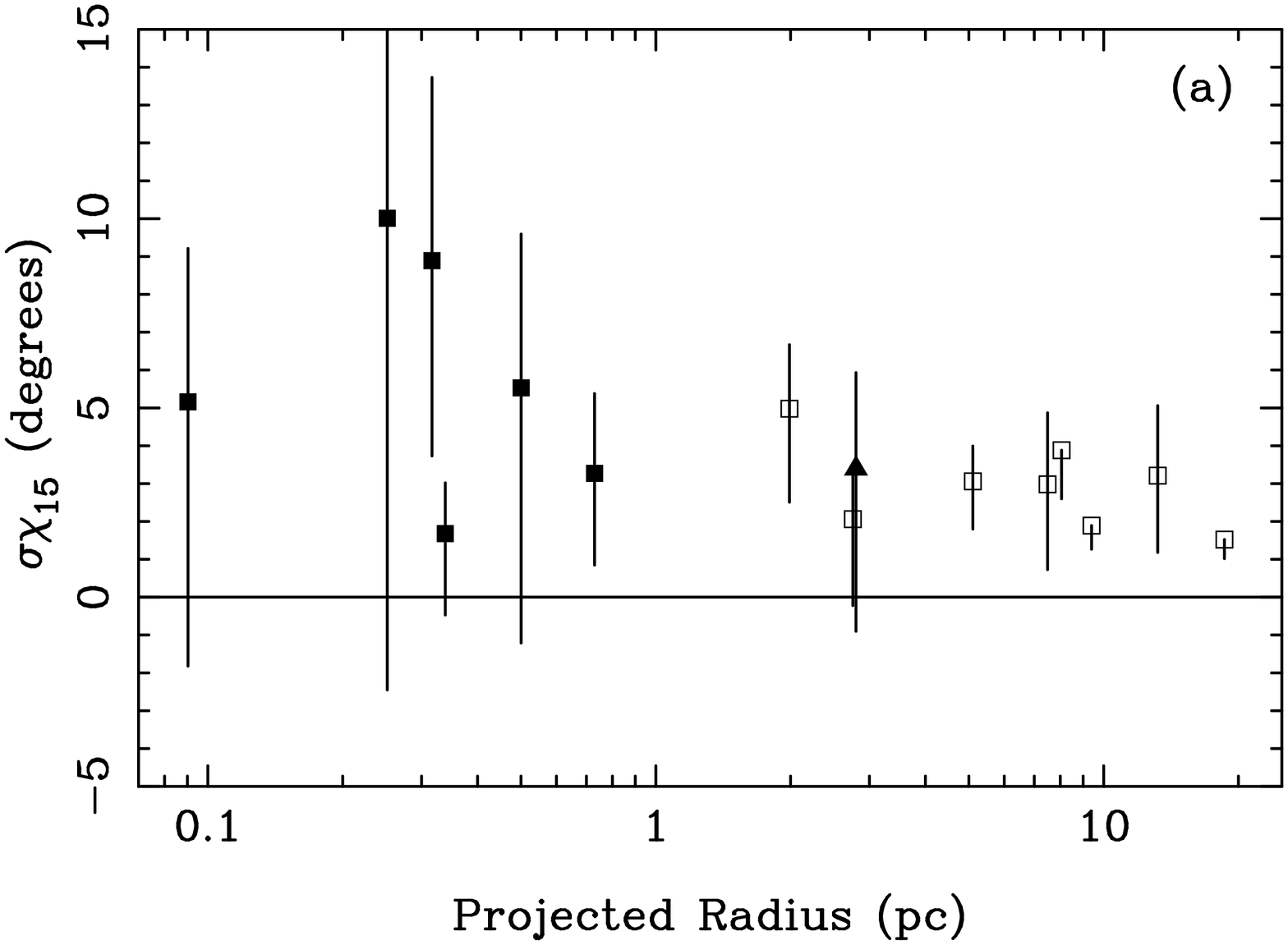,width=2.6in,angle=-90}
\epsfig{file=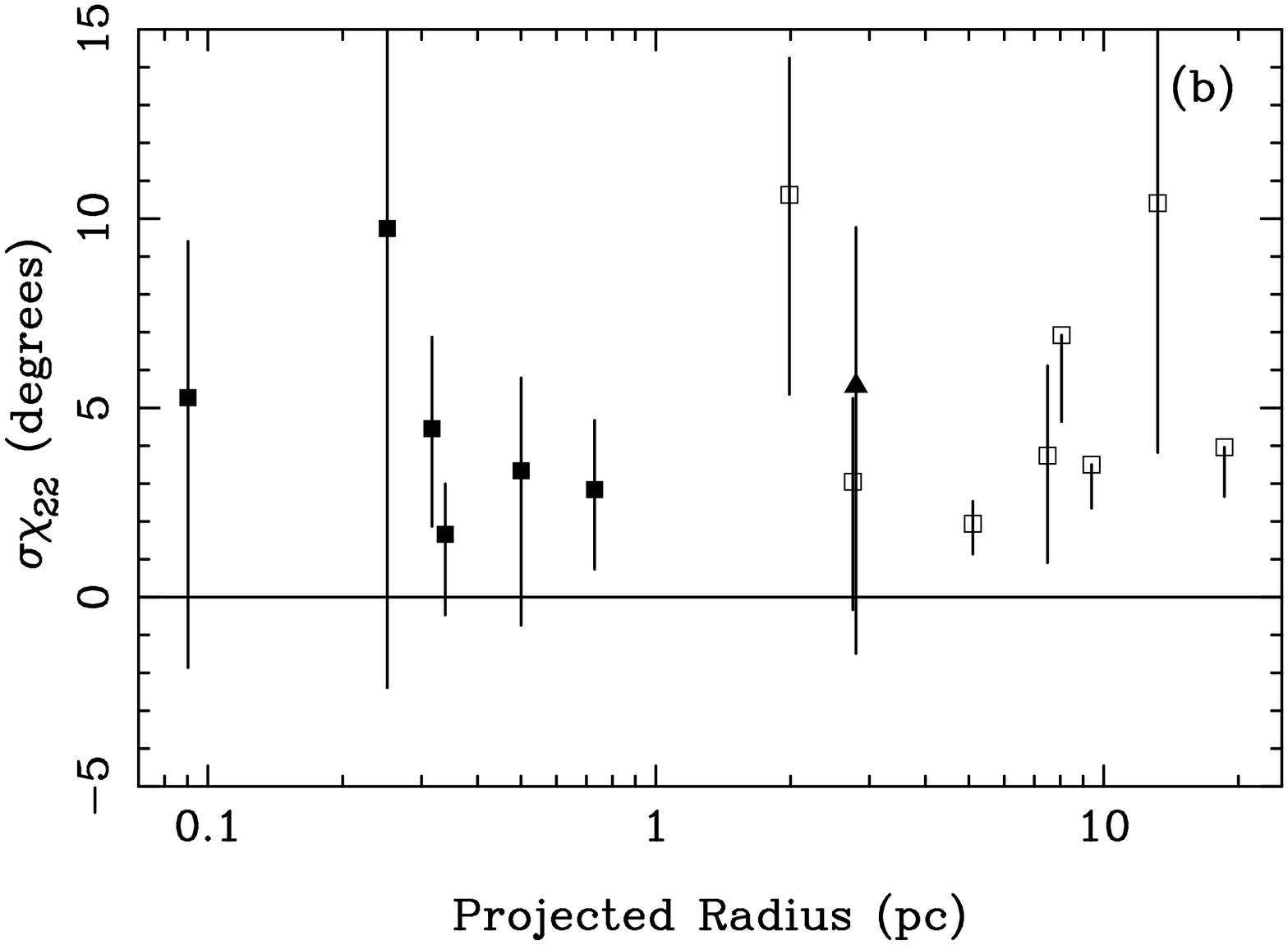,width=2.6in,angle=-90}
\end{center}
\figcaption[f30a.eps,f30b.eps]{\label{f:chi_err}
Polarization angle uncertainty estimates at 15 GHz (panel (a)) and
22 GHz (panel (b)) plotted against projected radius.
Filled squares represent core regions, and open squares 
represent jet features.  J1312$+$32, where we have analyzed the
total VLBI flux of the source, is represented by a filled triangle.
}
\end{figure}

Figures \ref{f:i_err}--\ref{f:chi_err} plot these uncertainty
estimates against projected radius for each quantity at
both frequencies.  For $I$ at 15 GHz most features have an
estimated uncertainty of less than $5$\% in a single epoch, with
several particularly good cases of $2-3$\% and a few poorer cases
of $\sim 10$\%.  For $I$ at 22 GHz, $5$\% or better is the case
for some of the more extended core regions (where nearby jet
components have been summed into the core flux), but $5-10$\% is
more typical for jet features with a couple cases approaching 
$20$\%. For fractional polarization, $0.5$\% is typical at 
15 GHz, and $0.5-1.0$\% is typical at 22 GHz.  For polarization
angle, uncertainties $\leq 5^\circ$ are typical at both 
frequencies with best case values $\leq 2^\circ$ and worse case
values up to $\sim 10^\circ$. 

%\placefigure{f:i_err} 
%\placefigure{f:m_err} 
%\placefigure{f:chi_err} 

%%%%%%%%%%%%%%%%%%%%%%
%%                  %%  
%%    References    %%
%%                  %%
%%%%%%%%%%%%%%%%%%%%%%

%%%%%%%%%%%%%%%%%%%%%%
%%                  %%  
%% Figure Captions  %%
%%                  %%
%%%%%%%%%%%%%%%%%%%%%%
\newpage

\clearpage

\clearpage

\clearpage

%%%%%%%%%%%%%%%%%%%%%%
%%                  %%  
%%     Tables       %%
%%                  %%
%%%%%%%%%%%%%%%%%%%%%%

\newpage
\singlespace

  %%%%%%%%%%%%%%%%%
  %%             %%   
  %%   Table 1   %% 
  %%             %%
  %%%%%%%%%%%%%%%%%

\begin{table}
\begin{small}
\begin{center}
\tablenum{1}
\caption[]{\label{t:Sources}Source Information\\}
\begin{tabular}{ccccc}
\tableline\tableline
J2000.0 & J1950.0 & Other Names & Redshift & Classification \\
\tableline 
J0433$+$053 & B0430$+$052 & 3C\,120, II Zw 14 & 0.033 & Sy 1\\
J0530$+$135 & B0528$+$134 & PKS 0528$+$134 & 2.060 & Quasar\\
J0738$+$177 & B0735$+$178 & OI\,158, DA\,237, PKS 0735$+$178 & 0.424 \tablenotemark{a} & BL\\
J0854$+$201 & B0851$+$202 & OJ\,287 & 0.306 & BL\\
J1224$+$212 & B1222$+$216 & 4C\,21.35 & 0.435 & Quasar\\
J1229$+$020 & B1226$+$023 & 3C\,273 & 0.158 & Quasar\\
J1256$-$057 & B1253$-$055 & 3C\,279 & 0.536 & Quasar\\
J1310$+$323 & B1308$+$326 & OP\,313 & 0.996 & Quasar/BL\\
J1512$-$090 & B1510$-$089 & OR\,$-$017 & 0.360 & Quasar\\
J1751$+$09  & B1749$+$096 & OT\,081, 4C\,09.56 & 0.322 & BL\\
J1927$+$739 & B1928$+$738 & 4C\,73.18 & 0.302 & Quasar\\
J2005$+$778 & B2007$+$777 &  & 0.342 & BL\\
\tableline 
\end{tabular} 

\tablenotetext{a}{Lower limit}
\end{center}
\end{small} 
\end{table}

  %%%%%%%%%%%%%%%%%
  %%             %%   
  %%   Table 2   %% 
  %%             %%
  %%%%%%%%%%%%%%%%%

\begin{deluxetable}{cccccc}
\tablewidth{0pc}
\tablecolumns{6}
\tabletypesize{\small}
\tablenum{2}
\tablecaption{Jet Features.\label{t:comp}}
\tablehead{Object & Jet Feature & $\langle R\rangle $ & $\langle \theta\rangle $ & $N_I$ & $N_P$ \\ 
& & (pc) & (deg.) & & }
\startdata
3C\,120 & KD$+$K3$+$K2/UD$+$U2 & $0.13$ & $-121.5$ & $6$ & $\ldots$ \\
        & K1A$+$K1B/U1A$+$U1B & $1.99$ & $-110.3$ & $6$& $6$ \\
\\
J0530$+$13 & KD$+$K2/UD$+$U2 & $0.50$ & $91.3$ & $6$& $5$ \\
\\
J0738$+$17 & KALL/UALL & $2.23$ & $65.4$ & $6$ & $\ldots$ \\
\\
OJ287 & KD$+$K4/UD$+$U4 & $0.09$ & $-106.1$ & $6$& $6$ \\
      & K3/U3 & $2.76$ & $-94.7$ & $6$& $5$ \\
\\
J1224$+$21 & KD$+$K4$+$K3/UD$+$U3 & $0.34$ & $-14.2$ & $5$& $5$ \\
\\
3C\,273 & KD$+$K10/UD$+$U10 & $0.35$ & $-118.2$ & $6$ & $\ldots$ \\
        & K9$+$K8$+$K7/U9$+$U8$+$U7 & $5.11$ & $-118.2$ & $6$& $6$ \\
        & K4$+$K5/U4$+$U5 & $13.2$ & $-111.4$ & $6$& $6$ \\
\\
3C\,279 & KD$+$K4/UD$+$U4 & $0.32$ & $-119.0$ & $6$& $6$ \\
        & K1$+$K2/U1$+$U2 & $18.6$ & $-114.6$ & $6$& $6$ \\
\\
J1310$+$32 & KALL/UALL & $2.80$ & $-63.1$ & $6$& $5$ \\
\\
J1512$-$09 & KD$+$K2/UD$+$U2 & $0.25$ & $-32.7$ & $6$& $5$ \\
           & K1/U1 & $7.49$ & $-28.6$ & $6$& $6$ \\
\\
J1751$+$09 & KD/UD$+$U3 & $0.09$ & $35.1$ & $6$ & $\ldots$ \\
\\
J1927$+$73 & KD$+$K3/UD$+$U3 & $1.12$ & $150.8$ & $6$ & $\ldots$ \\
           & K2/U2 & $8.06$ & $157.3$ & $6$& $5$ \\
           & K1/U1 & $9.41$ & $172.9$ & $6$& $6$ \\
\\
J2005$+$77 & KD$+$K2/UD$+$U3$+$U2 & $0.73$ & $-91.0$ & $6$& $6$ \\
           & K1/U1 & $8.13$ & $-94.1$ & $6$ & $\ldots$ \\
\enddata
\tablecomments{The mean radial position of the jet feature is given 
by $\langle R \rangle$ at a mean structural position angle of $\langle
\theta \rangle$ (measured counter clockwise from north).  $N_I$ is the 
number of epochs available for total intensity variability analysis, 
and $N_P$ is the number of epochs available for polarization 
variability analysis.  Not all features have adequate polarization
strength in at least five epochs (at both frequencies) to be used for 
variability analysis; these cases are indicated by $\ldots$.
See \S{\ref{s:model}} for a description of our selection criteria.}
\end{deluxetable}

  %%%%%%%%%%%%%%%%%
  %%             %%   
  %%   Table 3   %% 
  %%             %%
  %%%%%%%%%%%%%%%%%

\begin{deluxetable}{cccccccc}
\tablewidth{0pc}
\tablecolumns{8}
\tabletypesize{\small}
\tablenum{3}
\tablecaption{Mean Properties of Jet Features.\label{t:mean}}
\tablehead{Object & Jet Feature & $\langle I_{15}\rangle $ & $\langle I_{22}\rangle $ & $\langle m_{15}\rangle $ 
& $\langle m_{22}\rangle $ & $\langle \chi_{15}\rangle $ & $\langle \chi_{22}\rangle $ \\ 
& & (Jy) & (Jy) & (\%) & (\%) & (deg.) & (deg.) }
\startdata
3C\,120 & KD$+$K3$+$K2/UD$+$U2 & $1.54\pm0.19$ & $1.75\pm0.22$ & $\ldots$ & $\ldots$ & $\ldots$ & $\ldots$ \\
        & K1A$+$K1B/U1A$+$U1B & $0.73\pm0.05$ & $0.59\pm0.05$ & $7.7\pm0.8$ & $5.8\pm0.6$ & $-59.2\pm7.8$ & $-52.4\pm8.4$ \\
\\
J0530$+$13 & KD$+$K2/UD$+$U2 & $8.01\pm0.40$ & $7.90\pm0.51$ & $2.0\pm0.4$ & $3.1\pm0.5$ & $-98.6\pm8.2$ & $-104.9\pm5.0$ \\
\\
J0738$+$17 & KALL/UALL & $0.97\pm0.07$ & $0.89\pm0.07$ & $\ldots$ & $\ldots$ & $\ldots$ & $\ldots$ \\
\\
OJ287 & KD$+$K4/UD$+$U4 & $1.15\pm0.09$ & $1.29\pm0.11$ & $3.2\pm0.6$ & $3.3\pm0.5$ & $-18.3\pm26.3$ & $-14.6\pm26.1$ \\
      & K3/U3 & $0.48\pm0.13$ & $0.41\pm0.10$ & $9.7\pm2.0$ & $8.3\pm1.7$ & $-26.6\pm2.7$ & $-25.1\pm2.3$ \\
\\
J1224$+$21 & KD$+$K4$+$K3/UD$+$U3 & $1.47\pm0.08$ & $1.49\pm0.08$ & $5.5\pm0.4$ & $4.7\pm0.2$ & $-38.8\pm1.4$ & $-36.6\pm2.1$ \\
\\
3C\,273 & KD$+$K10/UD$+$U10 & $10.38\pm1.81$ & $14.22\pm1.79$ & $\ldots$ & $\ldots$ & $\ldots$ & $\ldots$ \\
        & K9$+$K8$+$K7/U9$+$U8$+$U7 & $5.24\pm0.73$ & $4.16\pm0.53$ & $8.4\pm1.6$ & $10.4\pm1.0$ & $-35.7\pm1.6$ & $-41.4\pm1.0$ \\
        & K4$+$K5/U4$+$U5 & $1.41\pm0.06$ & $1.32\pm0.08$ & $6.2\pm0.5$ & $4.9\pm0.7$ & $123.2\pm5.1$ & $109.4\pm7.7$ \\
\\
3C\,279 & KD$+$K4/UD$+$U4 & $15.03\pm0.85$ & $19.35\pm0.96$ & $2.7\pm0.6$ & $4.4\pm0.8$ & $50.0\pm8.4$ & $48.2\pm4.2$ \\
        & K1$+$K2/U1$+$U2 & $2.48\pm0.06$ & $2.19\pm0.02$ & $10.6\pm0.6$ & $11.1\pm0.6$ & $65.2\pm0.6$ & $66.8\pm1.6$ \\
\\
J1310$+$32 & KALL/UALL & $2.79\pm0.13$ & $2.48\pm0.13$ & $1.8\pm0.3$ & $1.3\pm0.2$ & $26.7\pm5.5$ & $27.3\pm9.0$ \\
\\
J1512$-$09 & KD$+$K2/UD$+$U2 & $1.19\pm0.14$ & $1.42\pm0.16$ & $1.9\pm0.3$ & $2.8\pm0.4$ & $-93.8\pm42.9$ & $-102.3\pm43.4$ \\
           & K1/U1 & $0.43\pm0.06$ & $0.31\pm0.04$ & $6.0\pm0.7$ & $6.6\pm0.8$ & $67.3\pm5.6$ & $63.2\pm7.1$ \\
\\
J1751$+$09 & KD/UD$+$U3 & $1.34\pm0.32$ & $1.50\pm0.33$ & $\ldots$ & $\ldots$ & $\ldots$ & $\ldots$ \\
\\
J1927$+$73 & KD$+$K3/UD$+$U3 & $2.28\pm0.06$ & $2.37\pm0.03$ & $\ldots$ & $\ldots$ & $\ldots$ & $\ldots$ \\
           & K2/U2 & $0.39\pm0.01$ & $0.31\pm0.01$ & $5.4\pm0.6$ & $6.4\pm0.9$ & $-75.7\pm1.7$ & $-79.7\pm3.1$ \\
           & K1/U1 & $0.39\pm0.02$ & $0.29\pm0.01$ & $15.2\pm0.4$ & $15.5\pm0.9$ & $48.1\pm1.6$ & $49.9\pm1.6$ \\
\\
J2005$+$77 & KD$+$K2/UD$+$U3$+$U2 & $0.82\pm0.04$ & $0.88\pm0.04$ & $5.2\pm0.5$ & $4.7\pm0.9$ & $83.3\pm5.3$ & $83.1\pm4.5$ \\
           & K1/U1 & $0.11\pm0.01$ & $0.07\pm0.01$ & $\ldots$ & $\ldots$ & $\ldots$ & $\ldots$ \\
\enddata
\tablecomments{Total intensity is denoted by $I$, fractional linear polarization by $m$, and
polarization position angle by $\chi$ (measured counterclockwise from
north).  Mean values are averages across all epochs used for
variability analysis.
Subscripts 
denote the frequency of observation in GHz.}
\end{deluxetable}

  %%%%%%%%%%%%%%%%%
  %%             %%   
  %%   Table 4   %% 
  %%             %%
  %%%%%%%%%%%%%%%%%

\begin{deluxetable}{cccccccc}
\tablewidth{0pc}
\tablecolumns{8}
\tabletypesize{\small}
\tablenum{4}
\tablecaption{Flux Variability.\label{t:i_var}}
\tablehead{
\multicolumn{2}{c}{} &
\multicolumn{3}{c}{Linear Slopes with Time} &
\multicolumn{3}{c}{ Correlated Fluctuations} \\
Object & Jet Feature & $\dot{I}_{15}/\langle I_{15}\rangle$ & $\dot{I}_{22}/\langle I_{22}\rangle$ 
& $\langle \dot{I}/\langle I\rangle \rangle $ & $r$ & $p_r$ & $\delta I/\langle I\rangle$ \\ 
&& (yr$^{-1}$) & (yr$^{-1}$) & (yr$^{-1}$) &&&  }
\startdata
3C\,120 & KD$+$K3$+$K2/UD$+$U2 & $0.74\pm0.30$ & $0.63\pm0.35$ & $0.69\pm0.23$
                      & $0.90$ & $0.02$ & $0.22$ $(+0.02,-0.03)$  \\
        & K1A$+$K1B/U1A$+$U1B & $-0.13\pm0.25$ & $0.07\pm0.30$ & $-0.03\pm0.20$
                      & $0.73$ & $0.05$ & $0.16$ $(+0.02,-0.04)$  \\
\\
J0530$+$13 & KD$+$K2/UD$+$U2 & $-0.31\pm0.11$ & $-0.43\pm0.12$ & $-0.37\pm0.08$
                      & $0.89$ & $0.02$ & $0.08$ $(+0.01,-0.01)$  \\
\\
J0738$+$17 & KALL/UALL & $-0.56\pm0.06$ & $-0.47\pm0.20$ & $-0.51\pm0.10$
                      & $0.58$ & $0.15$ & $0.05$ $(+0.01,-0.04)$  \\
\\
OJ287 & KD$+$K4/UD$+$U4 & $0.42\pm0.19$ & $0.18\pm0.32$ & $0.30\pm0.18$
                      & $0.82$ & $0.05$ & $0.16$ $(+0.02,-0.04)$  \\
      & K3/U3 & $-1.65\pm0.71$ & $-1.65\pm0.52$ & $-1.65\pm0.44$
                      & $0.91$ & $0.02$ & $0.30$ $(+0.06,-0.06)$  \\
\\
J1224$+$21 & KD$+$K4$+$K3/UD$+$U3 & $-0.28\pm0.18$ & $-0.40\pm0.11$ & $-0.34\pm0.11$
                      & $0.74$ & $0.13$ & $0.07$ $(+0.01,-0.03)$  \\
\\
3C\,273 & KD$+$K10/UD$+$U10 & $1.14\pm0.37$ & $0.66\pm0.35$ & $0.90\pm0.26$
                      & $0.77$ & $0.06$ & $0.21$ $(+0.03,-0.06)$  \\
        & K9$+$K8$+$K7/U9$+$U8$+$U7 & $-1.03\pm0.16$ & $-0.76\pm0.30$ & $-0.90\pm0.17$
                      & $0.35$ & $0.28$ & $0.06$ $(+0.03,-0.11)$  \\
        & K4$+$K5/U4$+$U5 & $-0.29\pm0.07$ & $-0.28\pm0.17$ & $-0.29\pm0.09$
                      & $0.69$ & $0.10$ & $0.07$ $(+0.01,-0.03)$  \\
\\
3C\,279 & KD$+$K4/UD$+$U4 & $0.30\pm0.15$ & $0.04\pm0.19$ & $0.17\pm0.12$
                      & $0.94$ & $0.01$ & $0.12$ $(+0.01,-0.01)$  \\
        & K1$+$K2/U1$+$U2 & $0.09\pm0.08$ & $-0.02\pm0.04$ & $0.03\pm0.04$
                      & $0.33$ & $0.26$ & $0.02$ $(+0.01,-0.04)$  \\
\\
J1310$+$32 & KALL/UALL & $-0.33\pm0.05$ & $-0.25\pm0.16$ & $-0.29\pm0.08$
                      & $-0.40$ & $0.26$ & $-0.04$ $(+0.07,-0.02)$  \\
\\
J1512$-$09 & KD$+$K2/UD$+$U2 & $-0.27\pm0.41$ & $-0.31\pm0.40$ & $-0.29\pm0.29$
                      & $0.96$ & $0.00$ & $0.27$ $(+0.02,-0.02)$  \\
           & K1/U1 & $-0.97\pm0.15$ & $-0.93\pm0.19$ & $-0.95\pm0.12$
                      & $0.08$ & $0.45$ & $0.02$ $(+0.04,-0.08)$  \\
\\
J1751$+$09 & KD/UD$+$U3 & $-0.61\pm0.86$ & $-0.18\pm0.82$ & $-0.39\pm0.60$
                      & $0.97$ & $0.00$ & $0.55$ $(+0.09,-0.09)$  \\
\\
J1927$+$73 & KD$+$K3/UD$+$U3 & $0.18\pm0.05$ & $0.06\pm0.03$ & $0.12\pm0.03$
                      & $-0.20$ & $0.37$ & $-0.01$ $(+0.03,-0.01)$  \\
           & K2/U2 & $-0.25\pm0.08$ & $0.04\pm0.14$ & $-0.10\pm0.08$
                      & $0.52$ & $0.18$ & $0.05$ $(+0.02,-0.07)$  \\
           & K1/U1 & $-0.30\pm0.08$ & $0.15\pm0.16$ & $-0.08\pm0.09$
                      & $0.83$ & $0.04$ & $0.08$ $(+0.01,-0.02)$  \\
\\
J2005$+$77 & KD$+$K2/UD$+$U3$+$U2 & $-0.03\pm0.18$ & $0.25\pm0.11$ & $0.11\pm0.11$
                      & $0.97$ & $0.00$ & $0.10$ $(+0.00,-0.00)$  \\
           & K1/U1 & $-0.61\pm0.10$ & $-0.55\pm0.27$ & $-0.58\pm0.14$
                      & $-0.24$ & $0.35$ & $-0.05$ $(+0.12,-0.04)$  \\
\enddata
\tablecomments{$\dot{I}_{15}/\langle I_{15} \rangle$ and $\dot{I}_{22}/\langle I_{22} \rangle$ 
are the fractional flux slopes with time for 15 and 22 GHz respectively.  
$\langle \dot{I}/\langle I\rangle \rangle$ is the mean of these
slopes.  
\S{\ref{s:tech}}
describes our analysis for  fluctuations that correlate between 15 and 22 GHz: $r$ is 
the correlation coefficient, and $p_r$ is the probability of obtaining 
a correlation this strong by pure chance.  $\delta I/\langle I \rangle$ is the
standard deviation of the correlated fluctuations divided by the mean flux 
(across epoch and frequency).}
\end{deluxetable}

  %%%%%%%%%%%%%%%%%
  %%             %%   
  %%   Table 5   %% 
  %%             %%
  %%%%%%%%%%%%%%%%%

\begin{deluxetable}{cccccccc}
\tablewidth{0pc}
\tablecolumns{8}
\tabletypesize{\small}
\tablenum{5}
\tablecaption{Fractional Polarization Variability.\label{t:m_var}}
\tablehead{
\multicolumn{2}{c}{} &
\multicolumn{3}{c}{Linear Slopes with Time} &
\multicolumn{3}{c}{ Correlated Fluctuations} \\
Object & Jet Feature & $\dot{m}_{15}$ & $\dot{m}_{22}$ & $\langle \dot{m}\rangle $ 
& $r$ & $p_r$ & $\delta m$ \\ 
&& (\%/yr) & (\%/yr) & (\%/yr) &&& (\%) }
\startdata
3C\,120 & K1A$+$K1B/U1A$+$U1B & $ 5.7\pm 1.0$ & $ 2.2\pm 2.2$ & $ 4.0\pm 1.2$
                      & $0.84$ & $0.04$ & $1.0$ $(+0.1,-0.2)$   \\
\\
J0530$+$13 & KD$+$K2/UD$+$U2 & $ 1.1\pm 1.3$ & $-0.2\pm 1.8$ & $ 0.4\pm 1.1$
                      & $0.69$ & $0.10$ & $0.8$ $(+0.2,-0.3)$   \\
\\
OJ287 & KD$+$K4/UD$+$U4 & $-4.1\pm 1.3$ & $-3.4\pm 1.0$ & $-3.7\pm 0.8$
                      & $0.82$ & $0.05$ & $0.7$ $(+0.1,-0.2)$   \\
      & K3/U3 & $ 9.6\pm 7.7$ & $10.0\pm 5.8$ & $ 9.8\pm 4.8$
                      & $0.98$ & $0.00$ & $4.1$ $(+0.0,-0.1)$   \\
\\
J1224$+$21 & KD$+$K4$+$K3/UD$+$U3 & $-0.8\pm 1.9$ & $ 0.1\pm 0.8$ & $-0.3\pm 1.0$
                      & $0.56$ & $0.16$ & $0.4$ $(+0.1,-0.3)$   \\
\\
3C\,273 & K9$+$K8$+$K7/U9$+$U8$+$U7 & $11.1\pm 1.8$ & $ 7.6\pm 0.7$ & $ 9.4\pm 0.9$
                      & $-0.41$ & $0.25$ & $-0.5$ $(+0.8,-0.2)$   \\
        & K4$+$K5/U4$+$U5 & $ 2.0\pm 1.8$ & $ 1.7\pm 2.6$ & $ 1.9\pm 1.6$
                      & $0.92$ & $0.01$ & $1.5$ $(+0.1,-0.1)$   \\
\\
3C\,279 & KD$+$K4/UD$+$U4 & $ 0.4\pm 2.1$ & $-1.0\pm 3.0$ & $-0.3\pm 1.9$
                      & $0.84$ & $0.02$ & $1.5$ $(+0.1,-0.2)$   \\
        & K1$+$K2/U1$+$U2 & $-2.5\pm 1.9$ & $ 0.9\pm 2.3$ & $-0.8\pm 1.5$
                      & $0.43$ & $0.20$ & $1.0$ $(+0.4,-1.3)$   \\
\\
J1310$+$32 & KALL/UALL & $-1.0\pm 0.7$ & $-0.7\pm 0.6$ & $-0.8\pm 0.5$
                      & $0.62$ & $0.13$ & $0.4$ $(+0.1,-0.2)$   \\
\\
J1512$-$09 & KD$+$K2/UD$+$U2 & $ 1.2\pm 1.0$ & $ 1.0\pm 1.8$ & $ 1.1\pm 1.0$
                      & $0.27$ & $0.33$ & $0.4$ $(+0.3,-0.8)$   \\
           & K1/U1 & $ 4.3\pm 1.4$ & $ 5.6\pm 1.4$ & $ 5.0\pm 1.0$
                      & $-0.32$ & $0.30$ & $-0.6$ $(+1.1,-0.3)$   \\
\\
J1927$+$73 & K2/U2 & $ 3.9\pm 0.6$ & $ 3.4\pm 2.6$ & $ 3.6\pm 1.4$
                      & $0.23$ & $0.39$ & $0.4$ $(+0.4,-1.0)$   \\
           & K1/U1 & $ 2.8\pm 1.0$ & $-0.0\pm 3.3$ & $ 1.4\pm 1.7$
                      & $0.42$ & $0.24$ & $0.8$ $(+0.4,-1.4)$   \\
\\
J2005$+$77 & KD$+$K2/UD$+$U3$+$U2 & $-2.6\pm 1.5$ & $-2.1\pm 3.0$ & $-2.4\pm 1.7$
                      & $0.82$ & $0.02$ & $1.5$ $(+0.1,-0.2)$   \\
\enddata
\tablecomments{$\dot{m}_{15}$ and $\dot{m}_{22}$ are the fractional 
polarization slopes with time for 15 and 22 GHz respectively.  
$\langle \dot{m} \rangle$ is the mean of these slopes.  \S{\ref{s:tech}}
describes our analysis for  fluctuations that correlate between 15 and 22 GHz: $r$ is 
the correlation coefficient, and $p_r$ is the probability of obtaining 
a correlation this strong by pure chance.  $\delta m$ is the
standard deviation of the correlated fluctuations.}
\end{deluxetable}

  %%%%%%%%%%%%%%%%%
  %%             %%   
  %%   Table 6   %% 
  %%             %%
  %%%%%%%%%%%%%%%%%

\begin{deluxetable}{cccccccc}
\tablewidth{0pc}
\tablecolumns{8}
\tabletypesize{\small}
\tablenum{6}
\tablecaption{Polarization Angle  Variability.\label{t:chi_var}}
\tablehead{
\multicolumn{2}{c}{} &
\multicolumn{3}{c}{Linear Slopes with Time} &
\multicolumn{3}{c}{ Correlated Fluctuations} \\
Object & Jet Feature & $\dot{\chi}_{15}$ & $\dot{\chi}_{22}$ & $\langle \dot{\chi}\rangle $ 
& $r$ & $p_r$ & $\delta\chi$ \\ 
&& (deg./yr) & (deg./yr) & (deg./yr) &&& (deg.) }
\startdata
3C\,120 & K1A$+$K1B/U1A$+$U1B & $55.6\pm 9.2$ & $49.4\pm19.6$ & $52.5\pm10.8$
                      & $0.44$ & $0.23$ & $6.5$ $(+2.5,-9.9)$  \\
\\
J0530$+$13 & KD$+$K2/UD$+$U2 & $ 1.0\pm29.3$ & $-3.8\pm17.6$ & $-1.4\pm17.1$
                      & $0.91$ & $0.02$ & $13.6$ $(+0.7,-1.4)$  \\
\\
OJ287 & KD$+$K4/UD$+$U4 & $-177.3\pm44.3$ & $-174.1\pm45.2$ & $-175.7\pm31.6$
                      & $0.97$ & $0.00$ & $32.1$ $(+0.5,-0.9)$  \\
      & K3/U3 & $-18.6\pm 6.7$ & $-8.6\pm 9.9$ & $-13.6\pm 6.0$
                      & $0.68$ & $0.16$ & $3.7$ $(+0.8,-2.6)$  \\
\\
J1224$+$21 & KD$+$K4$+$K3/UD$+$U3 & $ 2.8\pm 6.4$ & $13.1\pm 6.3$ & $ 8.0\pm 4.5$
                      & $0.77$ & $0.12$ & $3.0$ $(+0.5,-1.3)$  \\
\\
3C\,273 & K9$+$K8$+$K7/U9$+$U8$+$U7 & $ 4.2\pm 5.7$ & $ 1.5\pm 3.8$ & $ 2.9\pm 3.4$
                      & $0.41$ & $0.21$ & $2.0$ $(+0.8,-2.9)$  \\
        & K4$+$K5/U4$+$U5 & $34.7\pm 8.2$ & $24.3\pm26.5$ & $29.5\pm13.8$
                      & $0.71$ & $0.09$ & $9.0$ $(+1.5,-3.4)$  \\
\\
3C\,279 & KD$+$K4/UD$+$U4 & $15.2\pm30.6$ & $17.9\pm13.0$ & $16.5\pm16.6$
                      & $0.81$ & $0.02$ & $13.1$ $(+1.2,-2.3)$  \\
        & K1$+$K2/U1$+$U2 & $ 2.7\pm 1.9$ & $ 5.1\pm 5.5$ & $ 3.9\pm 2.9$
                      & $-0.32$ & $0.27$ & $-1.4$ $(+2.5,-0.7)$  \\
\\
J1310$+$32 & KALL/UALL & $22.9\pm14.4$ & $20.8\pm29.9$ & $21.8\pm16.6$
                      & $0.92$ & $0.01$ & $15.1$ $(+0.7,-1.4)$  \\
\\
J1512$-$09 & KD$+$K2/UD$+$U2 & $-347.0\pm35.3$ & $-351.9\pm34.4$ & $-349.4\pm24.7$
                      & $0.73$ & $0.14$ & $16.2$ $(+2.9,-8.7)$  \\
           & K1/U1 & $-25.7\pm16.8$ & $-33.0\pm20.8$ & $-29.3\pm13.4$
                      & $0.95$ & $0.00$ & $15.1$ $(+0.3,-0.6)$  \\
\\
J1927$+$73 & K2/U2 & $-7.8\pm 4.9$ & $ 4.1\pm11.7$ & $-1.8\pm 6.3$
                      & $-0.73$ & $0.08$ & $-4.4$ $(+1.5,-0.7)$  \\
           & K1/U1 & $10.6\pm 2.6$ & $ 7.3\pm 4.8$ & $ 9.0\pm 2.7$
                      & $-0.08$ & $0.45$ & $-0.7$ $(+2.6,-1.3)$  \\
\\
J2005$+$77 & KD$+$K2/UD$+$U3$+$U2 & $34.6\pm 9.9$ & $29.2\pm 8.6$ & $31.9\pm 6.6$
                      & $0.80$ & $0.05$ & $6.0$ $(+0.7,-1.5)$  \\
\enddata
\tablecomments{$\dot{\chi}_{15}$ and $\dot{\chi}_{22}$ are the 
polarization position angle slopes with time for 15 and 22 GHz respectively.  
$\langle \dot{\chi} \rangle$ is the mean of these slopes.  \S{\ref{s:tech}}
describes our analysis for  fluctuations that correlate 
between 15 and 22 GHz: $r$ is the correlation coefficient, and $p_r$ is the 
probability of obtaining a correlation this strong by pure chance.  
$\delta\chi$ is the standard deviation of the correlated 
fluctuations.}
\end{deluxetable}

  %%%%%%%%%%%%%%%%%
  %%             %%   
  %%   Table 7   %% 
  %%             %%
  %%%%%%%%%%%%%%%%%

\begin{deluxetable}{cccccccc}
\tablewidth{0pc}
\tablecolumns{8}
\tabletypesize{\small}
\tablenum{7}
\tablecaption{Spectral Properties.\label{t:comb}}
\tablehead{Object & Jet Feature & $\langle \alpha\rangle $ & $\dot{\alpha}$ & $\langle m_{ratio}\rangle$ 
& $\dot{m}_{ratio}$ & $\langle\Delta\chi\rangle$ & $\dot{\Delta\chi}$ \\ 
&& & (yr$^{-1}$) & & (yr$^{-1}$) & (deg.) & (deg./yr) }
\startdata
3C\,120 & KD$+$K3$+$K2/UD$+$U2 & $0.35\pm0.12$ & $-0.29\pm0.41$ & $\ldots$ & $\ldots$ & $\ldots$ & $\ldots$ \\
        & K1A$+$K1B/U1A$+$U1B & $-0.62\pm0.18$ & $0.76\pm0.56$ & $1.37\pm0.14$ & $0.49\pm0.46$ & $-6.8\pm 4.7$ & $ 6.2\pm17.5$  \\
\\
J0530$+$13 & KD$+$K2/UD$+$U2 & $-0.05\pm0.06$ & $-0.35\pm0.14$ & $0.67\pm0.14$ & $0.42\pm0.43$ & $ 6.4\pm 4.2$ & $ 4.8\pm14.9$  \\
\\
J0738$+$17 & KALL/UALL & $-0.23\pm0.13$ & $0.28\pm0.45$ & $\ldots$ & $\ldots$ & $\ldots$ & $\ldots$ \\
\\
OJ287 & KD$+$K4/UD$+$U4 & $0.31\pm0.17$ & $-0.56\pm0.57$ & $0.95\pm0.06$ & $-0.17\pm0.21$ & $-3.7\pm 2.7$ & $-3.2\pm10.2$  \\
      & K3/U3 & $-0.37\pm0.20$ & $-0.46\pm0.72$ & $1.19\pm0.06$ & $-0.28\pm0.23$ & $-1.5\pm 2.0$ & $-10.1\pm 7.3$  \\
\\
J1224$+$21 & KD$+$K4$+$K3/UD$+$U3 & $0.03\pm0.08$ & $-0.33\pm0.33$ & $1.19\pm0.08$ & $-0.24\pm0.33$ & $-2.2\pm 1.6$ & $-10.3\pm 4.3$  \\
\\
3C\,273 & KD$+$K10/UD$+$U10 & $1.01\pm0.27$ & $-1.75\pm0.54$ & $\ldots$ & $\ldots$ & $\ldots$ & $\ldots$ \\
        & K9$+$K8$+$K7/U9$+$U8$+$U7 & $-0.58\pm0.26$ & $1.06\pm0.81$ & $0.78\pm0.08$ & $0.49\pm0.19$ & $ 5.8\pm 1.5$ & $ 2.7\pm 5.6$  \\
        & K4$+$K5/U4$+$U5 & $-0.18\pm0.09$ & $0.02\pm0.33$ & $1.33\pm0.11$ & $0.14\pm0.41$ & $13.8\pm 5.9$ & $10.4\pm21.5$  \\
\\
3C\,279 & KD$+$K4/UD$+$U4 & $0.69\pm0.11$ & $-0.75\pm0.15$ & $0.62\pm0.07$ & $0.29\pm0.24$ & $ 1.8\pm 5.5$ & $-2.7\pm20.8$  \\
        & K1$+$K2/U1$+$U2 & $-0.33\pm0.06$ & $-0.29\pm0.17$ & $0.96\pm0.05$ & $-0.29\pm0.15$ & $-1.6\pm 1.9$ & $-2.3\pm 7.1$  \\
\\
J1310$+$32 & KALL/UALL & $-0.32\pm0.14$ & $0.29\pm0.49$ & $1.43\pm0.19$ & $-0.08\pm0.69$ & $-0.7\pm 4.5$ & $ 2.1\pm16.0$  \\
\\
J1512$-$09 & KD$+$K2/UD$+$U2 & $0.48\pm0.09$ & $0.01\pm0.34$ & $0.71\pm0.10$ & $0.27\pm0.45$ & $ 8.5\pm 5.4$ & $ 4.9\pm25.7$  \\
           & K1/U1 & $-0.93\pm0.13$ & $0.04\pm0.48$ & $0.94\pm0.10$ & $-0.03\pm0.39$ & $ 4.1\pm 2.4$ & $ 7.3\pm 8.3$  \\
\\
J1751$+$09 & KD/UD$+$U3 & $0.34\pm0.14$ & $0.92\pm0.28$ & $\ldots$ & $\ldots$ & $\ldots$ & $\ldots$ \\
\\
J1927$+$73 & KD$+$K3/UD$+$U3 & $0.11\pm0.06$ & $-0.32\pm0.18$ & $\ldots$ & $\ldots$ & $\ldots$ & $\ldots$ \\
           & K2/U2 & $-0.59\pm0.14$ & $0.79\pm0.34$ & $0.87\pm0.08$ & $0.02\pm0.31$ & $ 4.0\pm 4.5$ & $-12.0\pm16.0$  \\
           & K1/U1 & $-0.79\pm0.18$ & $1.24\pm0.24$ & $0.99\pm0.05$ & $0.17\pm0.17$ & $-1.8\pm 1.6$ & $ 3.3\pm 5.6$  \\
\\
J2005$+$77 & KD$+$K2/UD$+$U3$+$U2 & $0.17\pm0.11$ & $0.71\pm0.21$ & $1.20\pm0.15$ & $-0.18\pm0.57$ & $ 0.2\pm 1.8$ & $ 5.4\pm 6.1$  \\
           & K1/U1 & $-1.21\pm0.24$ & $0.31\pm0.87$ & $\ldots$ & $\ldots$ & $\ldots$ & $\ldots$ \\
\enddata
\tablecomments{Spectral properties taken between 15 and 22 GHz.  Mean
values across epoch and slopes versus
time are given for the spectral index, $\alpha$ ($S\propto\nu^{+\alpha}$), the fractional
ratio, $m_{ratio} = m_{15}/m_{22}$, and the polarization position angle difference, 
$\Delta\chi = \chi_{15} - \chi_{22}$.}
\end{deluxetable}

  %%%%%%%%%%%%%%%%%
  %%             %%   
  %%   Table 8   %% 
  %%             %%
  %%%%%%%%%%%%%%%%%

\begin{deluxetable}{cccccccc}
\rotate
\tablewidth{0pc}
\tablecolumns{8}
\tabletypesize{\scriptsize}
\tablenum{8}
\tablecaption{Empirical Uncertainty Estimates.\label{t:err}}
\tablehead{Object & Jet Feature & $\sigma I_{15}/\langle I_{15} \rangle$ 
& $\sigma I_{22}/\langle I_{22} \rangle$ & $\sigma m_{15}$ 
& $\sigma m_{22}$ & $\sigma \chi_{15}$ & $\sigma \chi_{22}$ \\ 
&&&& (\%) & (\%) & (deg.) & (deg.) }
\startdata
3C\,120 & KD$+$K3$+$K2/UD$+$U2 & $0.07$ $(+0.05,-0.08)$ & $0.08$ $(+0.06,-0.09)$ & $\ldots$ & $\ldots$ & $\ldots$ & $\ldots$ \\
        & K1A$+$K1B/U1A$+$U1B & $0.09$ $(+0.04,-0.05)$ & $0.10$ $(+0.05,-0.06)$ & $0.3$ $(+0.2,-0.2)$ & $0.6$ $(+0.4,-0.6)$ & $5.0$ $(+1.7,-2.5)$ & $10.6$ $(+3.6,-5.3)$ \\
\\
J0530$+$13 & KD$+$K2/UD$+$U2 & $0.03$ $(+0.02,-0.03)$ & $0.03$ $(+0.02,-0.03)$ & $0.5$ $(+0.3,-0.3)$ & $0.6$ $(+0.4,-0.4)$ & $5.5$ $(+4.1,-6.7)$ & $3.3$ $(+2.5,-4.1)$ \\
\\
J0738$+$17 & KALL/UALL & $0.02$ $(+0.01,-0.01)$ & $0.09$ $(+0.05,-0.05)$ & $\ldots$ & $\ldots$ & $\ldots$ & $\ldots$ \\
\\
OJ287 & KD$+$K4/UD$+$U4 & $0.06$ $(+0.04,-0.05)$ & $0.10$ $(+0.07,-0.08)$ & $0.4$ $(+0.3,-0.3)$ & $0.3$ $(+0.2,-0.3)$ & $5.2$ $(+4.0,-7.0)$ & $5.3$ $(+4.1,-7.1)$ \\
      & K3/U3 & $0.12$ $(+0.09,-0.15)$ & $0.08$ $(+0.06,-0.09)$ & $0.6$ $(+0.5,-0.9)$ & $0.5$ $(+0.4,-0.8)$ & $2.1$ $(+1.5,-2.3)$ & $3.1$ $(+2.2,-3.4)$ \\
\\
J1224$+$21 & KD$+$K4$+$K3/UD$+$U3 & $0.05$ $(+0.04,-0.06)$ & $0.03$ $(+0.02,-0.04)$ & $0.6$ $(+0.3,-0.3)$ & $0.2$ $(+0.1,-0.1)$ & $1.7$ $(+1.3,-2.2)$ & $1.7$ $(+1.3,-2.1)$ \\
\\
3C\,273 & KD$+$K10/UD$+$U10 & $0.11$ $(+0.07,-0.08)$ & $0.12$ $(+0.08,-0.08)$ & $\ldots$ & $\ldots$ & $\ldots$ & $\ldots$ \\
        & K9$+$K8$+$K7/U9$+$U8$+$U7 & $0.04$ $(+0.01,-0.02)$ & $0.17$ $(+0.05,-0.08)$ & $1.3$ $(+0.0,-0.4)$ & $0.5$ $(+0.0,-0.2)$ & $3.1$ $(+0.9,-1.3)$ & $1.9$ $(+0.6,-0.8)$ \\
        & K4$+$K5/U4$+$U5 & $0.03$ $(+0.02,-0.02)$ & $0.07$ $(+0.04,-0.04)$ & $0.4$ $(+0.2,-0.3)$ & $0.5$ $(+0.3,-0.4)$ & $3.2$ $(+1.9,-2.0)$ & $10.4$ $(+6.0,-6.6)$ \\
\\
3C\,279 & KD$+$K4/UD$+$U4 & $0.03$ $(+0.02,-0.03)$ & $0.03$ $(+0.02,-0.04)$ & $0.5$ $(+0.3,-0.3)$ & $0.8$ $(+0.4,-0.5)$ & $8.9$ $(+4.8,-5.2)$ & $4.4$ $(+2.4,-2.6)$ \\
        & K1$+$K2/U1$+$U2 & $0.05$ $(+0.01,-0.02)$ & $0.02$ $(+0.00,-0.01)$ & $1.1$ $(+0.4,-0.5)$ & $1.1$ $(+0.4,-0.5)$ & $1.5$ $(+0.0,-0.5)$ & $4.0$ $(+0.0,-1.3)$ \\
\\
J1310$+$32 & KALL/UALL & $0.03$ $(+0.00,-0.01)$ & $0.12$ $(+0.00,-0.04)$ & $0.4$ $(+0.2,-0.2)$ & $0.3$ $(+0.1,-0.2)$ & $3.4$ $(+2.5,-4.3)$ & $5.6$ $(+4.2,-7.1)$ \\
\\
J1512$-$09 & KD$+$K2/UD$+$U2 & $0.05$ $(+0.03,-0.04)$ & $0.05$ $(+0.03,-0.04)$ & $0.5$ $(+0.1,-0.2)$ & $0.8$ $(+0.1,-0.3)$ & $10.0$ $(+7.6,-12.5)$ & $9.7$ $(+7.4,-12.1)$ \\
           & K1/U1 & $0.06$ $(+0.01,-0.02)$ & $0.10$ $(+0.01,-0.04)$ & $1.1$ $(+0.0,-0.3)$ & $1.0$ $(+0.0,-0.3)$ & $3.0$ $(+1.9,-2.2)$ & $3.7$ $(+2.4,-2.8)$ \\
\\
J1751$+$09 & KD/UD$+$U3 & $0.10$ $(+0.07,-0.08)$ & $0.09$ $(+0.06,-0.07)$ & $\ldots$ & $\ldots$ & $\ldots$ & $\ldots$ \\
\\
J1927$+$73 & KD$+$K3/UD$+$U3 & $0.04$ $(+0.00,-0.01)$ & $0.02$ $(+0.00,-0.01)$ & $\ldots$ & $\ldots$ & $\ldots$ & $\ldots$ \\
           & K2/U2 & $0.04$ $(+0.02,-0.02)$ & $0.07$ $(+0.03,-0.04)$ & $0.3$ $(+0.0,-0.2)$ & $1.5$ $(+0.2,-0.8)$ & $3.9$ $(+0.0,-1.3)$ & $6.9$ $(+0.0,-2.3)$ \\
           & K1/U1 & $0.02$ $(+0.02,-0.02)$ & $0.05$ $(+0.03,-0.04)$ & $0.5$ $(+0.2,-0.3)$ & $1.8$ $(+0.6,-0.9)$ & $1.9$ $(+0.0,-0.6)$ & $3.5$ $(+0.0,-1.2)$ \\
\\
J2005$+$77 & KD$+$K2/UD$+$U3$+$U2 & $0.02$ $(+0.02,-0.03)$ & $0.01$ $(+0.01,-0.02)$ & $0.6$ $(+0.3,-0.3)$ & $0.9$ $(+0.5,-0.5)$ & $3.3$ $(+2.1,-2.4)$ & $2.8$ $(+1.8,-2.1)$ \\
           & K1/U1 & $0.06$ $(+0.00,-0.02)$ & $0.19$ $(+0.00,-0.07)$ & $\ldots$ & $\ldots$ & $\ldots$ & $\ldots$
\enddata
\tablecomments{These represent empirical estimates of the uncertainty in single-epoch measurements 
of the flux ($I$), the fractional polarization ($m$), and polarization position angle ($\chi$) of jet
features in our survey.  Estimates are made for both 15 and 22 GHz.}
\end{deluxetable}

\end{document}